\documentclass{aa}
\usepackage{graphicx}
\def\eqnum#1{}
\begin{document}

   \title{Robust quantitative measures of cluster X-ray morphology, and comparisons between cluster characteristics}

   \subtitle{}

   \author{Yasuhiro Hashimoto\inst{1}
           Hans B\"ohringer\inst{1}
           J. Patrick Henry \inst{1,2}
          G. Hasinger\inst{1}
          \and G. Szokoly \inst{1}
          }

   \offprints{Y.\ Hashimoto,\\ e-mail: hashimot@mpe.mpg.de}

    \institute{Max-Planck-Institut f\"ur extraterrestrische Physik,
              Giessenbachstrasse
              D-85748 Garching, Germany
          \and
              Institute for Astronomy, University of Hawaii, 2680 Woodlawn Drive, Honolulu, Hawaii 96822, USA
          }

   \date{Received ; accepted}

   \authorrunning{Hashimoto et al.}
   \titlerunning{Morphology measures}

\abstract
{}
{To investigate the possible relationships between dynamical status 
and other important characteristics of galaxy clusters, 
we conducted a study of X-ray cluster morphology
using a sample of 101 clusters  
at redshift z$\sim$0.05-1 taken from the Chandra archive.}
{The X-ray morphology is quantitatively characterized by 
a series of objectively measured simple statistics of
the X-ray surface brightness distribution,
which are designed to be  robust against variations of image quality
caused by various exposure times and various cluster redshifts.
Using these measures, we quantitatively investigated the relationships between
the cluster X-ray morphology and various other cluster characteristics.}
{We found:
   (1) Our measures are robust against various image quality
        effects introduced by exposure time difference,
        and various cluster redshifts.
   (2) 
      The distorted and non-distorted clusters occupy well-defined loci in the L-T plane,
      demonstrating the measurements of the global luminosity and temperature  for distorted
   clusters should be interpreted with caution, or alternatively, a rigorous morphological 
  characterization is  necessary when we use a sample of clusters with heterogeneous morphological 
  characteristics to investigate the L-T or other related scaling relations.
(3) 
Ellipticity and Off-center show no evolutionary effects between high and low redshift cluster subsets,
 while there may be a hint of weak evolutions for the Concentration and Asymmetry,
 in such a way that high-z clusters show more distorted morphology.
   (4) 
       No correlation is found between X-ray morphology and X-ray luminosity or
       X-ray morphology and
       X-ray temperature of clusters, implying that  interaction of clusters
       may not enhance or decrease the luminosity or temperature of clusters
       for extended period of time.
   (5) 
        Clusters are scattered and occupying various places in the plane composed of
       two X-ray morphological measures, showing a wide variety of characteristics.
   (6) 
       Relatively strong correlations in Asymmetry-Concentration 
      and Offcenter-Concentration plots indicate that low concentration clusters generally 
      show high degree of asymmetry or skewness, illustrating the fact that there are not 
      many highly-extended smooth symmetric clusters. Similarly, a correlation between 
      Asymmetry and Ellipticity may imply that there are not many highly-elongated but 
      otherwise smooth symmetric clusters.
}
{}


 \maketitle

\section{Introduction}

 Over the past decade, studies have provided evidence that a significant fraction
  of galaxy clusters have undergone recent mergers (e.g. Geller \& Beers 1982; 
Dressler \& Shectman 1988). 
These mergers are observed as disturbed cluster morphologies.
The important connection between the morphologies of galaxy clusters and
   cosmological parameters has received much recent attention
   (Richstone, Loeb, \& Turner 1992; Evrard et al. 1993; Mohr et al. 1995).
   This connection has generally been formulated in terms of the frequency 
   of `substructure' in clusters and
from qualitative measures of the frequency of substructure in clusters, investigators have
    attempted to determine $\Omega_m$ (e.g. Richstone et al. 1992) and the 
    power spectrum of primordial
    density fluctuations (e.g. David et al. 1993) 
    by comparison to Press \& Schechter (1974)
    type predictions of the distribution of collapsed objects.

    Methods to quantify structures at optical wavelengths
      have mostly used both the
     distribution of cluster galaxies, and lensing.
      However, the distribution study  requires  
       a large number of galaxies,
     and is more susceptible to contamination
     from foreground and background objects.
     Lensing is also sensitive to this contamination, 
     and does not have good spatial resolution except for the
     central region of a cluster.
     An alternative method comes from X-ray wavelengths, because
      cluster mergers compress and heat the intracluster gas, 
        and this can be measured as distortions of the spatial
       distribution of X-ray surface brightness and temperature.
       Moreover,
      X-ray emissivity is proportional to the square
   of the electron density, and therefore less affected
   by the superposed structures than optical data.
     Jones \& Forman (1999) visually examined 208
      clusters observed with Einstein X-ray satellite and 
      separated these clusters into six morphological classes.
      They found that about 40\% of their clusters 
     displayed some type of `substructure'.
 
    However, a more quantitative measure of cluster structure
             at X-ray wavelengths
             is desirable to 
             quantitatively test various
             scenarios related to clusters, including cosmology.
   Using $Einstein$ images,
   Mohr et al. (1995) measured emission-weighted centroid variation, 
   axial ratio, orientation, and radial falloff for a sample of 65 clusters,
    while several other studies  used 
      ellipticity (e.g. McMillan et al. 1989; Gomez et al. 1997; 
      Kolokotronis et al.
      2001; Melott et al. 2001; Plionis 2002).
      Buote \& Tsai (1995, 1996) used a
      power ratio method for 59 clusters observed with $ROSAT$,
      while Schuecker et al. (2001) conducted a study of 470 clusters
      from $ROSAT$-ESO Flux-Limited X-ray (REFLEX) cluster survey (B\"ohringer et al. 2000),
      using sophisticated statistics, such as Fourier elongation test,
      Lee test, and $\beta$ test.

    Despite the success of these studies,
     all of them are unfortunately limited to clusters  
     in the nearby universe (z $<$ 0.3), where we may expect to see
     less frequent
     morphological distortions, and
     little evolutionary effect.  
    This is due to the fact that, 
    until recently, only a small number of high-z clusters have been 
     known, or
     observed with sufficient depth and sufficient spatial resolution.
     With the advent of big-aperture satellites 
     equipped with high spatial-resolution instruments,
   such as $Chandra$ and $XMM$-$Newton$,
     together with newly-available lists of distant clusters
     generated based on various deep cluster surveys,
     it is finally possible to extend the morphological study to 
     higher redshifts.
     Indeed, Jeltema et al. (2005) have recently extended the 
     power ratio method of 
     Buote \& Tsai 
      to 40 clusters at z=0.15-0.9 using Chandra data, and reported 
      the evolution of cluster morphology between two redshift bins (z$<$0.5 and z$>$0.5).

      Extending the morphological study to high redshift is important but a difficult task because
      of inevitable low data quality associated with high-z clusters.
      Conventional methods for characterizing the cluster X-ray morphology 
      are often sophisticated
      and some methods have an advantage of being more directly related  
      to a particular  characteristic of a cluster, such as mass, 
      dark matter content, or gravitational potential. 
      However, most of these conventional methods are
      originally developed to analyze the low redshift clusters 
      with high data quality,
      and, perhaps because of their intrinsic sophistication, ofter require 
      many
      photon counts, making the measures rapidly uncertain or unmeasurable 
      as the data quality decreases. 
      Moreover,
      these methods also often require
      some 
      interactive processes, and thus,
      are not suitable for the investigation involving a large dataset
      with a wide variety of image quality 
      where various systematics should be 
      treated and removed in a consistent manner.
      Although
      it is important to try to extend the sophisticated methods 
    to high redshift,
      a complementary  study using the robust measures of 
       the cluster morphology,  
       less sensitive to variation in the data quality and suitable for
       a large dataset, is much needed, 
      to enable us to study the low-z and high-z universe in a uniform manner.

Here we report our study of X-ray cluster morphology 
using a sample of 101 clusters at redshift z$\sim$0.05-1 taken from the
Chandra archive.
The X-ray morphology is quantitatively characterized by 
a series of 
objectively-measured simple statistics of 
X-ray surface brightness,
which are designed to be  robust against 
  variations of image quality
  caused by various exposure times
   and various cluster redshifts.
Using these measures, we quantitatively investigated the relationships between
   the cluster X-ray morphology and various other cluster characteristics.

This paper is organized as follows. In sec 2, we describe our sample and data preparation,
while in sec 3, details of our measures are described,
and in sec 4, uncertainty and systematics are investigated.
Sec. 5 summarizes our results.
Throughout the paper, we use $H_{o}$ = 70 km s$^{-1}$ Mpc$^{-1}$,
$\Omega_{m}$=0.3, and $\Omega_{\Lambda}$=0.7, unless otherwise stated. 

\section{Sample  \& data preparation}

\begin{table}
\tiny
\caption[]{The Sample}
\begin{tabular}{lcccccccc}
\hline
\hline
\noalign{\smallskip}
Name       & z   & Lbol           & Tx & r$_c$ & $\beta$ & ref \\
         &       & 10$^{44}erg/s$ &  keV &  kpc &     &  \\
\noalign{\smallskip}
\hline
\noalign{\smallskip}
A3562   &       0.050&3.56&5.2&99&0.472&a\\
A85     &       0.052&13.15&6.9&83&0.532&a\\
HydraA  &       0.052&6.19&4.3&50&0.573&a\\
A754    &       0.053&6.44&9.5&239&0.698&a\\
A2319   &       0.056&35.80&11.8&170&0.550&e\\
A3158   &       0.059&6.92&5.8&269&0.661&a\\
A3266   &       0.059&12.81&8.0&564&0.796&a\\
A2256   &       0.060&5.05&6.6&587&0.914&a\\
A1795   &       0.063&14.73&7.8&78&0.596&a\\
A399    &       0.072&9.75&7.0&450&0.713&a\\
A2065   &       0.072&6.72&5.5&690&1.162&a\\
A401    &       0.075&16.53&8.0&246&0.613&a\\
ZwCl1215+0400 &       0.075&--&--&--&--&\\
A2029   &       0.077&27.84&9.1&83&0.582&a\\
A2255   &       0.080&12.53&6.9&593&0.797&a\\
A1651   &       0.083&10.35&6.1&181&0.643&a\\
A478    &       0.088&27.49&8.4&98&0.613&a\\
RXJ1844+4533  &       0.091&--&--&--&--&\\
A2244   &       0.102&12.11&7.1&126&0.610&a\\
RXJ0820.9+0751  &       0.110&--&--&--&--&\\
A2034   &       0.110&12.51&7.9&290&0.690&d\\
A2069   &       0.115&--&--&--&--&\\
RXJ0819.6+6336  &       0.119&--&--&--&--&\\
A1068   &       0.139&7.79&3.6&25&0.520&b\\
A2409   &       0.147&--&--&--&--&\\
A2204   &       0.152&40.57&7.2&67&0.597&a\\
HerculesA       &       0.154&--&--&--&--&\\
A750    &       0.163&--&--&--&--&\\
A2259   &       0.164&--&--&--&--&\\
RXJ1720.1+2638  &       0.164&25.58&5.6&--&--&i\\
A1201   &       0.169&--&--&--&--&\\
A586    &       0.171&11.80&7.0&119&0.680&b\\
A2218   &       0.171&12.10&7.6&165&0.580&b\\
A1914   &       0.171&33.75&10.5&231&0.751&a\\
A2294   &       0.178&--&--&--&--&\\
A1689   &       0.184&36.62&9.2&163&0.690&a\\
A1204   &       0.190&--&--&--&--&\\
MS0839.8+2938   &       0.194&4.51&3.4&40&0.560&b\\
A115$^1$    &       0.197&13.50&5.8&16&0.400&b\\
A520    &       0.203&22.89&8.59&--&--&j\\
A963    &       0.206&12.00&6.8&71&0.500&b\\
RXJ0439.0+0520  &       0.208&--&--&--&--&\\
A2111   &       0.211&9.50&6.9&149&0.490&b\\
A1423   &       0.213&--&--&--&--&\\
ZwCl0949+5207 &       0.214&8.56&4.0&41&0.530&b\\
MS0735.6+7421   &       0.216&9.56&4.5&27&0.460&b\\
A773    &       0.217&15.60&8.1&190&0.660&b\\
A2261   &       0.224&23.90&6.6&62&0.510&b\\
A1682   &       0.226&11.00&6.4&384&0.750&b\\
A1763   &       0.228&18.40&8.1&168&0.490&b\\
A2219   &       0.228&38.90&9.2&189&0.560&b\\
A267    &       0.230&12.00&5.5&141&0.620&b\\
A2390   &       0.233&40.80&9.2&44&0.460&b\\
RXJ2129.6+0006  &       0.235&18.30&5.7&42&0.510&b\\
RXJ0439.0+0715  &       0.244&--&--&--&--&\\
A2125   &       0.247&5.96&3.2&--&--&l\\
A68     &       0.255&15.60&6.9&177&0.610&b\\
ZwCl1454+2233 &       0.258&18.30&4.4&43&0.590&b\\
A1835   &       0.258&48.10&7.4&46&0.550&b\\
A1758$^2$   &       0.280&23.60&9.0&1149&3.000&h\\
A697    &       0.282&30.90&8.2&198&0.580&b\\
ZwCl1021+0426 &       0.291&51.30&6.41&--&--&k\\
A781    &       0.298&--&--&--&--&\\
A2552   &       0.299&--&--&--&--&\\
A1722   &       0.327&11.10&5.8&92&0.510&b\\
MS1358.4+6245   &       0.328&9.43&5.5&40&0.460&b\\
RXJ1158.8+0129  &       0.352&--&--&--&--&\\
A370    &       0.357&10.80&6.6&231&0.540&b\\
RXJ1532.9+3021  &       0.361&32.90&4.9&47&0.590&b\\
MS1512.4+3647   &       0.372&4.10&2.8&42&0.540&b\\
RXJ0850.2+3603  &       0.374&--&--&--&--&\\
RXJ0949.8+1708  &       0.382&--&--&--&--&\\
ZwCl0024+1652   &       0.390&3.54&4.5&59&0.410&f\\
RXJ1416+4446    &       0.400&5.43&3.7&26&0.438&c\\
RXJ2228.5+2036  &       0.412&--&--&--&--&\\
MS1621+2640     &       0.426&10.92&6.8&185&0.563&c\\
RXJ1347-1145    &       0.451&116.75&10.3&38&0.571&c\\
RXJ1701+6412    &       0.453&6.42&4.5&13&0.396&c\\
3c295   &       0.460&14.07&4.3&31&0.553&c\\
\noalign{\smallskip}
\hline
\end{tabular}
\end{table}

\begin{table}
\tiny
{\bf Table 1} (Continued)
\\~
\\~
\begin{tabular}{lcccccccc}
\hline
\hline
\noalign{\smallskip}
Name       & z   & Lbol           & Tx & r$_c$ & $\beta$ & ref \\
         &       & 10$^{44}erg/s$ &  keV &  kpc &     &  \\
\noalign{\smallskip}
\hline
\noalign{\smallskip}
RXJ1641.8+4001 &       0.464&--&--&--&--&\\
CRSSJ0030.5+26 &       0.500&--&--&--&--&\\
RXJ1525+0957    &       0.516&6.92&5.1&229&0.644&c\\
MS0451-0305     &       0.540&50.94&8.0&201&0.734&c\\
MS0016+1609     &       0.541&53.27&10.0&237&0.685&c\\
RXJ1121+2326    &       0.562&5.45&4.6&427&1.180&c\\
RXJ0848+4456    &       0.570&1.21&3.2&97&0.620&c\\
MS2053-0449     &       0.583&5.40&5.5&99&0.610&c\\
RXJ0542-4100    &       0.634&12.15&7.9&132&0.514&c\\
RXJ1221+4918    &       0.700&12.95&7.5&263&0.734&c\\
RXJ1113-2615    &       0.730&4.43&5.6&89&0.639&c\\
RXJ2302+0844    &       0.734&5.45&6.6&96&0.546&c\\
MS1137+6625     &       0.782&15.30&6.9&111&0.705&c\\
RXJ1350+6007    &       0.810&4.41&4.6&106&0.479&c\\
RXJ1716+6708    &       0.813&13.86&6.8&121&0.635&c\\
MS1054-0321     &       0.830&28.48&10.2&511&1.375&c\\
RXJ0152-1357$^3$   &       0.835&18.40&6.5&--&--&c\\
WGA1226+3333    &       0.890&54.63&11.2&123&0.692&c\\
RXJ0910+5422    &       1.106&2.83&6.6&147&0.843&c\\
RXJ1053.7+5735$^4$ &     1.134&2.80&3.9&--&--&g\\
RXJ1252-2927    &       1.235&5.99&5.2&77&0.525&c\\
RXJ0849+4452    &       1.260&2.83&5.2&128&0.773&c\\
\noalign{\smallskip}
\hline
\end{tabular}
References:
a: Reiprich \& B\"ohringer(2002); 
b: Ota \& Mitsuda(2004); 
c: Ettori et al(2004); 
d: Kempner et al(2003); e: O'Hara et al(2004);
f: Ota et al. (2004) (R$_c$ \& $\beta$ from Ota \& Mitsuda(2004); 
g: Hashimoto et al(2004);
h: David \& Kempner(2004);
i: Mazzota et al. (2001);
j: Wu et al. (1999);
k: Allen (2000);
l: Wang et al. (2004).
Comments:
(1): Distant southern component A115S is excluded;
(2): Distant southern component A1758S is excluded;
(3): RXJ0152, both north and south components are treaded as one cluster;
(4): RXJ1054, both east and west components are treaded as one cluster.
\end{table}

Almost all clusters are selected from flux-limited X-ray surveys, and data are
taken from the Chandra ACIS archive.
A lower limit of z = 0.05 or 0.1 is placed on the redshift to ensure that
a cluster is observed with sufficient field-of-view with ACIS-I or ACIS-S, respectively.
The majority of our sample comes from 
the $ROSAT$ Brightest Cluster Sample
    (BCS; Ebeling et al 1998), and the
Extended $ROSAT$ Brightest Cluster Sample (EBCS; Ebeling et al. 2000).
 The BCS sample includes 201 clusters, with the flux limit of
     4.4$\times$10$^{-12}$ erg s$^{-1} cm^{-2}$ (0.1-2.4 keV). The authors 
estimated a sample completeness of 90 \% for the 201 BCS clusters, 
and 75 \% for the EBCS clusters.
When combined with EBCS, the BCS clusters represent one of the 
largest and most complete X-ray selected
cluster samples, and they are currently the most frequently 
observed by $Chandra$.  
As of 2005 October, 55 BCS + 13 EBCS (hereafter BCS) clusters 
with z $>$ 0.05 (ACIS-I), 
or z $>$ 0.1 (ACIS-S), are publicly
available in the $Chandra$ archive. 
Additionally we included all clusters from the X-ray flux limited
sample of Edge et al. (1990) at z $>$ 0.05 or 0.1 not in the BCS that
were observed with the Chandra ACIS. 
This added 12 more clusters.
The Edge et al. sample 
is estimated to
be $\sim$90\% complete, and 
contains
the 55 brightest clusters from $EXOSAT$, $HEAO-1$, and $Einstein$. 

To extend our sample to higher redshifts, 
additional high-z clusters are selected from various deep  surveys;
10 of these clusters are selected from the
      $ROSAT$ Deep Cluster Survey (RDCS: Rosati et al. 1998),
10 from the $Einstein$ Extended Medium Sensitivity Survey 
(EMSS; Gioia et al. 1990; Henry et al. 1992), 
14 from the 160 Square Degrees $ROSAT$  Survey (Vikhlinin et al. 1998),
2 from the Wide Angle $ROSAT$ Pointed Survey (WARPS; Perlman et al. 2002),
and 1 from  the North Ecliptic Pole survey (NEP; Gioia et al 1999),
RXJ1054 was discovered by Hasinger et al. (1998), 
RXJ1347 was discovered in the $ROSAT$ All Sky Survey (Schindler et al. 1995),
and 3C295 has been mapped with $Einstein$ (Henry \& Henriksen 1986).

The resulting sample we processed contains 120 clusters.
At the final stage of our data processing, to employ our full analysis, 
we further applied a selection based on the total counts of cluster emission,
(for details, please see Sec. 4), eliminating clusters with very 
low signal-to-noise ratio.
Clusters whose center is too close to the edge of the CCD are also removed.
The resulting final sample contains
101 clusters with redshifts between 0.05 - 1.26 (median z = 0.226), 
and luminosity between 1.0 $\times$ 10$^{44}$ -- 1.2 $\times$ 10$^{46}$ erg/s 
(median 8.56 $\times$ 10$^{44}$ erg/s) (Fig.1).
The final cluster sample together with their published redshifts and 
bolometric luminosities, if available, as well as  $\beta$s and  core radii, 
are listed in Table 1.

We reprocessed the level=1 event file retried from the archive
 using CIAO v3.1. and CALDB v2.29.
For observations taken in the VFAINT mode, we ran the script
acis\_process\_events to flag probable 
background events, using the information in a 5 $\times$ 5
event island.
We also applied the charge transfer inefficiency (CTI) correction
and the time dependent gain correction
for ACIS-I data, 
when the temperature of the focal plane at the time of the observation
was 153 K. 
The data were filtered to include only the standard
event grades 0,2,3,4,6 and status 0,
then multiple pointings were merged, if any.
We eliminated time intervals of high background count rate
by performing a 3 $\sigma$ clipping of the
background level using the script analyze\_ltcrv. 
To prepare the images for analysis,
we selected photons in the observed-frame 0.7-8.0 keV 
and rest-frame 0.7-8 keV bands
initially binned into 0.5 " pixel (see Sec 4 for the binning scheme of
later analysis steps). 
We 
corrected the images for exposure variations across the field of view, detector response and telescope vignetting. 

We detected point sources using the CIAO routine 
celldetect with signal-to-noise threshold for source detection of three.
 An elliptical background annulus region was defined around each source such that 
 its outer major and minor axes became three times of the source region.
 We removed the detected sources, except for a source at the center of the cluster which was 
 mostly the peak of the surface brightness distribution rather than a real point source, 
and filled the source regions
 using the CIAO tool dmfilth. 
The images were then smoothed  with Gaussian $\sigma$=5".
We have decided to perform the smoothing, as well as
    the total-count cut (see sec 4.2.3 for detail) to avoid
    the case where we have an image
    predominantly with zero count pixels,
    which
    makes
    the exposure map correction difficult,
     as well as the determination of the object region (see below),
    and for the investigation of various systematics (see sec 4).
    We found that
    the choice of smoothing-sigma
    hardly affects our robust morphological values
    (please see section 4.2.2. for detail).

 Some clusters have a chip gap, or bad column inside
             the extracted cluster region.
             Most of these clusters, however, 
             were observed with multiple pointings, thus those artifacts
             were reasonably corrected by exposure map.
             For those clusters with a single pointing,
             the artifacts were all crossing the cluster region
             far (typically more than 2 arcmin) from the cluster center.
             Using clusters with multiple pointing observations,
             we discovered that,
             the effect of these artifacts are negligible
             on our morphological measures,
             particularly after the smoothing.

We decided to use isophotal contours to characterize
an object region, instead of a conventional
circular aperture, 
because we did not want to introduce any
possible bias in the shape of an object. 
To define constant metric scale to all clusters, 
we adjusted an extracting threshold in such a way that
the square root of detected object area times a constant was 0.5 Mpc,
i.e. const$\sqrt{area}$ = 0.5 Mpc.
We chose to use the const =1.5, because
the isophotal limit of a detected object was best represented by
this value.

\begin{figure}
 \resizebox{\hsize}{!}{\includegraphics[height=2cm,width=4cm,clip,angle=0]{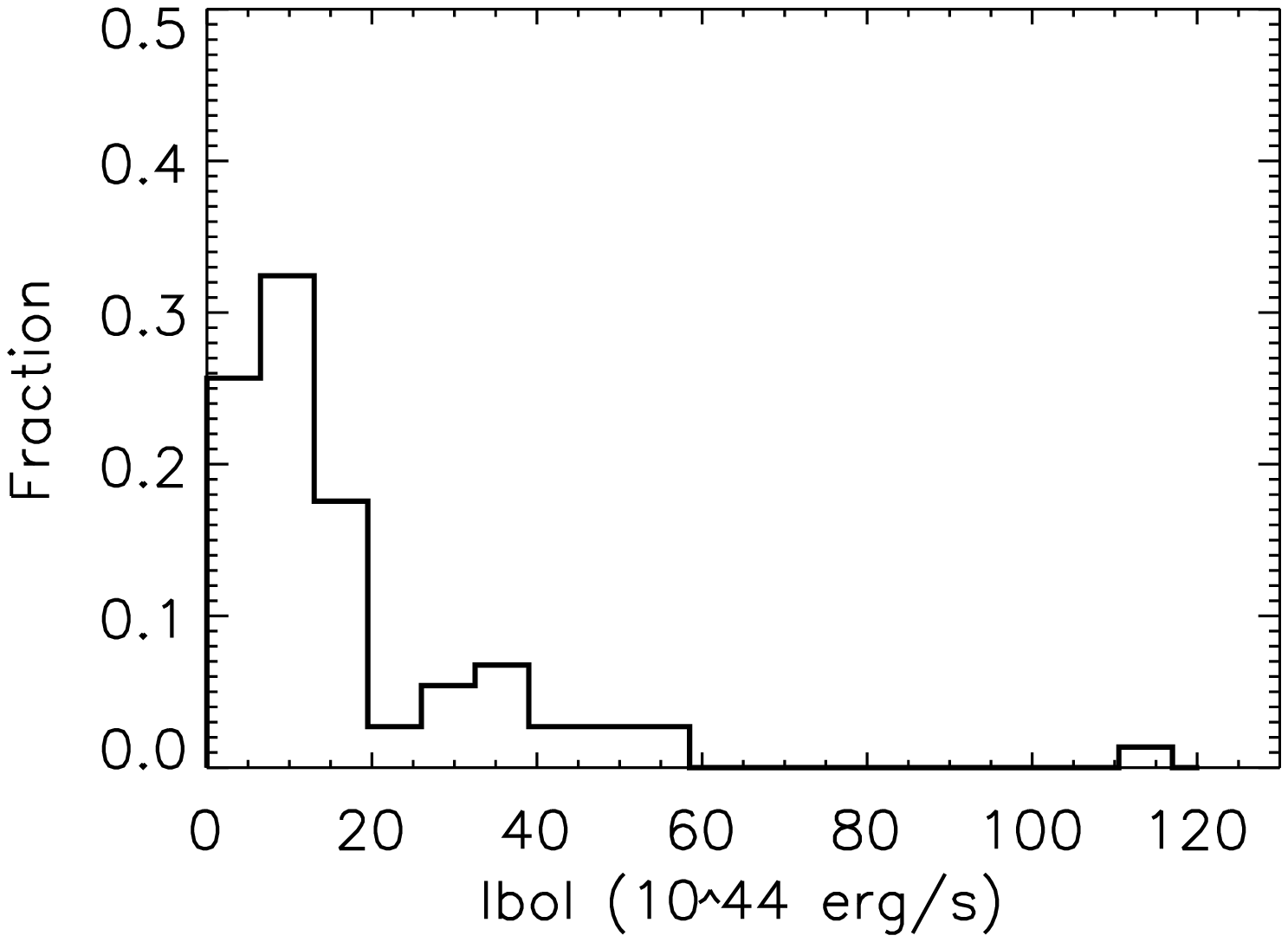}}
 \resizebox{\hsize}{!}{\includegraphics[height=2cm,width=4cm,clip,angle=0]{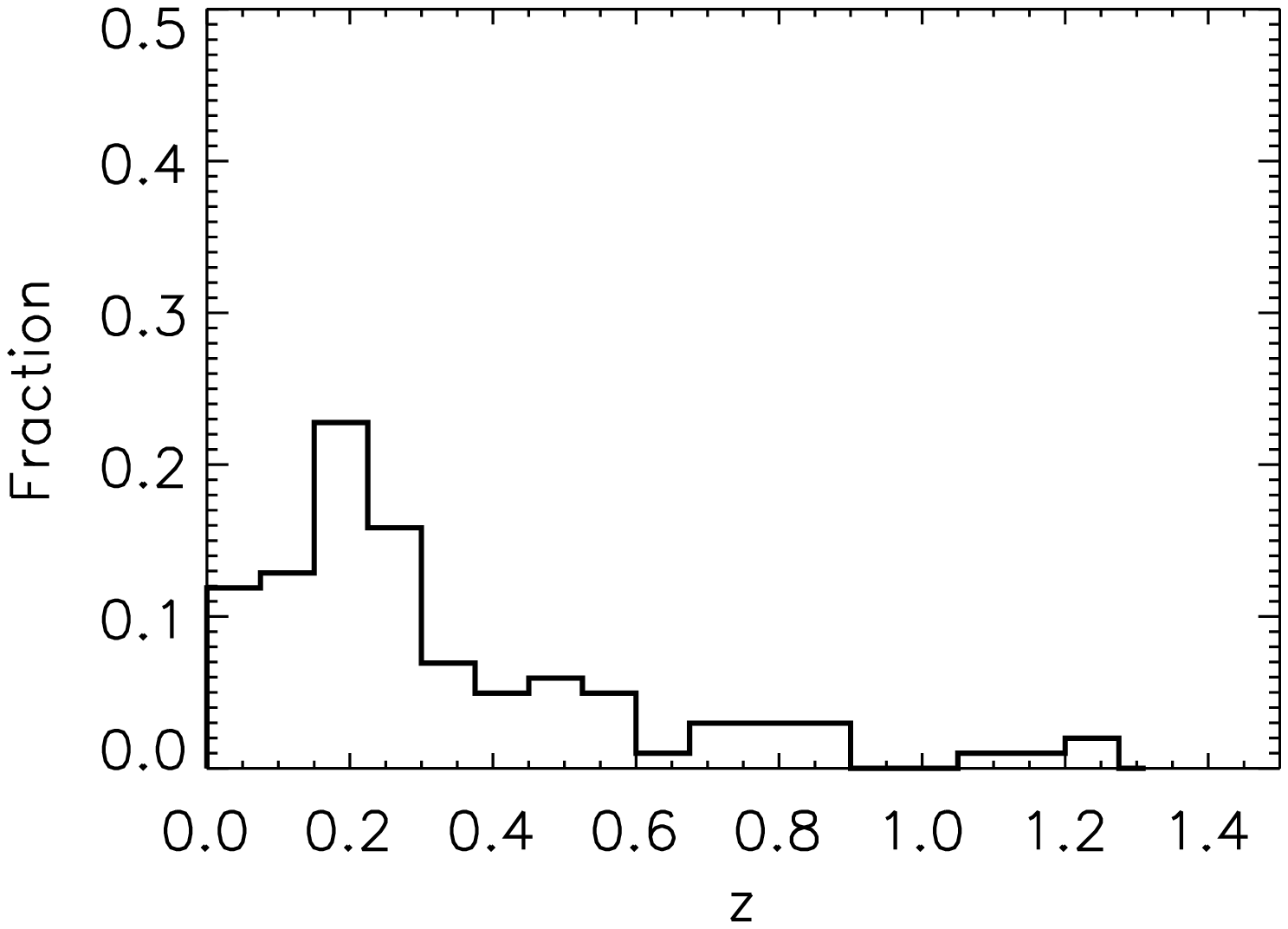}}
 \caption{
Distributions of
the X-ray bolometric luminosity (Lbol) and redshift (z) for our sample.
}
\label{FigTemp}
\end{figure}

\section{Morphological measures}
\subsection{Centroid \& second moments}
  Centroid and centered-second moments are 
 computed using the first and second order moments of the profile:\\
 
    \begin{eqnarray}
      \overline{x}=\frac{\sum\limits_{i \in S}I_i x_i}{\sum\limits_{i \in S}I_i}, \;\;      
      \overline{y}=\frac{\sum\limits_{i \in S}I_i y_i}{\sum\limits_{i \in S}I_i}
    \end{eqnarray}
    \begin{eqnarray}
      \overline{x^2}=\frac{\sum\limits_{i \in S}I_i x_i^2}{\sum\limits_{i \in S}I_i}-\overline{x}^2, \;\;      
      \overline{y^2}=\frac{\sum\limits_{i \in S}I_i y_i^2}{\sum\limits_{i \in S}I_i}-\overline{y}^2, \;\;      
      \overline{xy}=\frac{\sum\limits_{i \in S}I_i x_i y_i}{\sum\limits_{i \in S}I_i}-\overline{x}\, \overline{y} 
    \end{eqnarray}
    where $x_i$ and $y_i$ are the x-coordinate and y-coordinate of a pixel $i$ of value $I_i$ inside area $S$ of an object.

\subsection{Ellipticity }
Ellipticity is simply defined by the ratio of semi-major (A) and semi-minor
axis (B) lengths as: 
    \begin{eqnarray}
         Elli & = & 1  -B/A  
    \end{eqnarray}
   where 
   A and  B  are defined by the
   maximum and minimum spatial {\it rms} of the 
   object profile along any direction and computed by the formula: \\ 
    \begin{eqnarray}
         A^2 & = &\frac{\overline{x^2}+\overline{y^2}}{2}+\sqrt{\left(\frac{\overline{x^2}-\overline{y^2}}{2}\right)^2 + \overline{xy}^2}  \\
         B^2 & = &\frac{\overline{x^2}+\overline{y^2}}{2}-\sqrt{\left(\frac{\overline{x^2}-\overline{y^2}}{2}\right)^2 + \overline{xy}^2}  
    \end{eqnarray}

\subsection{Off-center}
The degree of off-center
is determined by the distance between the centroid and maximum intensity
peak:
    \begin{eqnarray}
      Offcen = \frac{\sqrt{(x_p-\overline{x})^2+(y_p-\overline{y})^2}}{3(A+B)} 
    \end{eqnarray}
    where, flux peaks in a pixel at x${_p}$, y${_p}$.

\subsection{Concentration}
  The degree of concentration of the surface brightness profile is
  measured using a method described in Hashimoto et al. (1998), 
  and is defined by the ratio
  between central 30\% and whole 100\% elliptical apertures as:
\begin{eqnarray}
     Conc & = & \frac{\sum\limits_{r_i <0.3}I(r_i)}{\sum\limits_{r_i<1.0}I(r_i)} 
\end{eqnarray}
   where, $r_i$ is a position of a pixel $i$ in a parameter which scales the ellipse, in unit of A (or B), and computed using the position angle 
of each pixel  ($\theta_i$): \\
\begin{eqnarray}
   r^2_i& =&[(\frac{cos^2\theta_i}{A} +\frac{sin^2\theta_i}{B})(x_i-\overline{x}) \nonumber \\ 
 & & +(\frac{sin\theta_i cos\theta_i}{A}-\frac{sin\theta_i cos\theta_i}{B})(y_i-\overline{y})]^2 \nonumber \\
& & +[(\frac{sin\theta_i cos\theta_i}{A} -\frac{sin\theta_i cos\theta_i}{B})(x_i-\overline{x}) \nonumber \\ 
& &  +(\frac{sin^2\theta_i}{A}+\frac{cos^2\theta_i}{B})(y_i-\overline{y})]^2
\end{eqnarray}

\subsection{Asymmetry}
   To measure the degree of  asymmetry of the profile around the centroid,
   an asymmetry index  is computed as:
\begin{equation}
     Asym=\frac{\frac{1}{2}\sum\limits_{i \in S}|I(x_i,y_i)-I(2x_i-\overline{x2},2y_i-\overline{y2})|}{\sum\limits_{i \in S} I(x_i,y_i)} 
\end{equation}
where,
\begin{eqnarray}
      \overline{x2}=\frac{\sum\limits_{i \in S}I_i^2 x_i}{\sum\limits_{i \in S}I_i^2}, \;\;      
      \overline{y2}=\frac{\sum\limits_{i \in S}I_i^2 y_i}{\sum\limits_{i \in S}I_i^2}
    \end{eqnarray}
  After  testing various centroids, we have chosen to use the second 
  order centroid $\overline{x2}$ and  $\overline{y2}$
  to make the asymmetry measure less sensitive to the 
  very faint outer structure than the case using simple 
  $\overline{x}$ and  $\overline{y}$. 

\section{Uncertainty and systematics}
\subsection{Uncertainty}
We applied a Monte Carlo simulation to estimate the
   uncertainties in our measures caused by point sources
   and Poisson noise.
  For each cluster image, starting from the image used for real analysis,
  we added random artificial point sources consistent with $Chandra$
  PSF and numbers consistent with the logN-logS given by 
  Campana et al. (2001).
   We chose to use the real image instead of $\beta$ model,
   since many of the clusters were not well descried by the $\beta$ model.
   Poisson noise was then added to the images. 
  We then excised the bright point sources again 
  exactly the same way as the real analysis, 
  followed by  
   smoothing.
   For each cluster we performed 100 such realizations, and
   the
   morphological measures were  computed for each realization.
   We then simply defined our 1 sigma error for each measure  to be 
   $rms$ of the distribution of each measure. 

\subsection{Systematics}
\subsubsection{Exposure time effect}
 To investigate the  systematic effect of various 
exposure times
on the morphological measures,
one of the  standard approaches is to 
simulate lower signal-to-noise data caused by a shorter 
integration time 
by scaling the real data  by the exposure time, and adding Poisson noise 
 taking each pixel value as the
mean for a Poisson distribution.
However, this simple rescaling and adding noise process will 
produce an excessive amount of Poisson noise, because of the intrinsic noise
already present in the initial real data.
Meanwhile,
using a model image with no intrinsic noise, 
instead of the real data, will not have this
problem, however, here we need to approximate the various characteristics of 
a model to  complicated characteristics of a real cluster and this is 
an almost impossible task, 
particularly for a dynamically unsettled distorted cluster.
To circumvent this problem, 
we decided to use the real cluster data and employed
a series of `adaptive scalings'  accompanied by a noise adding
process.
Namely, to simulate data with integration t=t1, an original unsmoothed image 
(including the background) taken with original
integration time t0 was at first rescaled by a factor R$_0$/(1-R$_0$), instead of simple R$_0$,  where R$_0$=t1/t0, t0$>$t1.
That is, an intermediate scaled image I$_1$ was created from the original
unsmoothed image I$_0$ by:
\begin{eqnarray}
  I_1   &=& I_0\frac{R_0}{(1-R_0)}     
\end{eqnarray}
 
Poisson noise was then added to this rescaled image by taking each pixel value as the
mean for a Poisson distribution and then randomly selecting a new pixel value from 
that distribution. This image was then rescaled again by a factor (1-R$_0$)
to produce an image whose {\it signal} is scaled by R$_0$ relative 
to the original image, but
its {\it noise} is approximately scaled by $\sqrt{R_0}$, 
assuming the intrinsic noise initially present in the real data is Poissonian.
(The derivation of this scaling is described in the Appendix.)
Finally, the image was smoothed with Gaussian of $\sigma$=5".
\begin{figure}
 \resizebox{\hsize}{!}{\includegraphics[height=4cm,width=2cm,clip,angle=90]{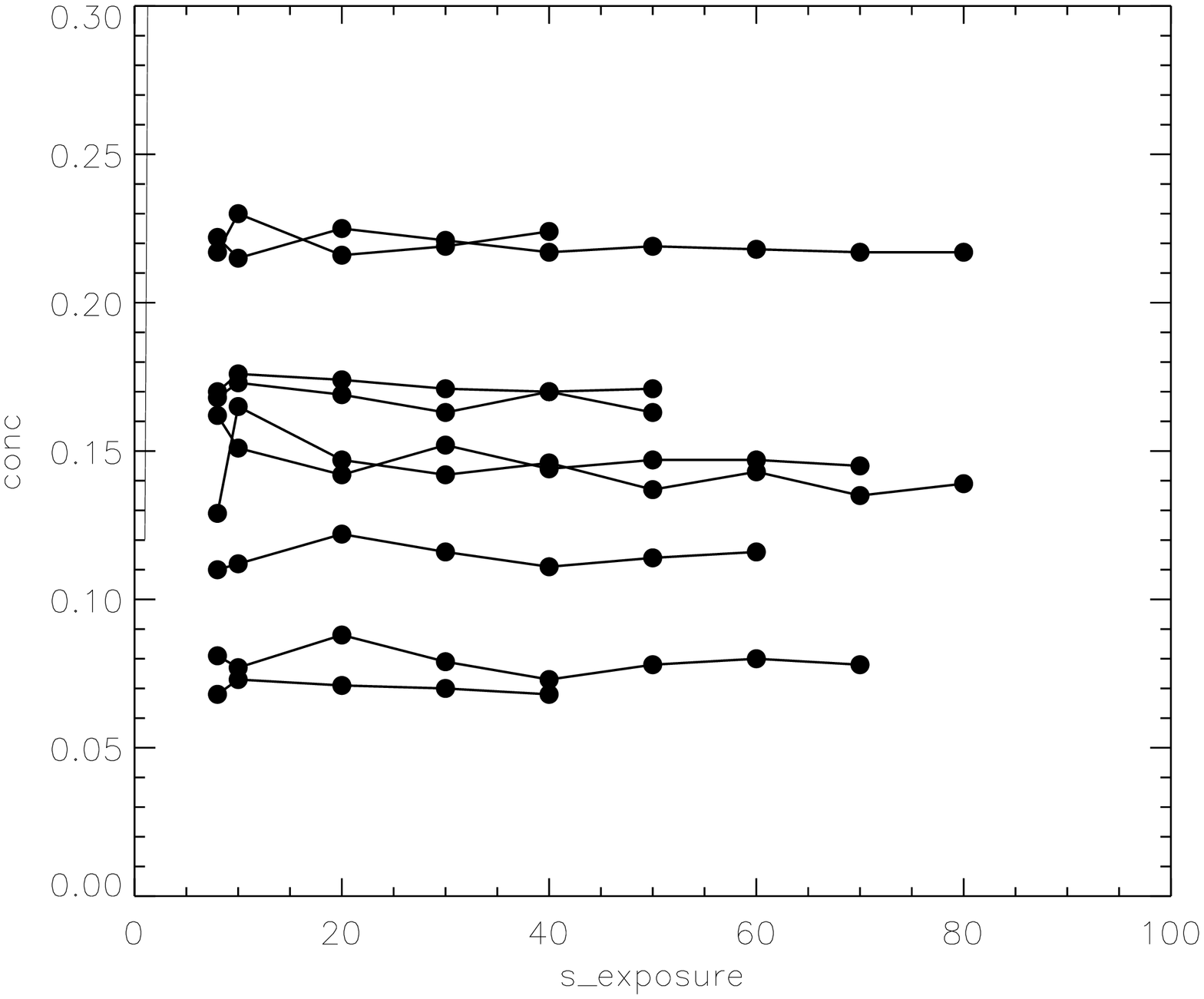}}
 \resizebox{\hsize}{!}{\includegraphics[height=4cm,width=2cm,clip,angle=90]{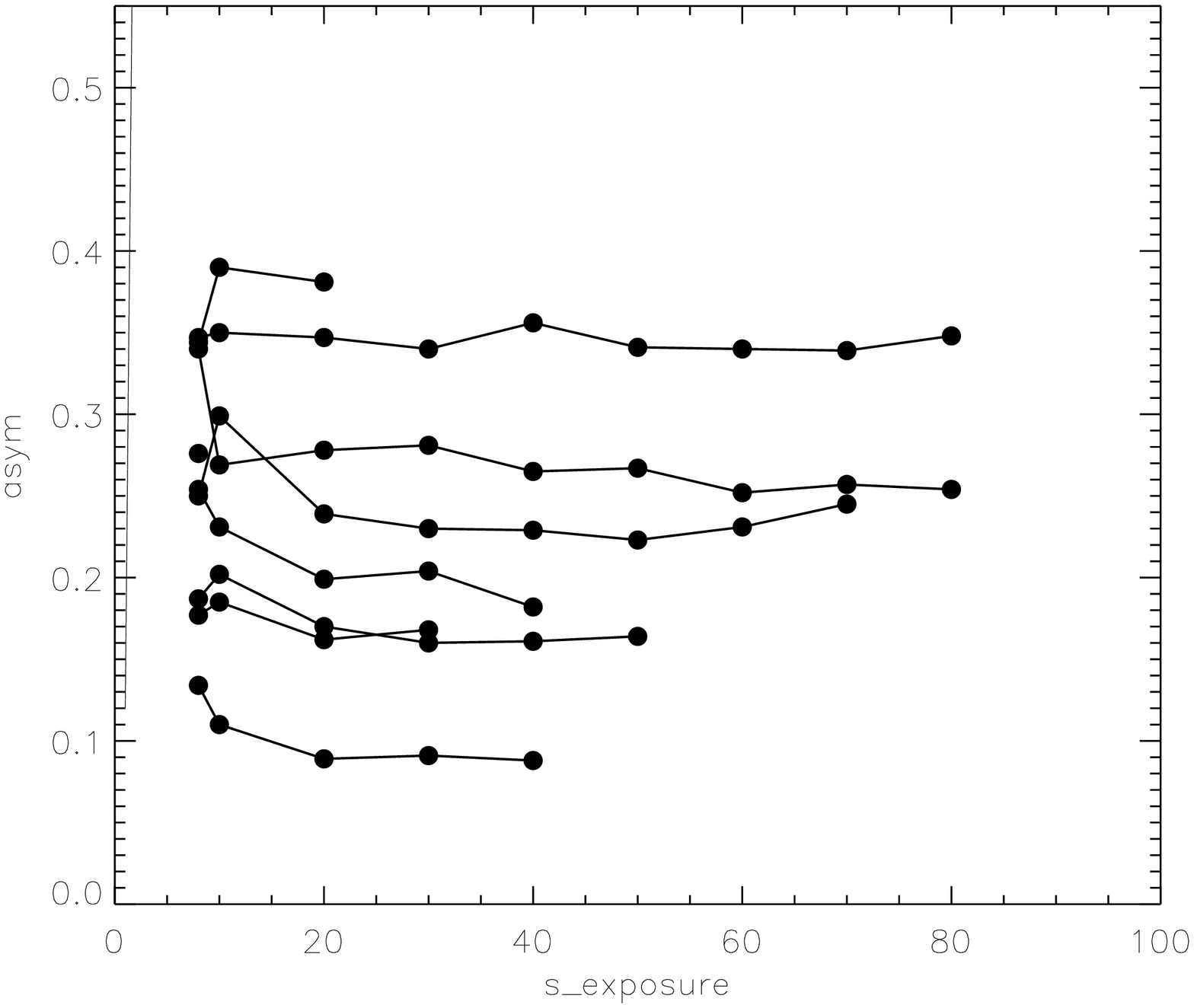}}
 \resizebox{\hsize}{!}{\includegraphics[height=4cm,width=2cm,clip,angle=90]{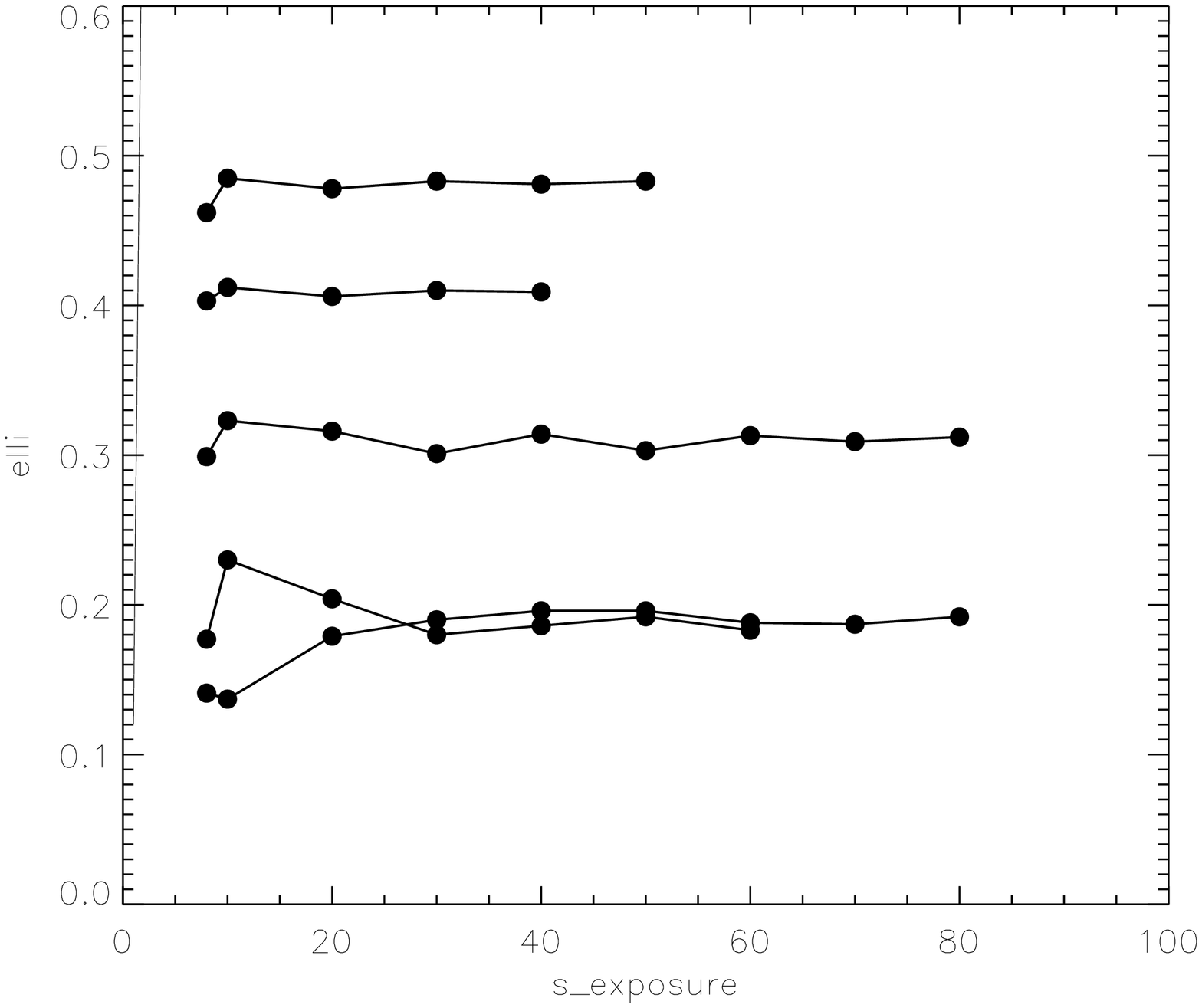}}
 \resizebox{\hsize}{!}{\includegraphics[height=4cm,width=2cm,clip,angle=90]{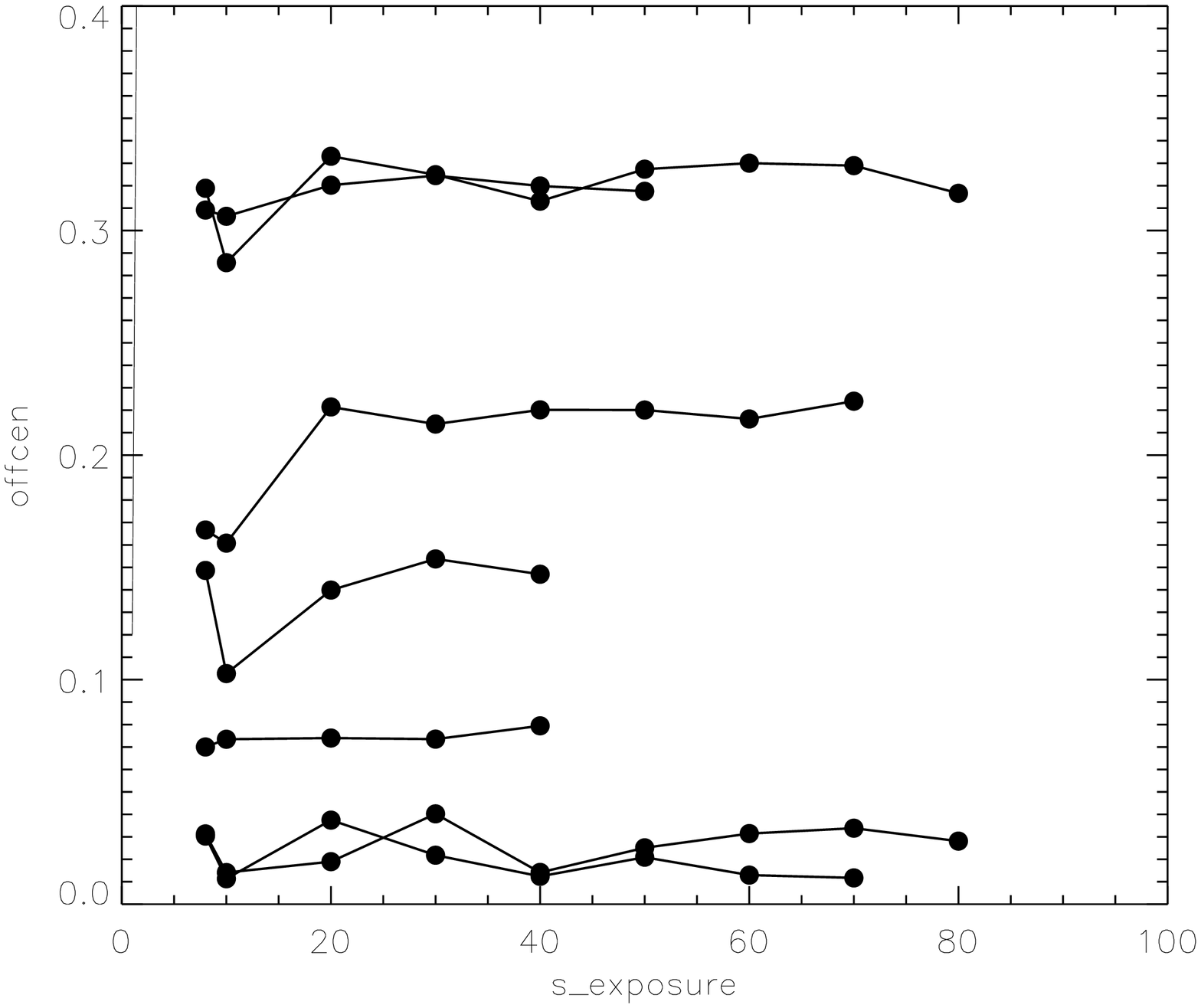}}
 \caption{
 Simulated exposure time (in ksec)  vs various morphological measures
 for several typical clusters.
}
\label{FigTemp}
\end{figure}

 Figure 2 shows the effect of exposure time on each morphological measure.
 For each cluster data, we simulated observations with several shorter exposure times
 using the method described above, and re-measured our morphological measures
 for each simulated observation. 
 In Figure 2, 
 we plotted the simulated exposure time (in ksec) against various morphological measures
 of our sample clusters. For brevity, we only plotted a handful of  typical clusters
 and the data points of the same cluster were connected with a line to illustrate the trend.
The figure shows that, for all of our measures, the systematics caused by
exposure time are small, 
demonstrating the relatively robust nature of our 
measures against the exposure differences.
 The morphological measures are generally constant 
 over a range of exposure time, except for
 some clusters at a very short exposure end, where the noise becomes dominant 
 and measures become uncertain.
 
\subsubsection{Redshift effect}
To investigate the systematic redshift effect on the morphological measures,
we simulated an observation of a cluster at a higher redshift than 
its actual redshift using the real data,
including the effect of waveband shift, a smaller angular size of the object
(also equivalent of having a bigger pixel scale and bigger 
smoothing scale), 
and dimming of the 
object signal
with respect to the sky. Namely, to simulate an observation at new redshift z=z1,
at first, we created an image of {\em restframe} 0.7-8 keV band at original redshift
z0 of each cluster (z1$>$z0).
The image was then corrected for 
detector response and telescope vignetting.
Because the dimming of surface brightness due to the redshift only occurs to
the cluster signal, and not to the background,
an object-only frame I$_0$ was created from this restframe 
image  by subtracting a constant background B. 
However, 
to approximate the dimming of cluster by the redshift,
we cannot simply scale I$_0$  by $1/(1+z)^4$ 
because the noise will not be correct (we {\it underestimate} it).
To properly scale I$_0$ with the proper amount of noise,
we employed an adaptive scaling technique similar to the exposure time
case in sec 4.2.1.
Unlike the exposure time case, however, the intrinsic noise contained
in I$_0$ is not proportional only to I$_0$
(it is proportional to I$_0$+B, instead, even if the background B is
already subtracted from the {\it signal}). 
This makes the adaptive scaling more complicated, and
we need a pixel-to-pixel scaling (or manipulation) rather than
just a simple whole-image scaling.
Namely, 
an intermediate scaled image I$_1$
was created from I$_0$ by a pixel-to-pixel manipulation:

\begin{eqnarray}
  I_{1}(x,y) &=& \frac{I_{0}(x,y)^2R_1^2}{[I_{0}(x,y)R_1+B-R_1^2(I_{0}(x,y)+B)]} \\
  where, & &\nonumber \\
  R_1&=&[(1+z0)/(1+z1)]^4 
\end{eqnarray}

Similarly to the exposure time effect in sec 4.2.1, 
Poisson noise was then added to I$_1$.
A new dimmed image I$_2$ whose {\it cluster signal} was 
scaled by R$_1$ with respect to the original {\it restframe} image
 with proper amount of Poisson noise was then
created from this noise-added image I$_1^{'}$ by a reverse 
pixel-to-pixel manipulation: 
\begin{eqnarray}
  I_{2}(x,y) &=& I_1^{'}(x,y)\frac{I_{0}(x,y)R_1+B-R_1^2[I_{0}(x,y)+B]}{I_{0}(x,y)R_1} 
\end{eqnarray}
Finally, adding back the background B gives,
\begin{eqnarray}
  I^{'}_{2}(x,y) &=& I_{2}(x,y) + B
\end{eqnarray}


This dimmed image I$^{'}_2$ should be then rebinned by a factor R$_2$ 
to account for the angular-size change due to the redshift difference 
between z0 and z1. 
However, this simple rebinning
again will not correctly reproduce the proper amount of noise
caused by the angular-size change due to the redshift effect.
To properly adjust, again {\it underestimated}  noise due to
the simple rebinning, the rebinned image was  rescaled by 
a factor 1/(R$^2_2$-1), then Poisson noise was added
by taking each pixel value as the
mean for a Poisson distribution.
The final image was created by rescaling back this noise-added image by a factor
(R$^2_2$-1)/R$_2^2$. The factor R$_2^2$ in the denominator is necessary
to rebin the image in such a way to conserve the surface brightness.
(The derivation of these scalings, or manipulations, are described
                   in the Appendix.)

\begin{figure}[h]
 \resizebox{\hsize}{!}{\includegraphics[height=4cm,width=2cm,clip,angle=90]{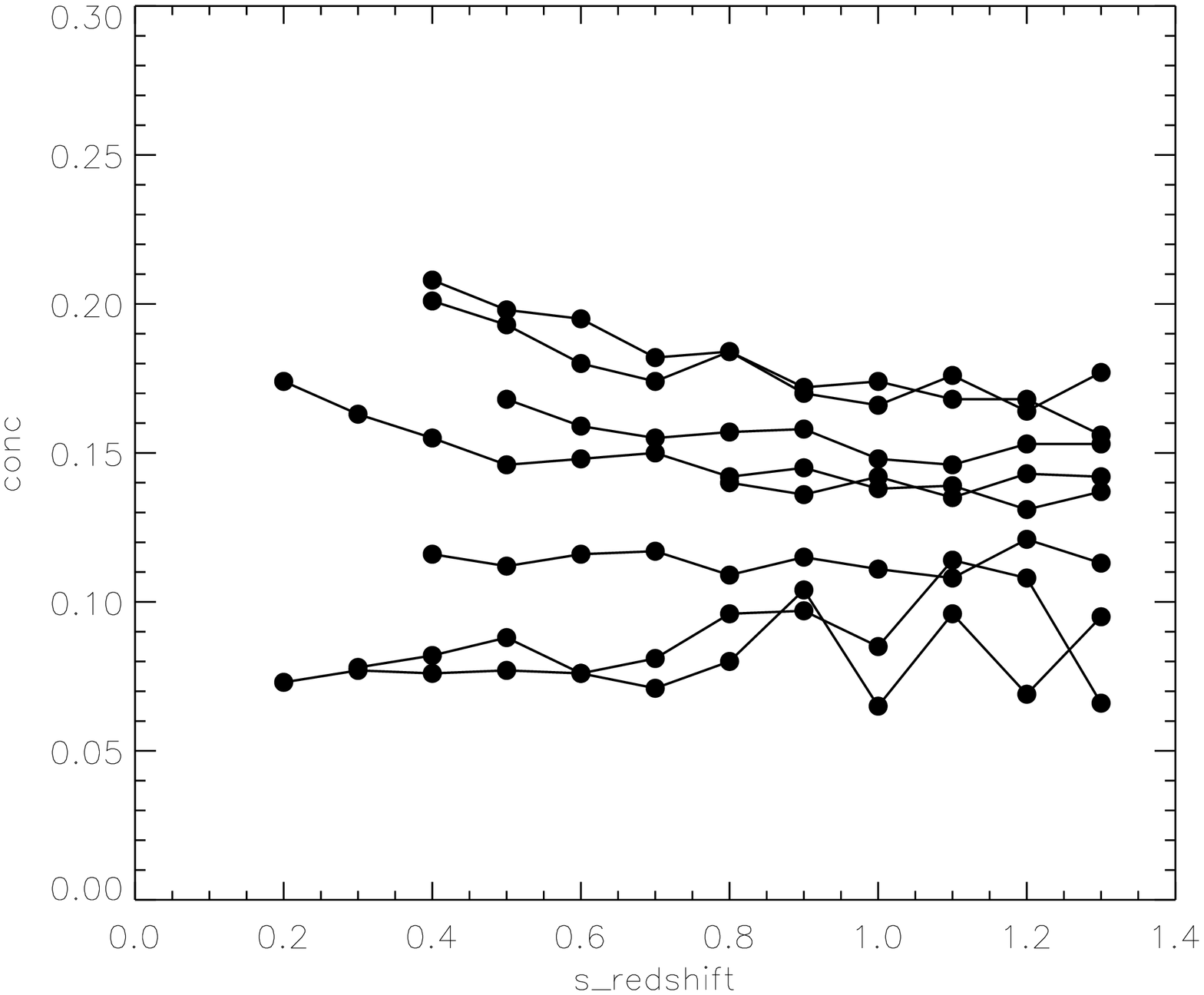}}
 \resizebox{\hsize}{!}{\includegraphics[height=4cm,width=2cm,clip,angle=90]{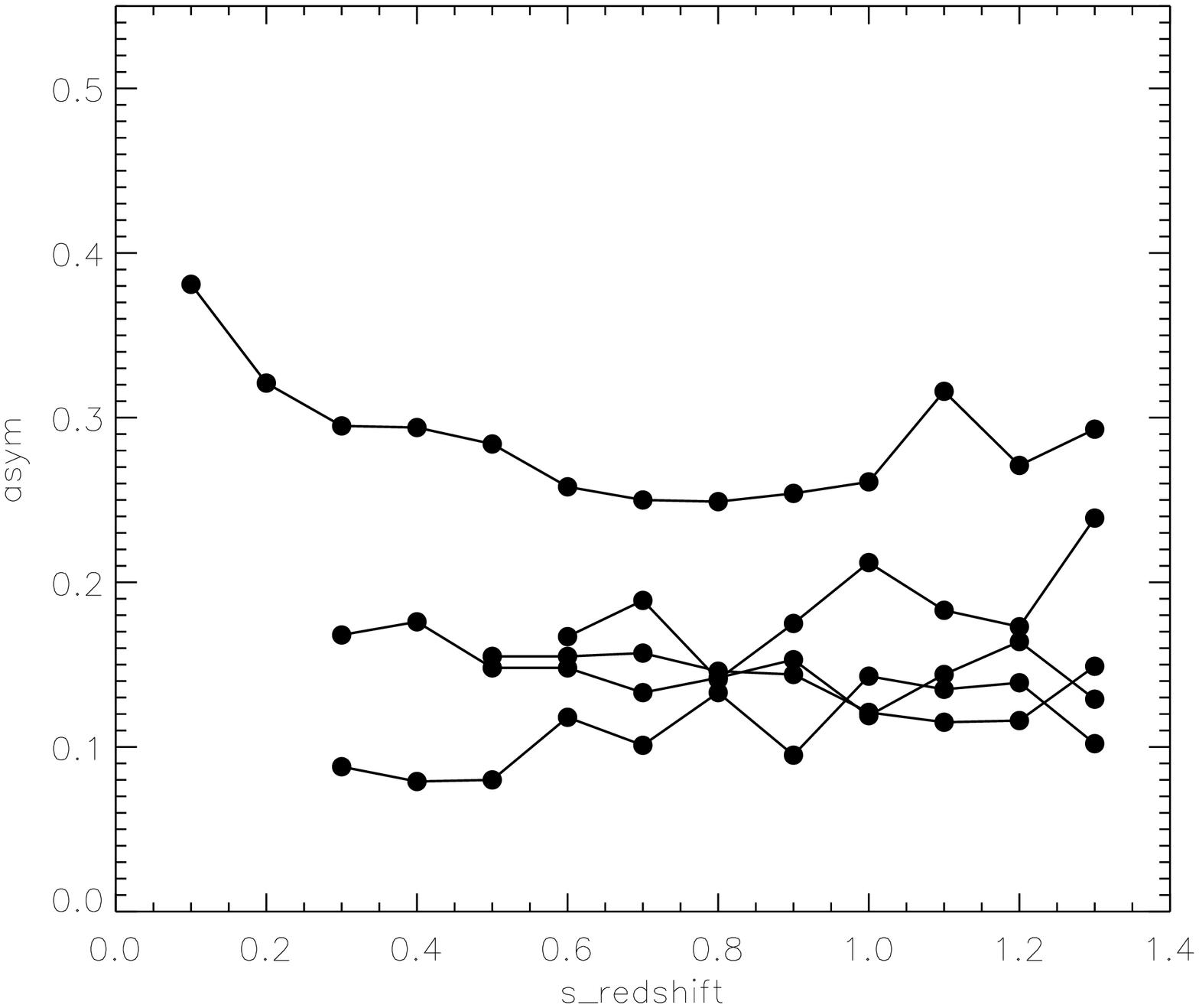}}
 \resizebox{\hsize}{!}{\includegraphics[height=4cm,width=2cm,clip,angle=90]{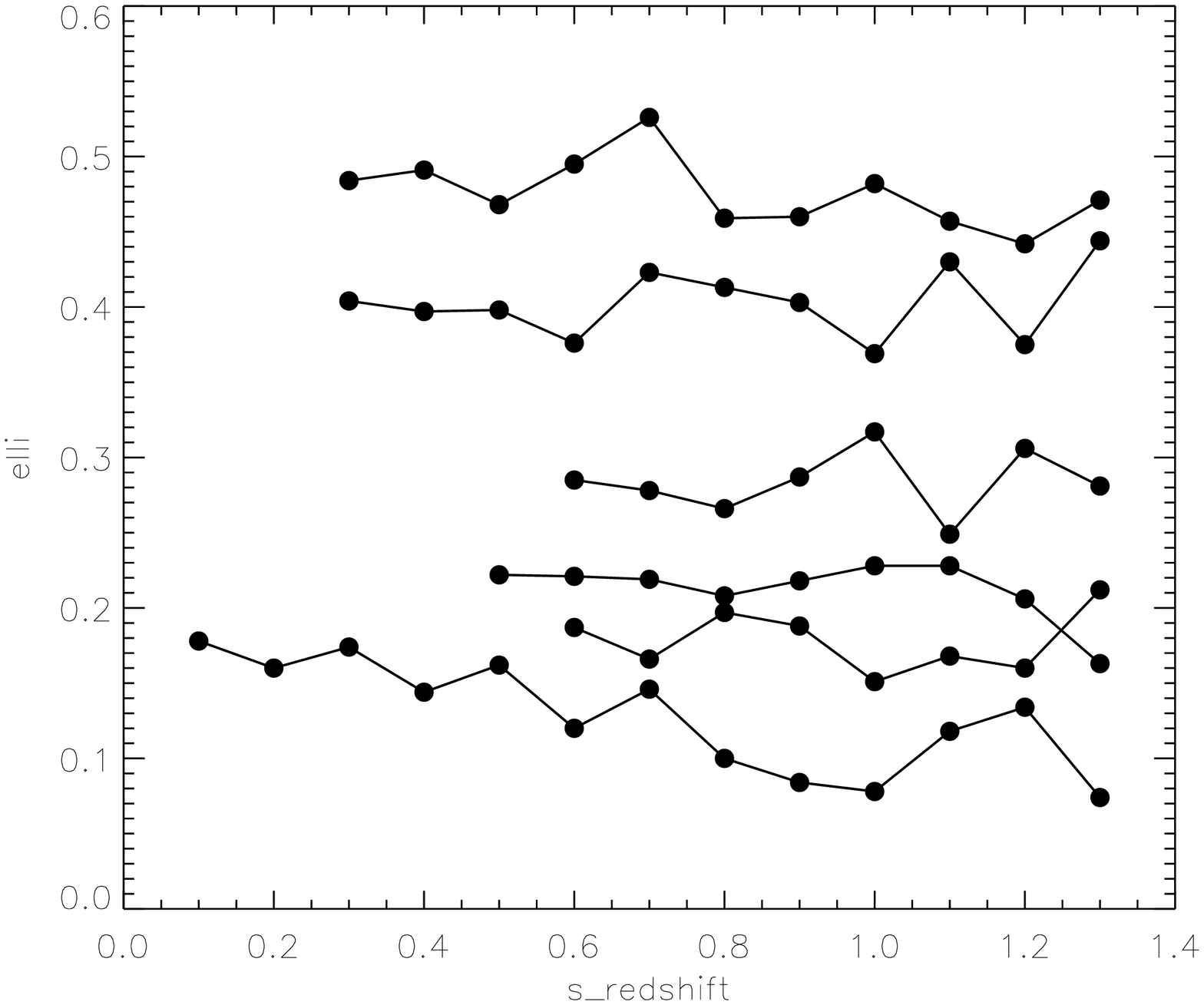}}
 \resizebox{\hsize}{!}{\includegraphics[height=4cm,width=2cm,clip,angle=90]{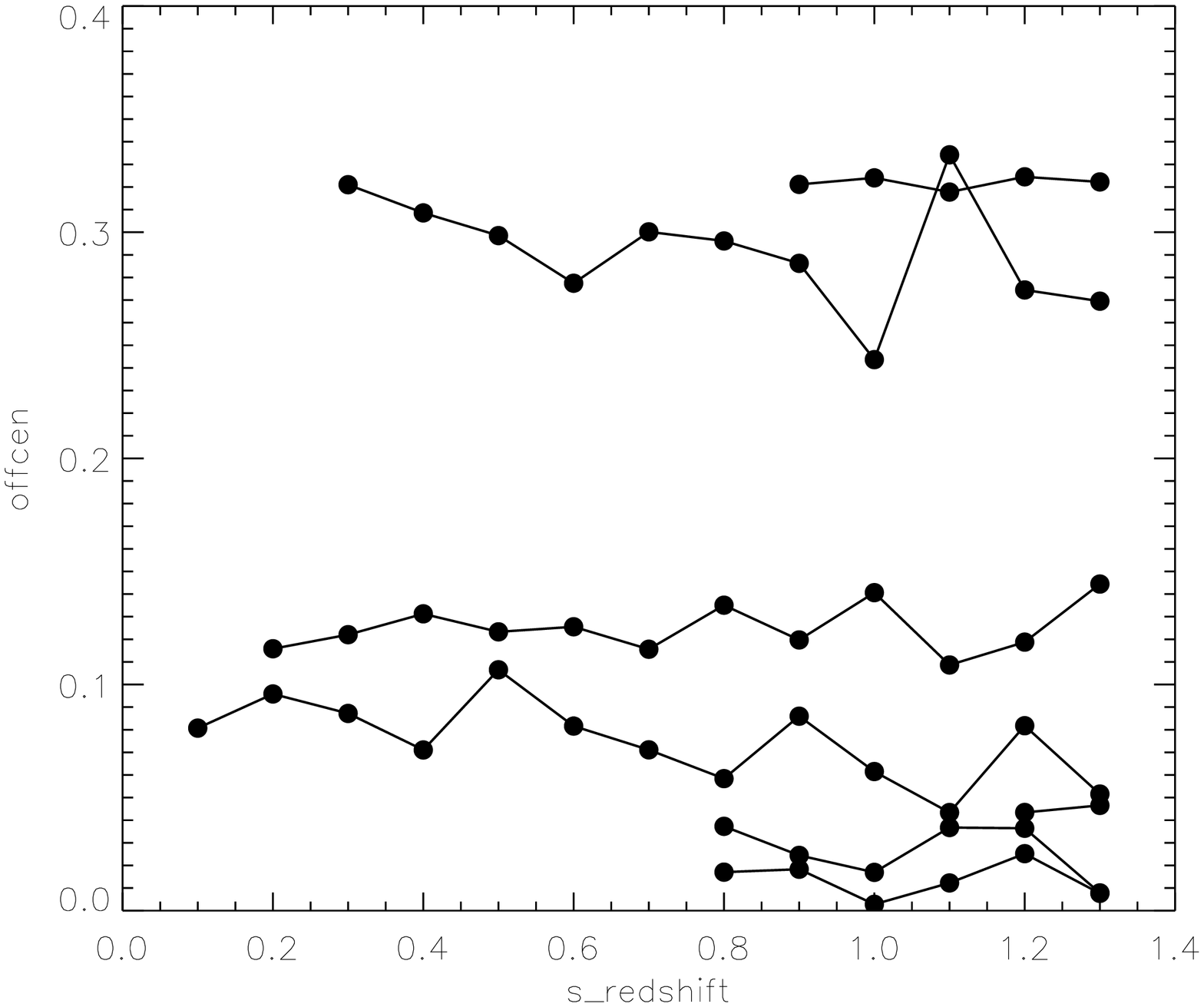}}
 \caption{
 Simulated redshift vs various morphological measures.
}
\label{FigTemp}
\end{figure}

Figure 3 shows the effect of redshift on each morphological measure.
 For one cluster data, we simulated observations with several higher redshifts 
 than its original redshift
 using the method described above, and re-measured our morphological measures
 for each simulated observation.
 In Figure 3, 
 we plotted the simulated redshift (s\_redshift) against various 
morphological measures
of our sample clusters. Again, for brevity, we only plotted a handful of  typical clusters
 and the data points of the same cluster were connected with a line to illustrate the trend.
 The figure shows that, for most of our measures, the systematics caused by
 redshift are very small. For the concentration index and asymmetry index,
 there may be  a slight trend that high concentration or high asymmetry objects
 tend to slightly decrease their values as we go to the higher redshifts,
 thus reducing the contrast between distorted and non-distorted morphologies.

\subsubsection{Combining the exposure and redshift effects}
Although the exposure time and redshift effects can be treated separately
as described in sec 4.2.1. and 4.2.2., 
these two effects are often coupled, 
because low redshift clusters are usually observed with 
shorter exposures
than high redshift clusters.
Under these conditions, it is often much more useful 
to treat the two effects together, because one can simulate
even 
a {\it longer } observation than its original exposure time, 
by intentionally reducing the amount of noise to be added for the 
{\it redshift-effect} part.
With this treatment, 
to compare clusters of various exposure times, 
we can simulate 
an observation with `increased' exposure time for the {\it low-z}
clusters,
in stead of standard  way of simulating an observation 
with `decreased' exposure time for the {\it high-z} clusters,
thus we can 
compare observations of various clusters
without greatly reducing precious signal-to-noise
ratio of the {\it high-z} cluster data.
In detail, 
we modified the final steps  described in sec 4.2.2. by 
introducing one more scaling parameter R$_3$ = t2/t0, t2$>$t0, where t2 is an 
{\it increased} exposure time, and t0 is an original integration time. 
After the rebinning (i.e. the rebinning after Eq. 15),
instead of simply rescaling by 1/(R$_2^2$-1) described
in sec 4.2.2.,
we rescaled the image by a factor
R$_3$/(R$_2^2$-R$_3$), where  
(R$_2^2$ -R$_3$) $>$ 0, namely,
\begin{eqnarray}
  I_3   &=& I_2^{''}\frac{R_3}{(R_2^2-R_3)}      
\end{eqnarray}
  where, 
  I$_3$ is the intermediate scaled image, and  
  I$_2^{''}$ is the  dimmed, background re-added, and rebinned image. 
Poisson noise was then added.
The noise added image was then rescaled back by a factor
(R$_2^2$-R$_3$)/R$_2^2$ to produce the final image whose {\it signal} is  scaled
by R$_3$ relative to 
I$^{'}_2$, with a proper amount of Poisson noise.
(Again, the derivation of this scaling is described in the Appendix.)
The maximum length of integration time we can `increase' (t2$_{max}$) 
is naturally
limited by the original  exposure time and how much
we increase the redshift for the redshift-effect part,
and determined by a relationship:
\begin{eqnarray}
 R_2^2-R_3 = 0, 
\end{eqnarray}
which is equivalent to the case when  no Poisson noise is added after 
the rebinning described in sec 4.2.2.
Thus,
\begin{eqnarray}
 t2_{max}=t0R_2^2. 
\end{eqnarray}
This t2$_{max}$ can be also used as a rough estimate of
effective image depth.
The t2$_{max}$ provides an estimate of the image depth  
much better than conventional simple exposure time,
because t2$_{max}$ is related to a quantity
which is affected both by exposure time and redshift, 
and thus
enabling us to quantitatively compare exposure times of observations 
involving targets at different redshifts (e.g. 100 ksec at z=0.1 and
100 ksec at z=0.9).
The distribution of t2$_{max}$ for our sample is plotted in Fig. 4
for the case z1 = 0.9.
Several clusters whose t2$_{max}$ is much bigger than  600 ksec are not shown 
in Fig. 4, for brevity.
The figure shows what the effective exposure times would be,
if all clusters were at z=0.9. 

Judging from figures 2, 3, and 4 altogether, 
we have decided to modify all of the observations
to be equivalent of z=0.9 and t=t2$_{max}$,
to make sure to eliminate even the small systematics 
in Fig. 3, but otherwise to maximize the image quality. 
Meanwhile, four clusters whose original redshifts above z=0.9 were 
modified only in the exposure time.
After this stage,
to ensure that t2$_{max}$ is well above the low signal-to-noise end,
we discarded clusters whose total counts are below 300.
The resulting sample size after this final data preparation is 101.

\begin{figure}
\resizebox{\hsize}{!}{\includegraphics[height=4cm,width=2cm,clip,angle=90]{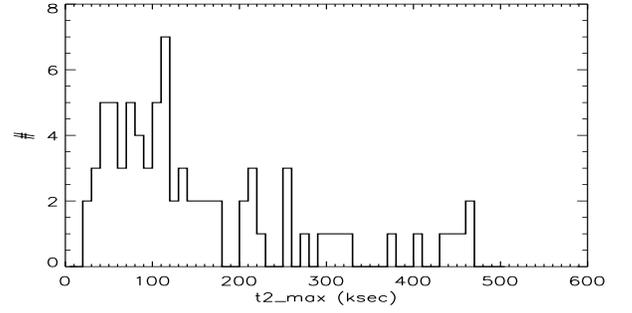}}
 \caption{
Distribution of the maximum integration time (t2$_{max}$)
for our sample. 
}
\label{FigTemp}
\end{figure}

 \begin{figure}[t]
 \resizebox{\hsize}{!}{
 \includegraphics[bb=118 110 600 700,height=4cm,width=2cm,clip,angle=90]{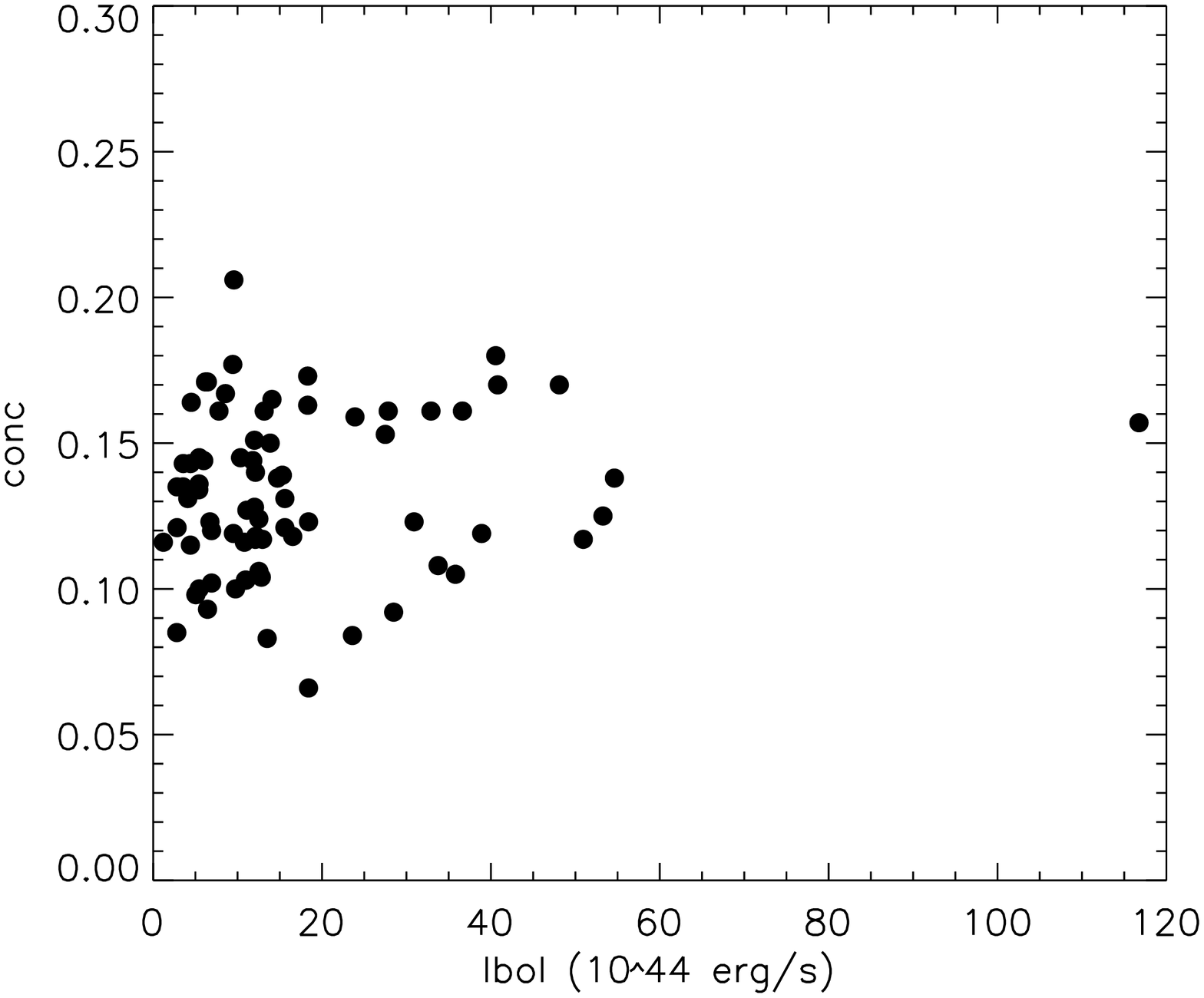}
 }
 \resizebox{\hsize}{!}{
 \includegraphics[bb=118 110 600 700,height=4cm,width=2cm,clip,angle=90]{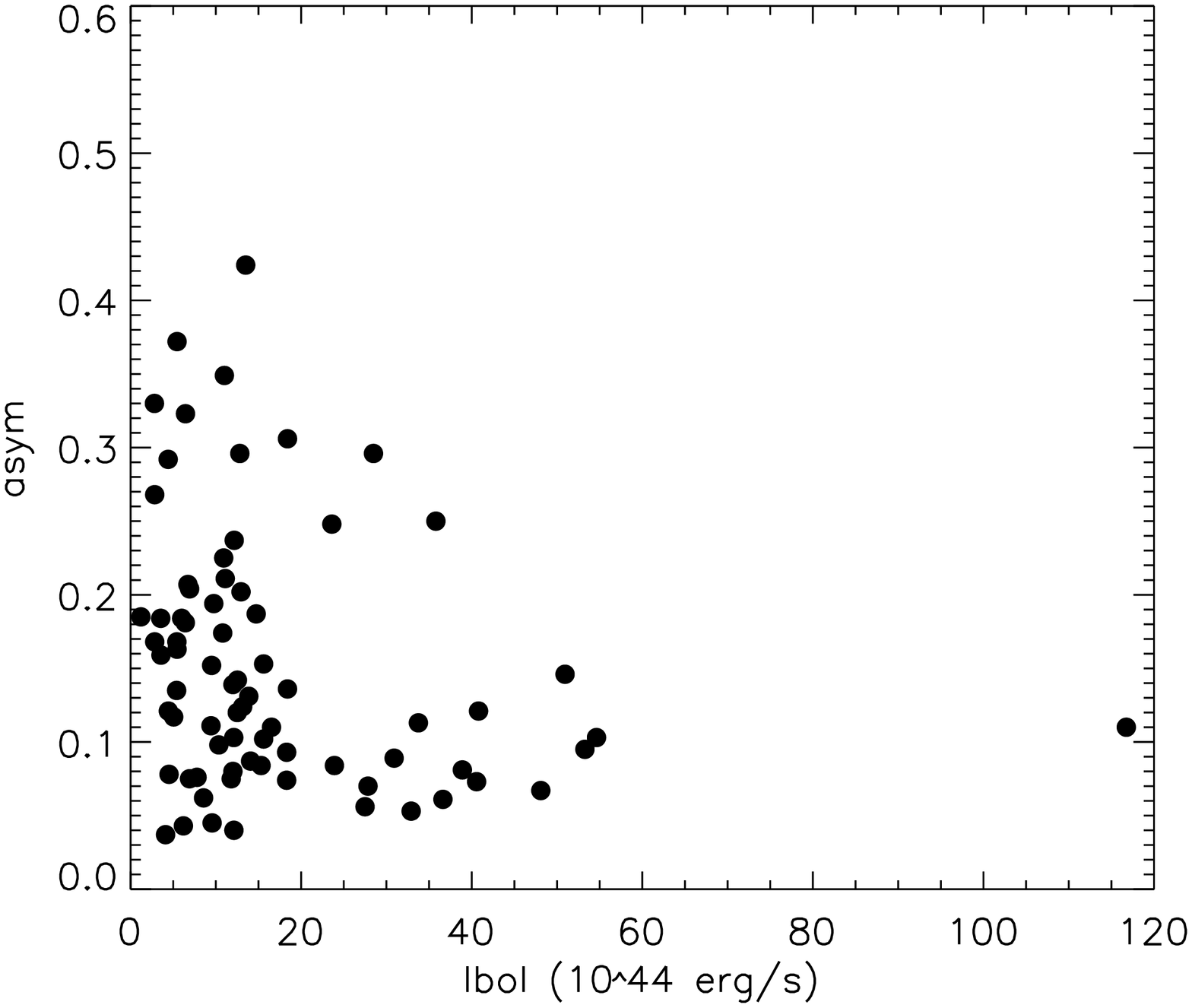}
  }
 \resizebox{\hsize}{!}{
 \includegraphics[bb=118 110 600 700,height=4cm,width=2cm,clip,angle=90]{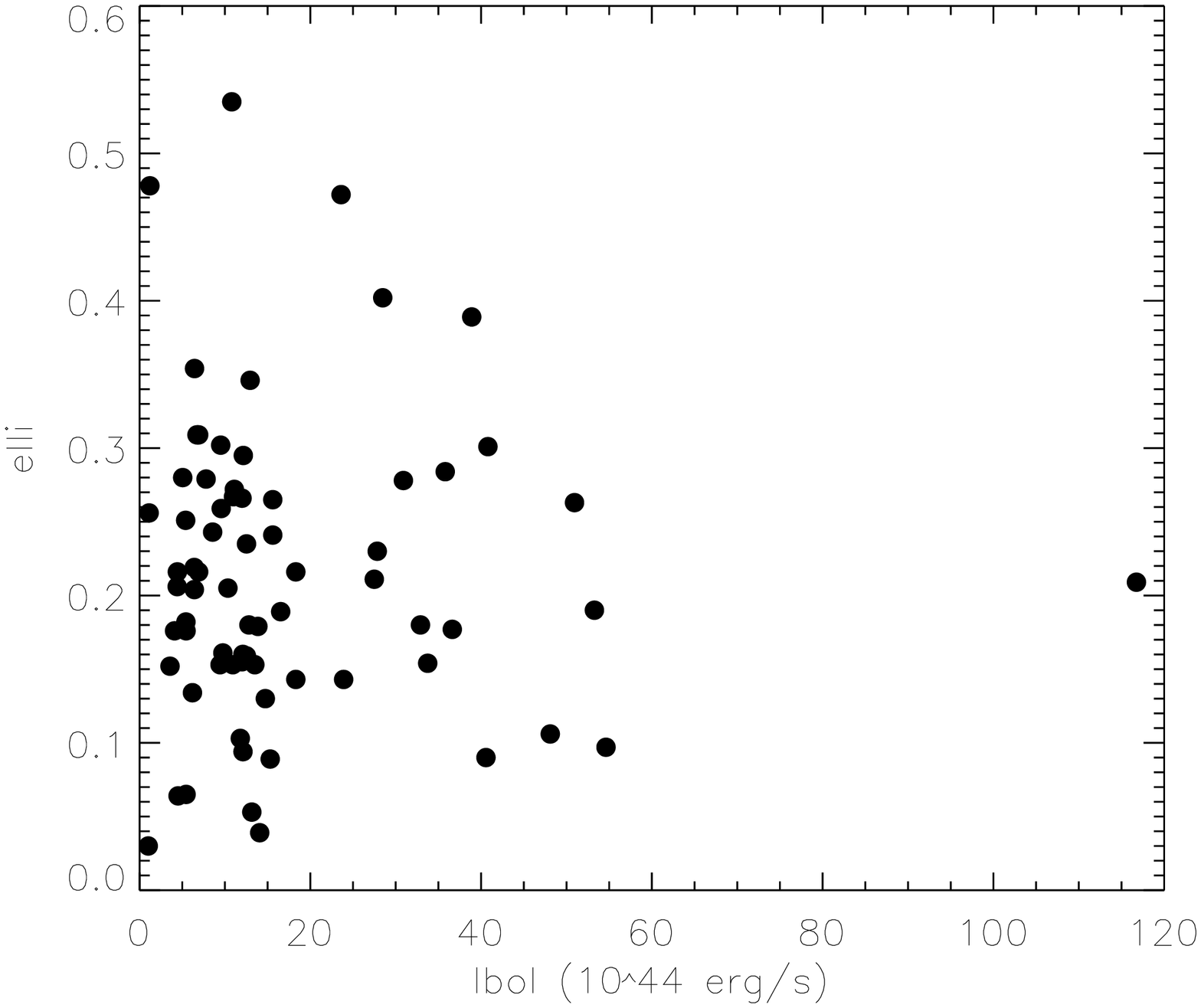}
  }
 \resizebox{\hsize}{!}{
 \includegraphics[bb=60 110 600 700,height=4cm,width=2.2cm,clip,angle=90]{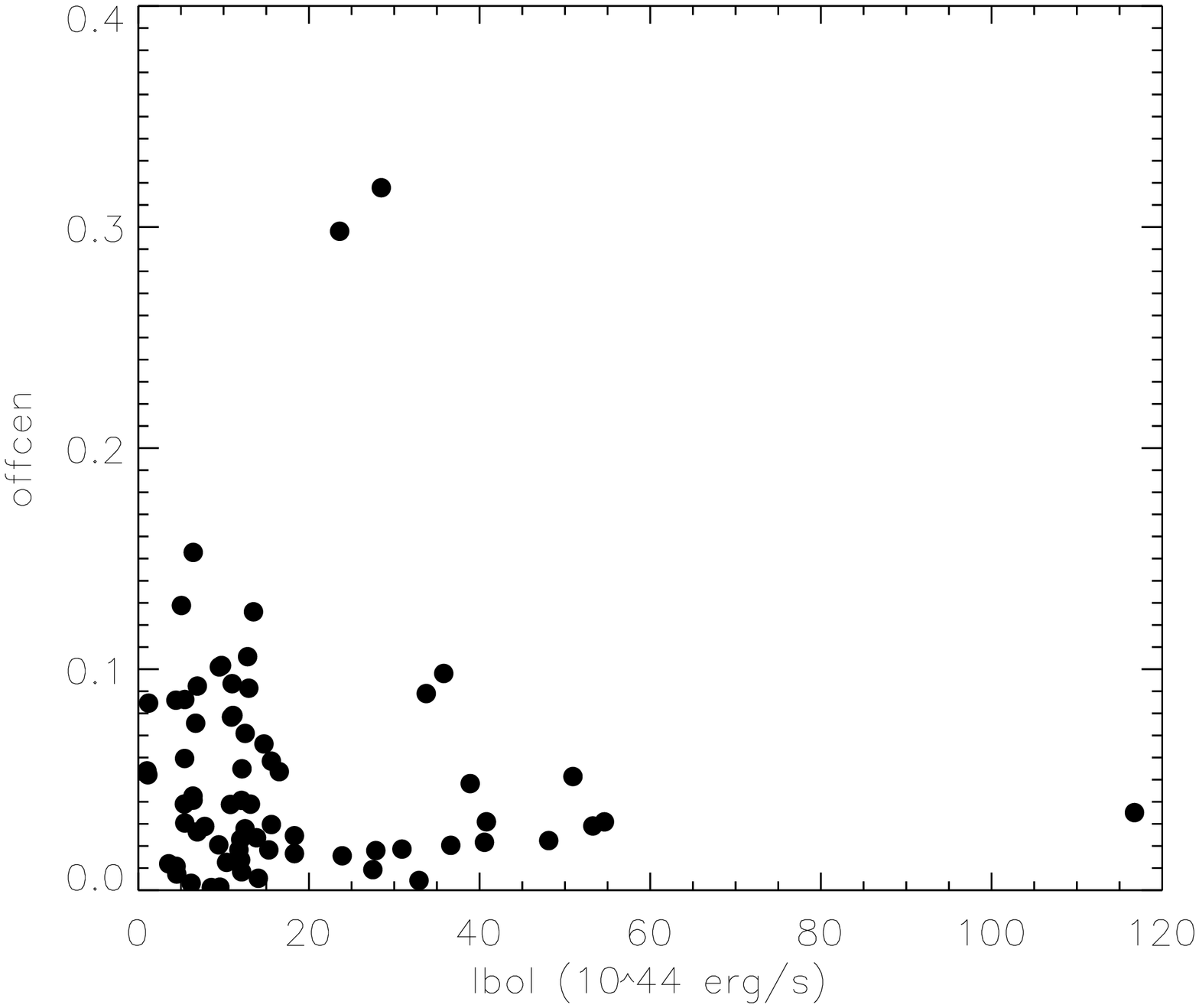}
  }
 \caption{
Comparisons between X-ray cluster morphology and  X-ray  bolometric
luminosity.
}
\label{FigTemp}
\end{figure}

\begin{figure}[t]
 \resizebox{\hsize}{!}{
 \includegraphics[bb=118 110 600 700,height=4cm,width=2cm,clip,angle=90]{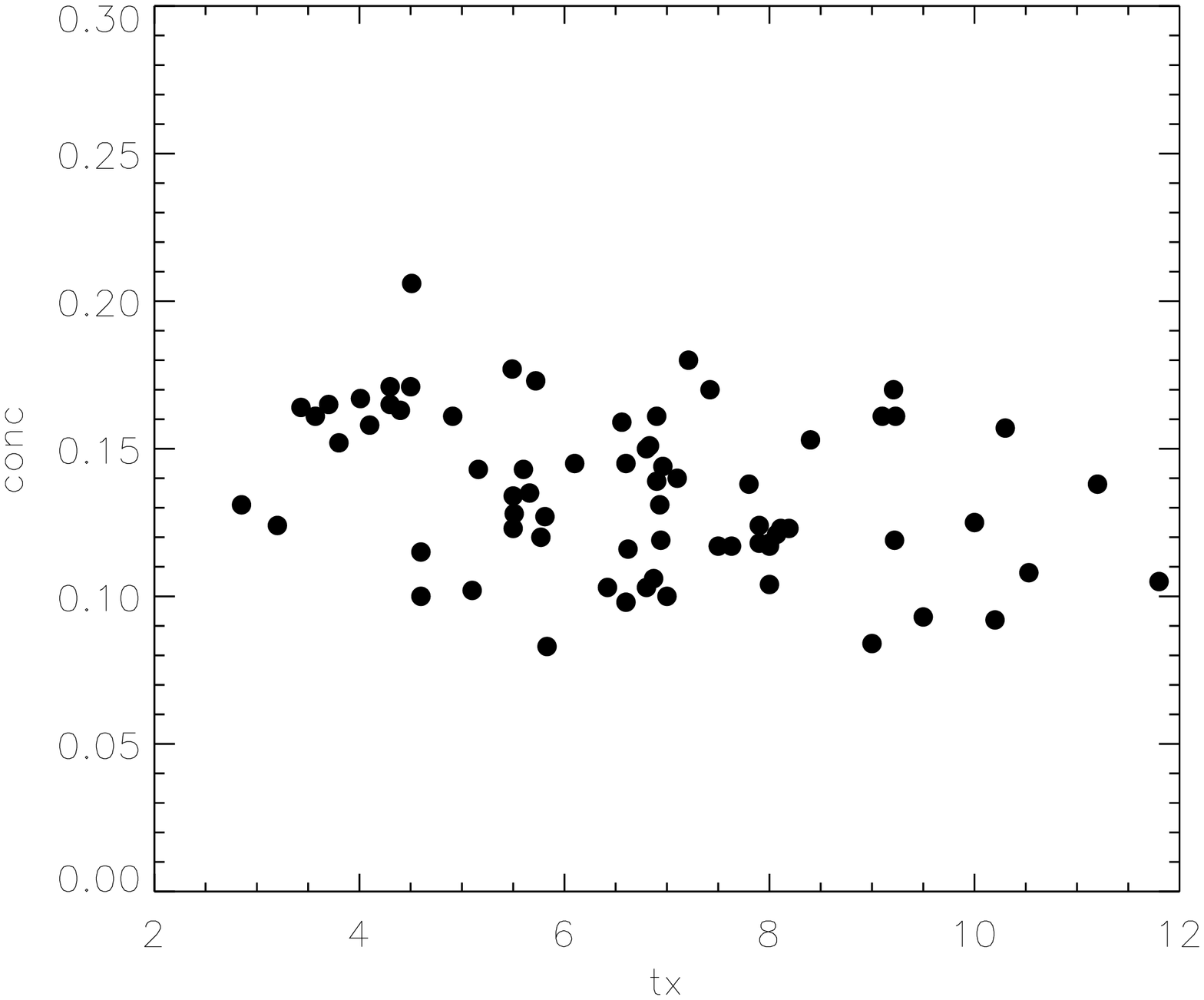}
 }
 \resizebox{\hsize}{!}{
 \includegraphics[bb=118 110 600 700,height=4cm,width=2cm,clip,angle=90]{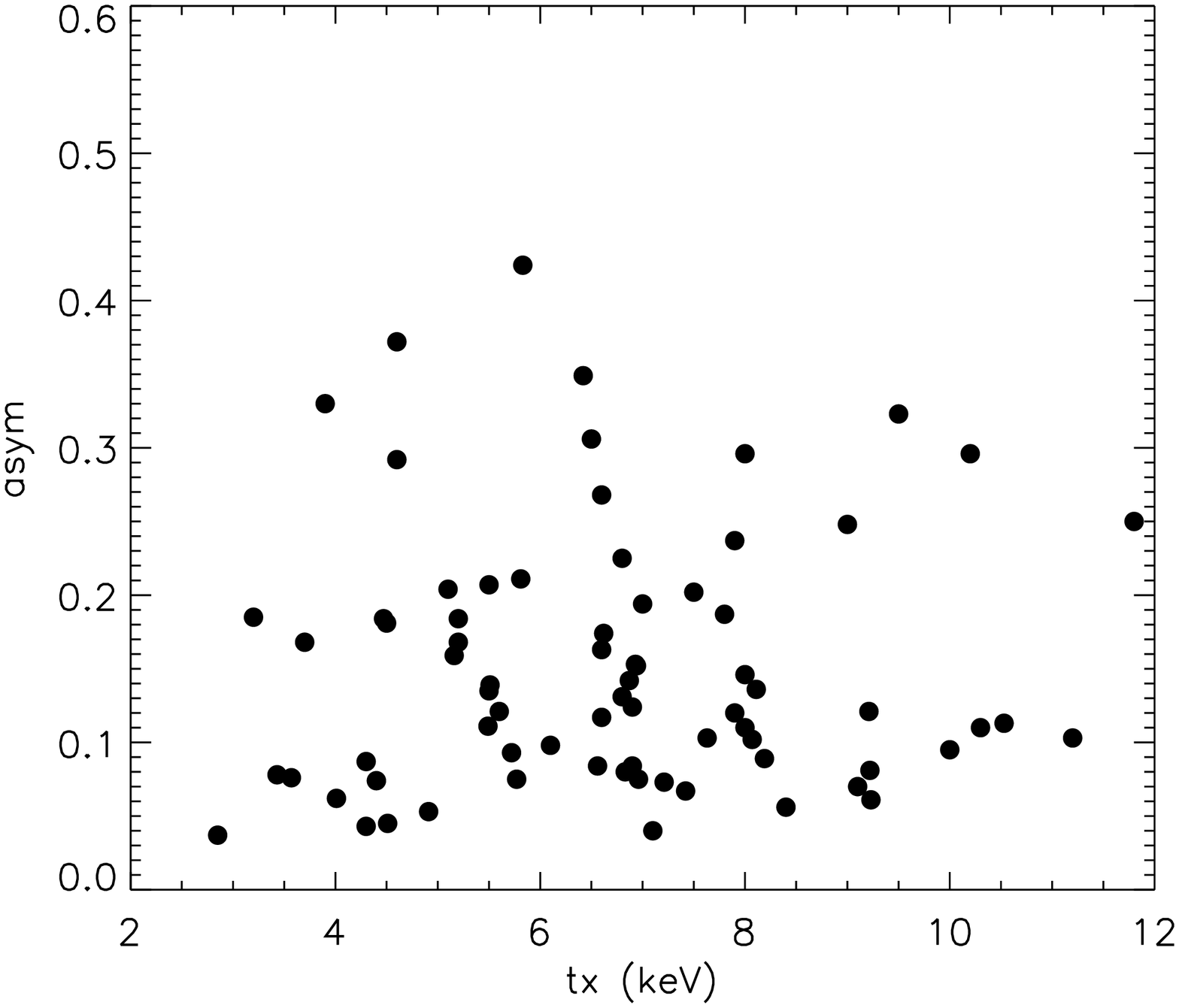}
  }
 \resizebox{\hsize}{!}{
 \includegraphics[bb=118 110 600 700,height=4cm,width=2cm,clip,angle=90]{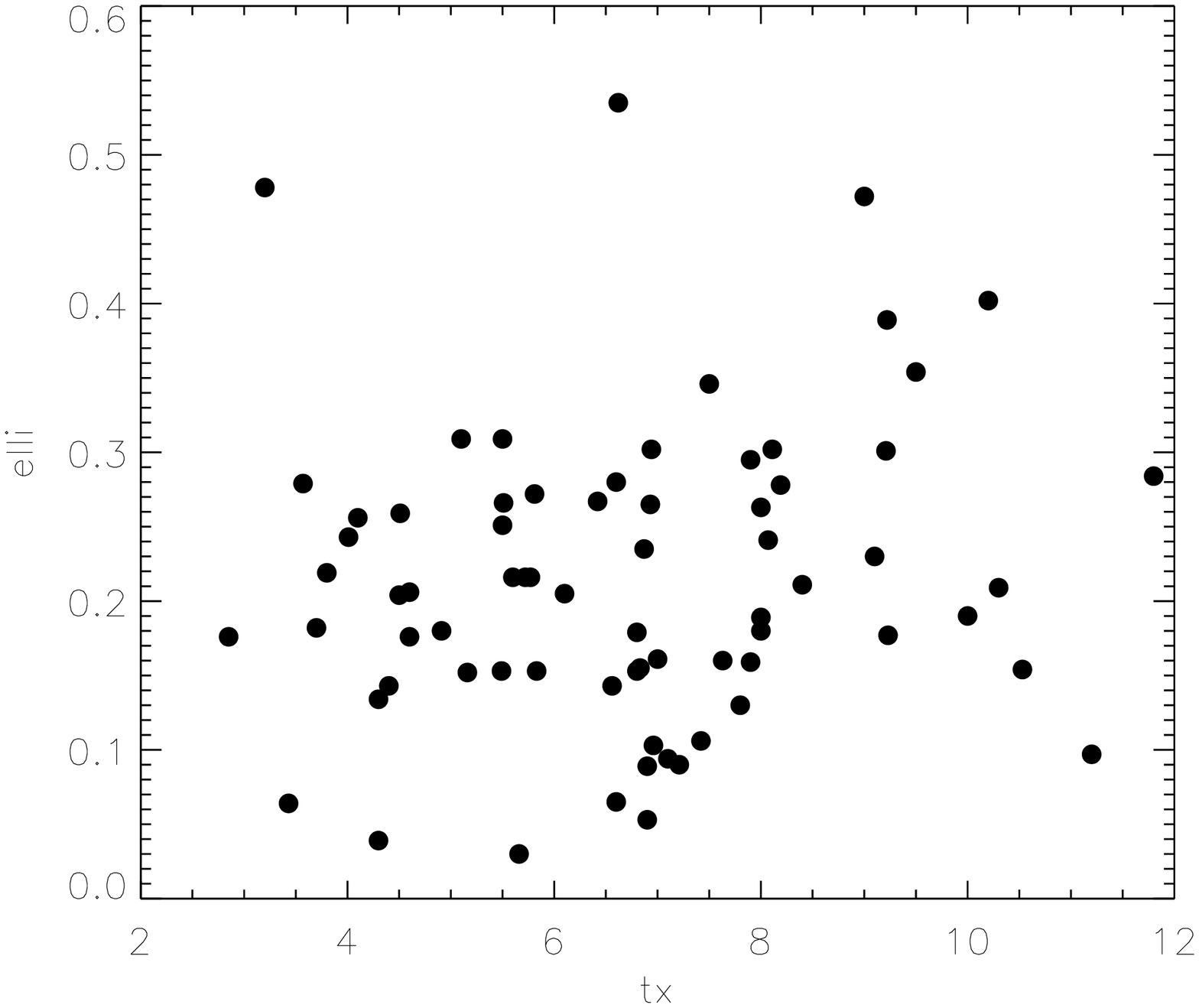}
  }
 \resizebox{\hsize}{!}{
 \includegraphics[bb=60 110 600 700,height=4cm,width=2.2cm,clip,angle=90]{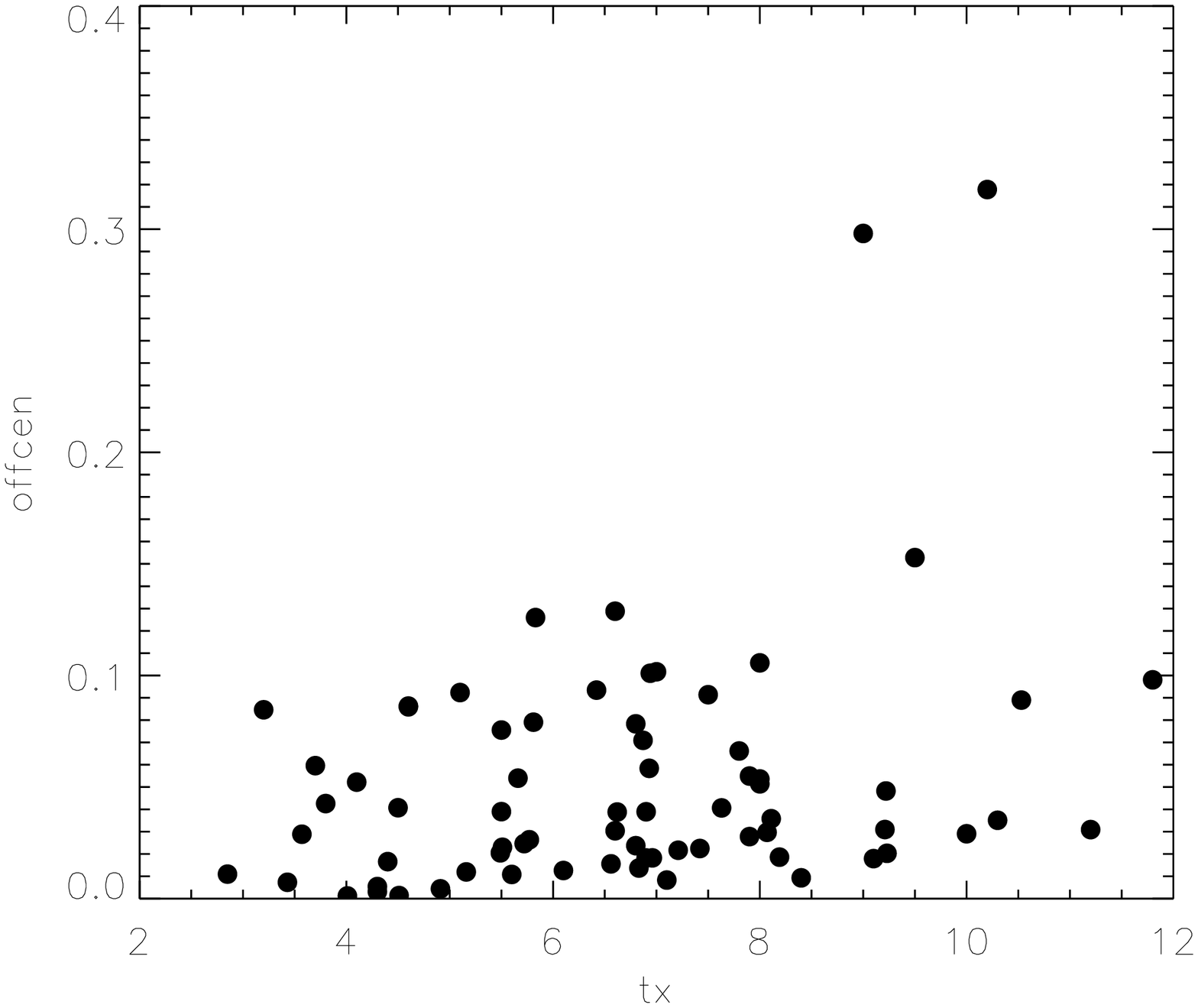}
  }
 \caption{
Comparisons between X-ray cluster morphology and cluster X-ray temperature.
}
\label{FigTemp}
\end{figure}

\section{RESULTS}

\subsection{Comparison with other cluster characteristics}

\subsubsection{X-ray luminosity and temperature}

Fig. 5 and 6 show the relation between our morphological measures and
X-ray bolometric luminosity (Fig. 5), or X-ray temperature (Fig. 6)
taken from the literature.
In Fig. 5 and 6, we see no obvious trend, demonstrating that there is 
no systematics on our measures
due to cluster luminosity, temperature, 
or possibly cluster mass.  
The lack of trend can also  mean that 
more massive clusters do {\it not} show more distortions,
inconsistent with a simple hierarchical structure scenario which 
predicts
more massive clusters are younger and thus showing more merger activity.
Conversely,
the lack of trend can also  mean that interaction  of clusters may {\it not} 
simply enhance or reduce the cluster global X-ray luminosity or temperature, 
although we cannot rule out the possibility of such change for a very brief period of time. 
The lack of correlation of X-ray morphology with X-ray luminosity or 
temperature is also reported by Buote \& Tsai (1996) based on their
power ratio analysis.
The value of 
the rank-order correlation coefficient,
Spearman $\rho$, 
where $\rho$ = 1 or -1 means a perfect linear correlation
of a positive or negative slope, respectively, while $\rho$ = 0 indicates that
two variables are uncorrelated, 
is 0.04, -0.28, -0.02, and -0.06 
for Conc, Asym, Elli, and Offcen,
respectively for Fig. 5, and 
 -0.35, -0.01, 0.14, and 0.26 for Conc, Asym, Elli, and Offcen,
respectively for Fig. 6.

\subsubsection{Visual X-ray Classification}
\begin{figure}[t]
 \resizebox{\hsize}{!}{
 \includegraphics[height=4cm,width=2cm,clip,angle=90]{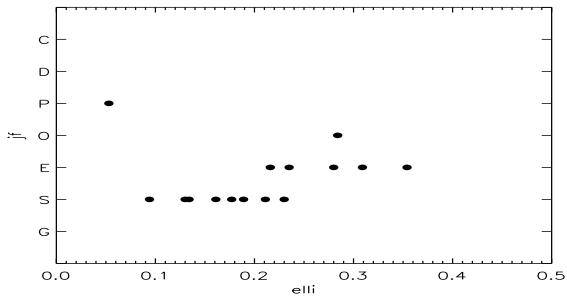}
  }
 \caption{
Comparison between our objective ``Ellipticity" measure and
visual X-ray classification (e.g. E=elliptical, S=single, P=primary with small secondary) by  Jones \& Forman (1999).
}
\label{FigTemp}
\end{figure}

We attempted to compare our measures with X-ray classification by
Jones \& Forman (JF:1999) in which they visually classified clusters into
classes:
 single(S), elliptical(E), primary with small secondary(P), double(D),
 offset center(O), and complex(C), base on inspection of
$Einstein$ X-ray images.
Unfortunately, JF sample mostly consists of low-z (z $<$ 0.05) clusters, 
and only 15 JF clusters were in our sample. Moreover, they were 
mostly of S or E type,
and therefore, a statistically significant  comparison was not possible.
However, a simple comparison between JF S/E type and our ellipticity parameter 
(Fig. 7) 
already indicates that `E' clusters in the JF sample tend to to show
the higher ellipticity parameters than `S' clusters, illustrating 
the consistency between visual classification 
and our measures, and between $Einstein$ and $Chandra$ observations.

\subsubsection{Beta model profile fitting}

\begin{figure}[h]
 \resizebox{\hsize}{!}{
 \includegraphics[bb=118 110 600 700,height=4cm,width=2cm,clip,angle=90]{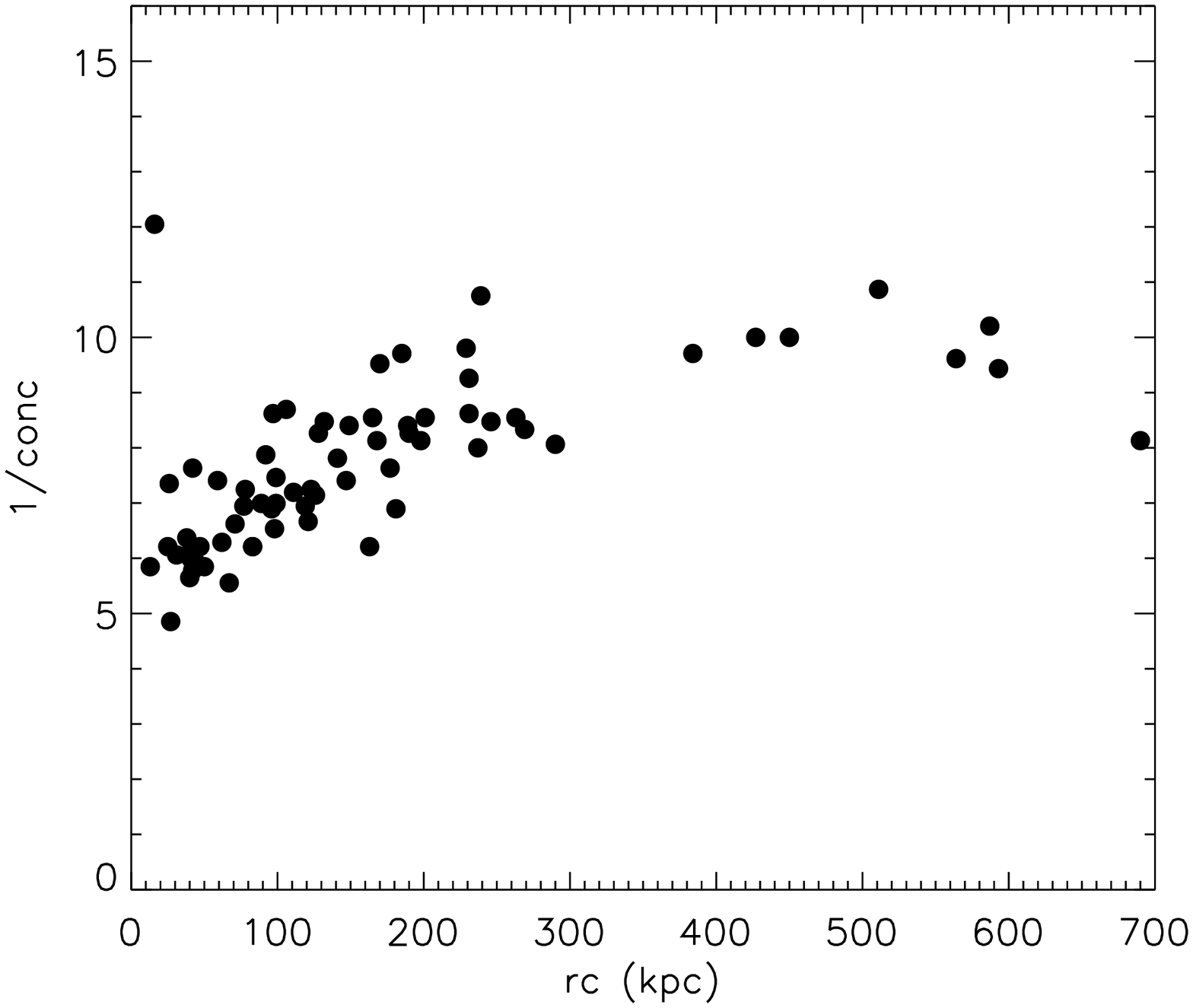}
 }
 \resizebox{\hsize}{!}{
 \includegraphics[bb=118 110 600 700,height=4cm,width=2cm,clip,angle=90]{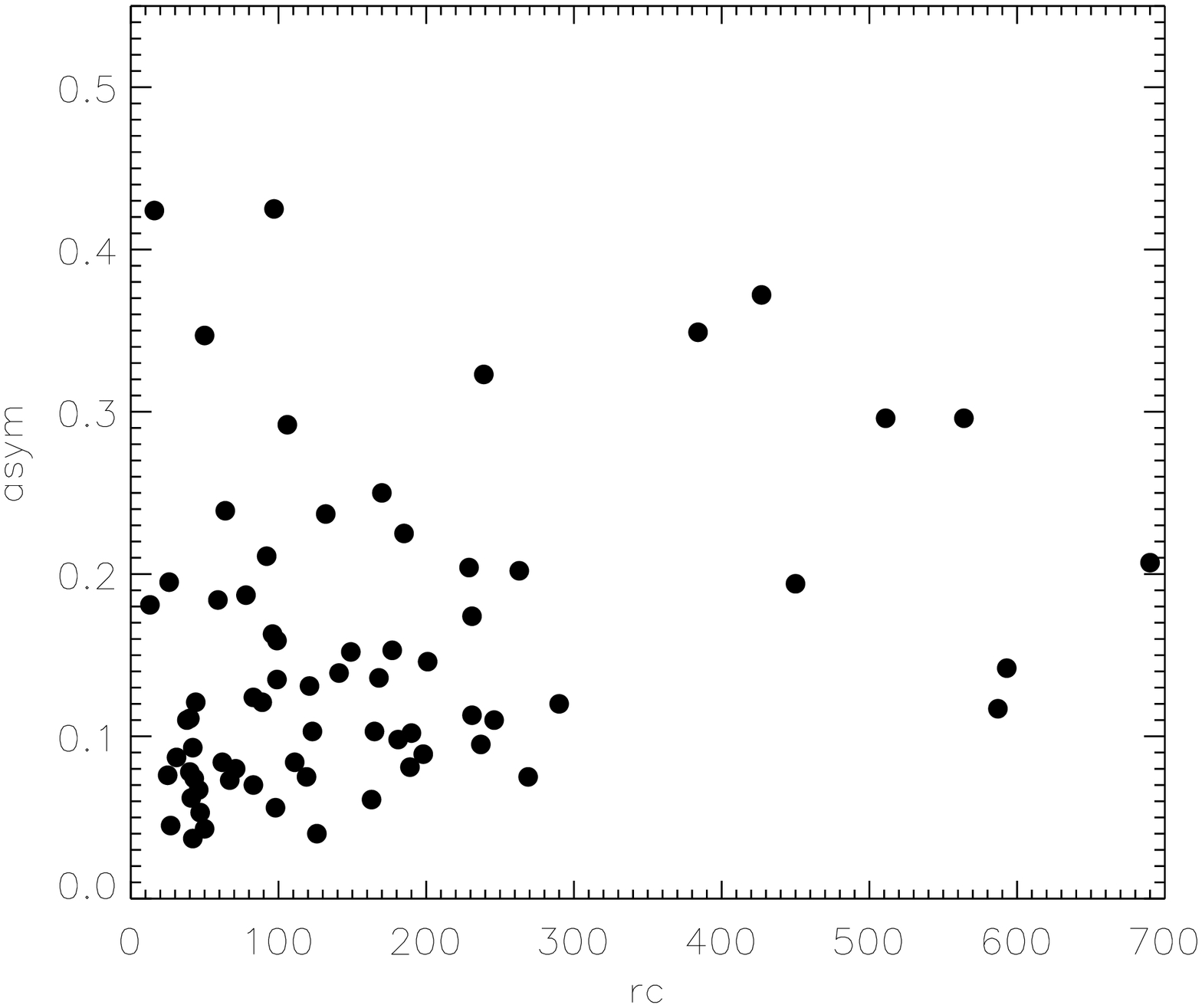}
  }
 \resizebox{\hsize}{!}{
 \includegraphics[bb=118 110 600 700,height=4cm,width=2cm,clip,angle=90]{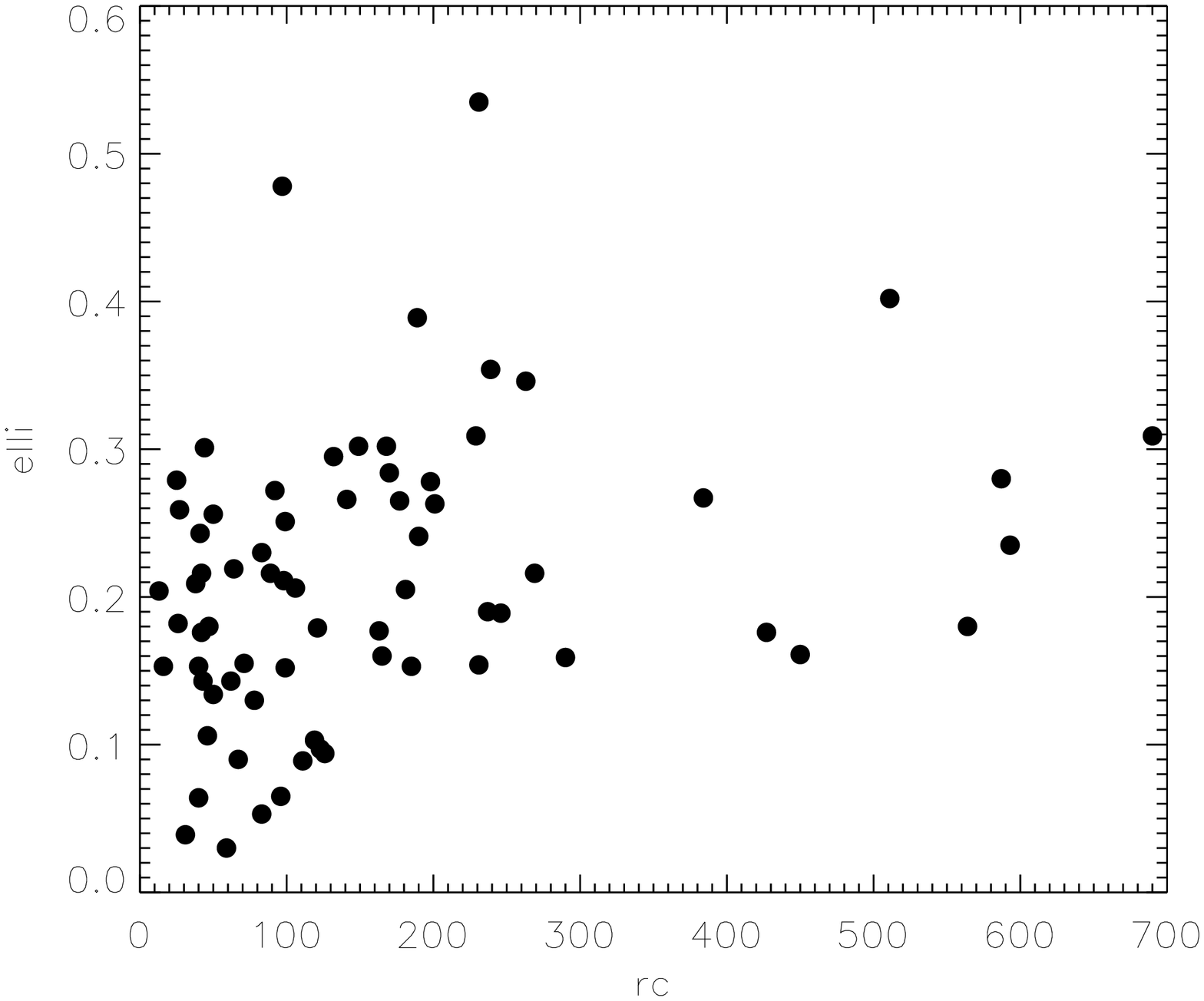}
  }
 \resizebox{\hsize}{!}{
 \includegraphics[bb=60 110 600 700,height=4cm,width=2.2cm,clip,angle=90]{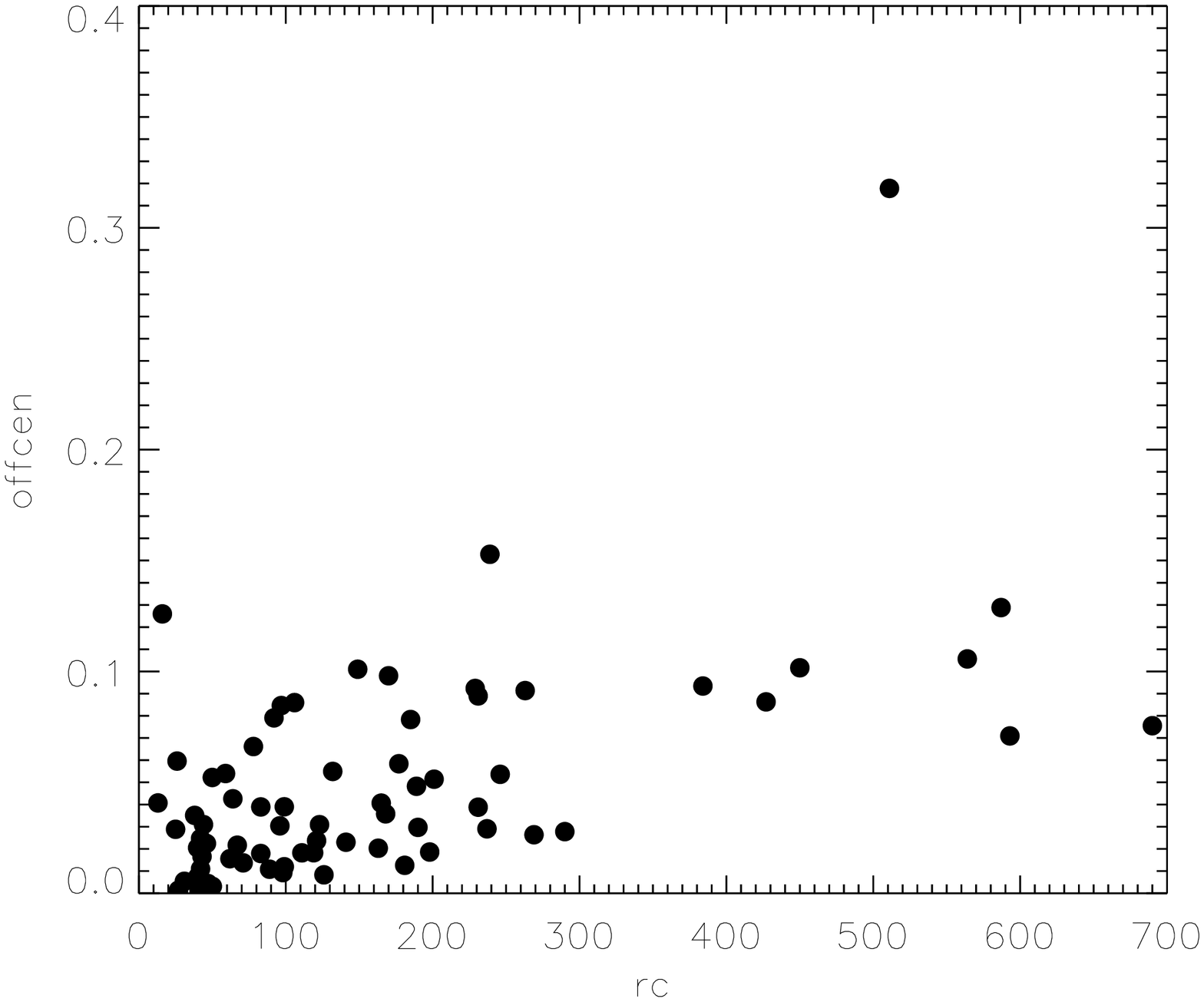}
  }
 \caption{
Our  X-ray morphological measures compared with
core radius (r$_{c}$, in kpc). 
}
\label{FigTemp}
\end{figure}

\begin{figure}[h]
 \resizebox{\hsize}{!}{
 \includegraphics[bb=118 110 600 700,height=4cm,width=2cm,clip,angle=90]{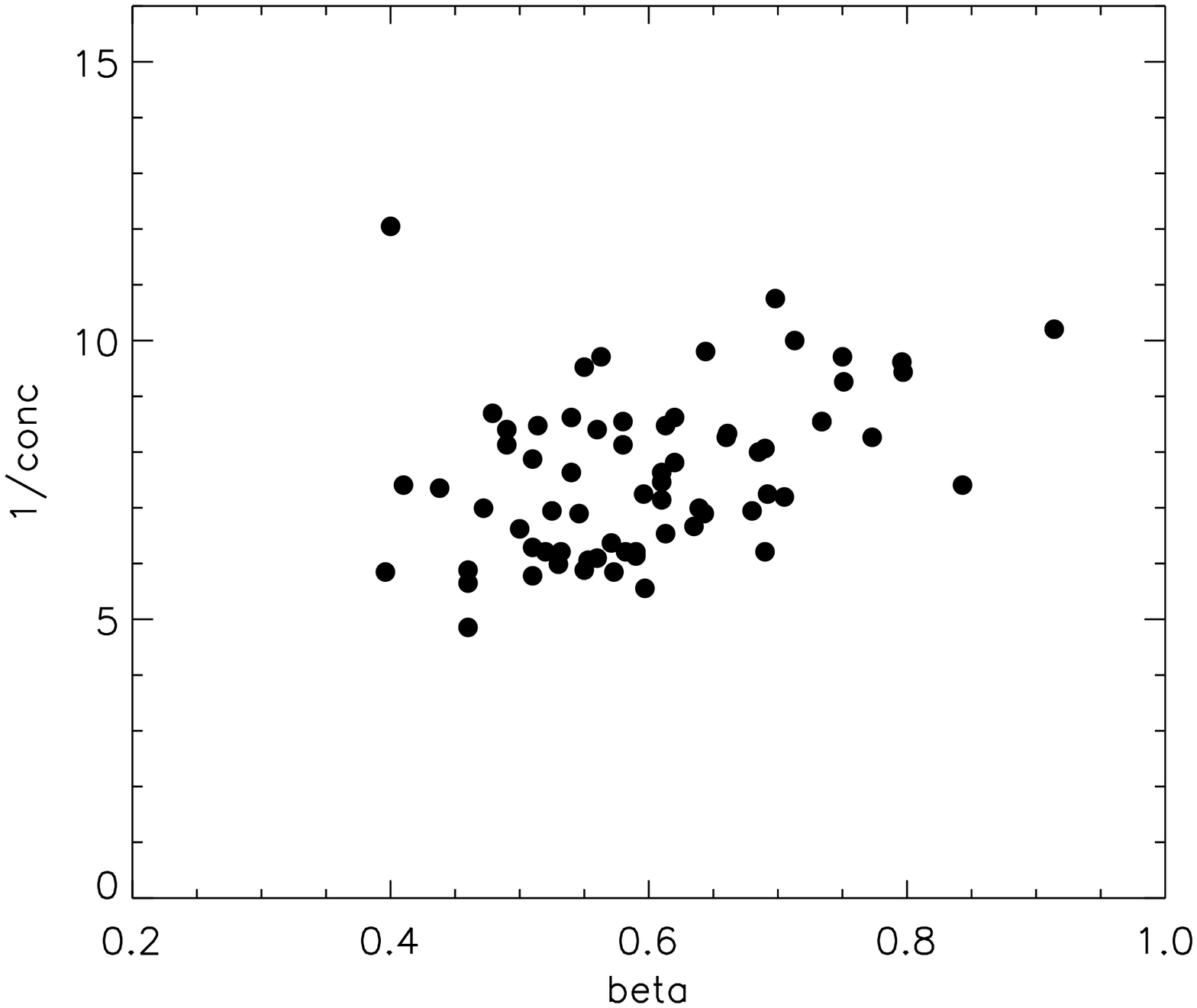}
  }
 \resizebox{\hsize}{!}{
 \includegraphics[bb=118 110 600 700,height=4cm,width=2cm,clip,angle=90]{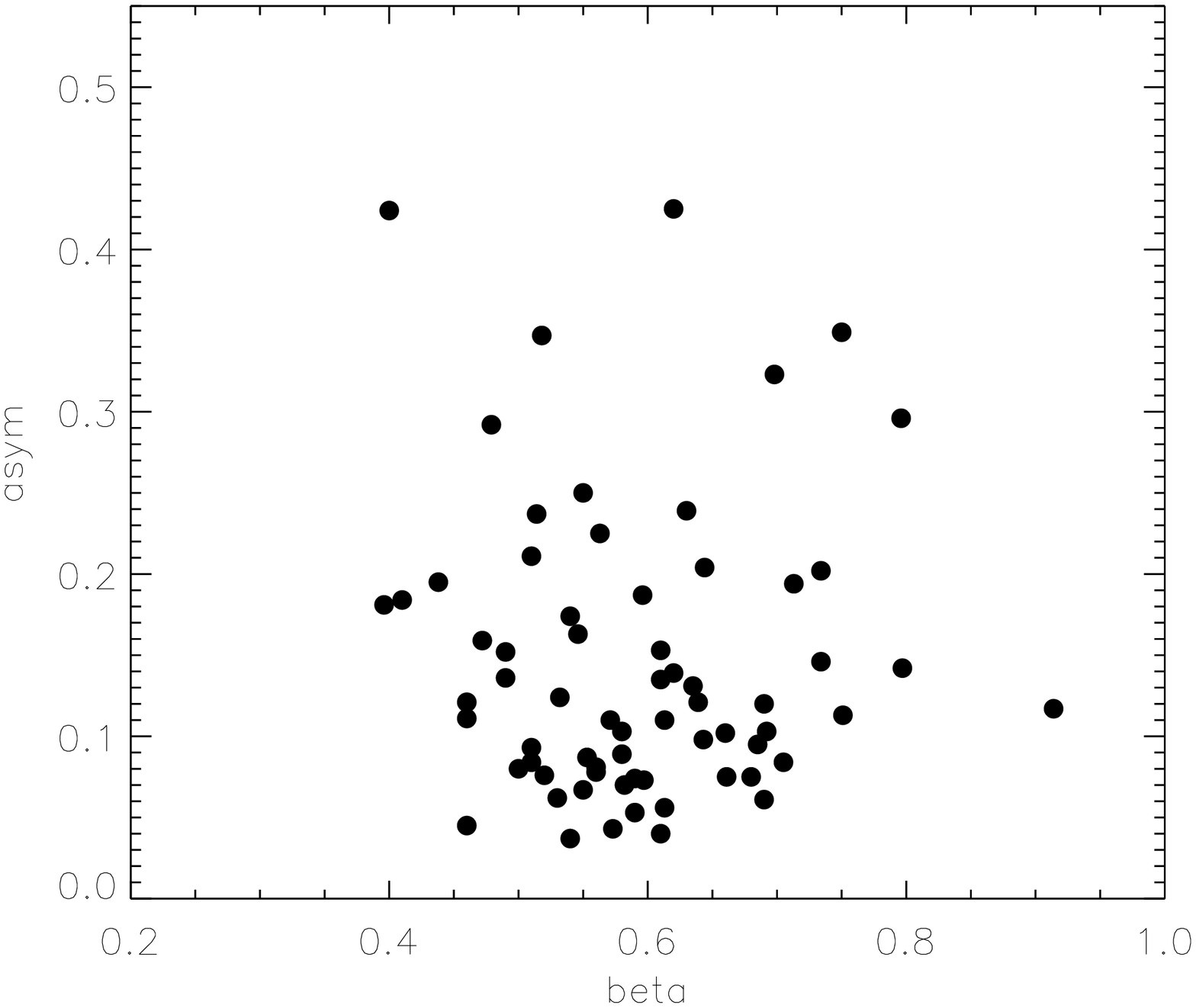}
  }
 \resizebox{\hsize}{!}{
 \includegraphics[bb=118 110 600 700,height=4cm,width=2cm,clip,angle=90]{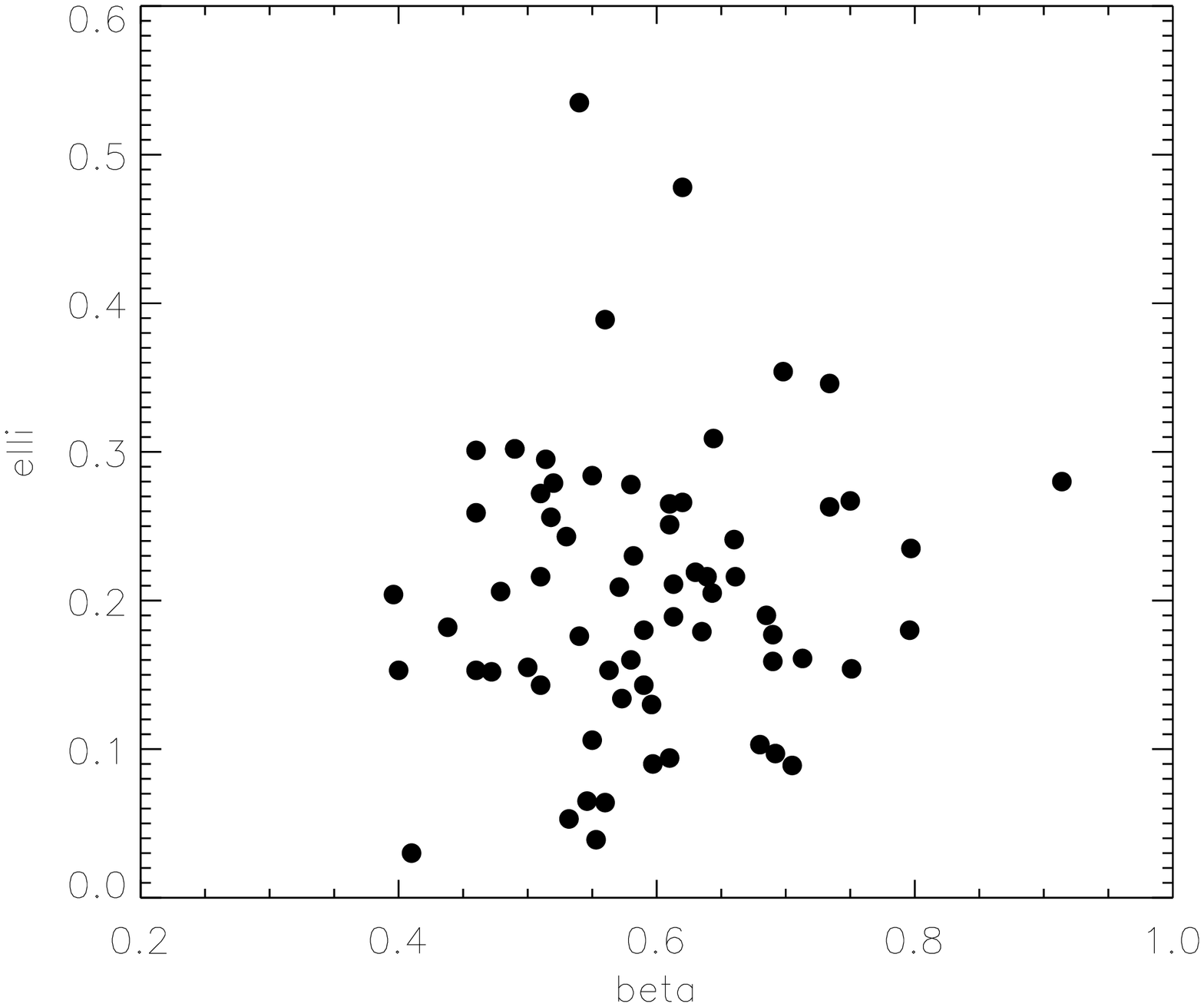}
  }
 \resizebox{\hsize}{!}{
 \includegraphics[bb=60 110 600 700,height=4cm,width=2.2cm,clip,angle=90]{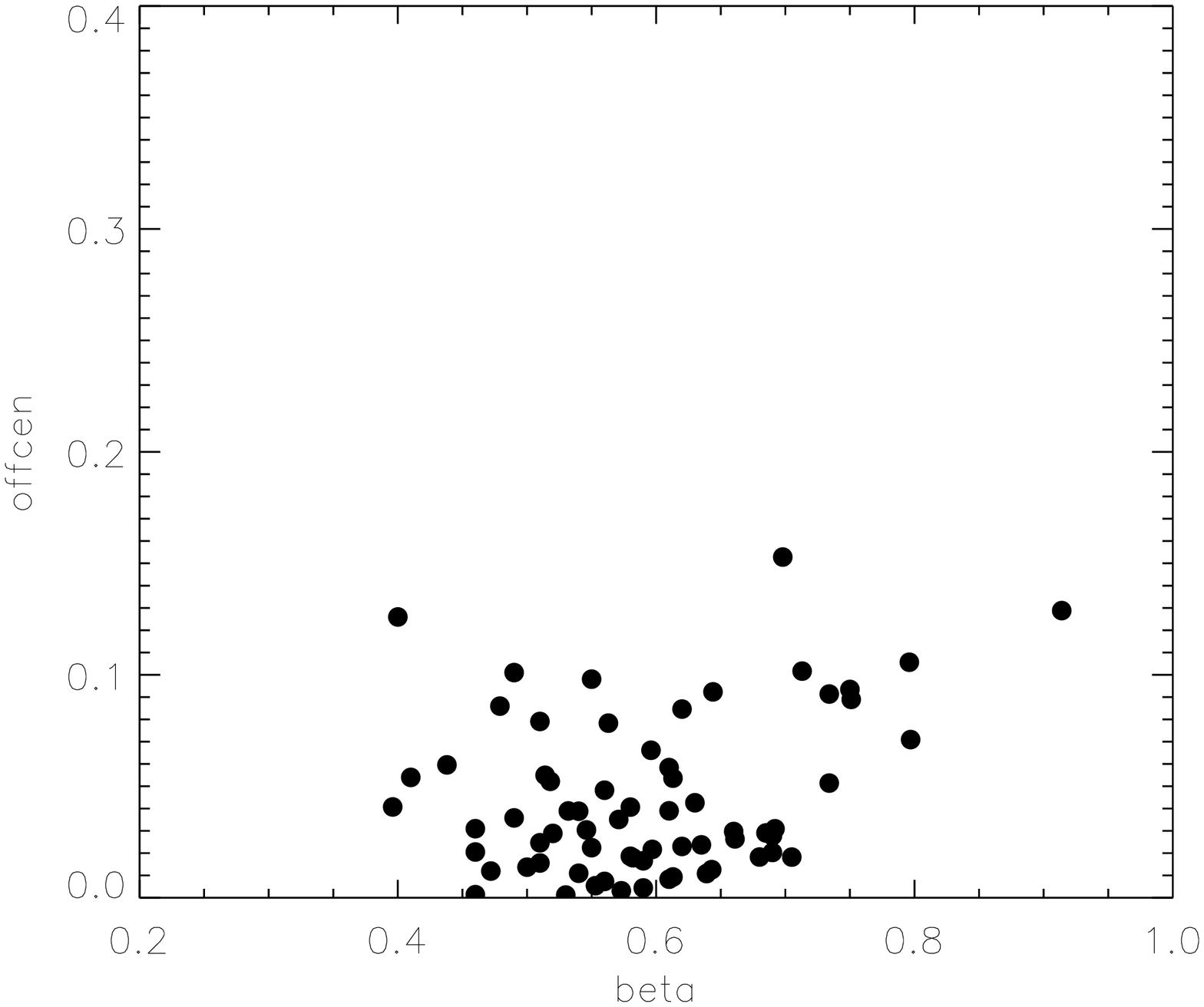}
  }
 \caption{
Comparisons between our  X-ray morphological measures and $\beta$.
}
\label{FigTemp}
\end{figure}

In Fig. 8 and 9, we compare our morphological measures with
the isothermal $\beta$-model (Cavaliere \& Fusco-Femiano 1976). 
The single $\beta$-model fitting function is written as
 \begin{eqnarray}
   S(r)=S0\left[1+\left(\frac{r}{r_c}\right)^2\right]^{-3\beta+1/2} + B
 \end{eqnarray}
where S0, r$_c$, $\beta$ and B are the central surface brightness,
core radius, the outer slope, and a constant background. 
Fig. 8 shows comparisons between r$_c$ and the measures, while
Fig. 9 shows comparisons between $\beta$ and the measures.

Again, we only used the values from 
the literature.
For our sample, 72 clusters have published
values for the single beta-model fitting, although some clusters have
very distorted X-ray morphology and cannot simply be described by the single or any beta-model. 
The value of 
Spearman $\rho$
is 0.79, 0.33, 0.37, and 0.51 for Conc, Asym, Elli, and Offcen,
respectively for r$_c$ plots, and
0.43, -0.01, 0.06 and 0.17 for Conc, Asym, Elli, and Offcen,
respectively for $\beta$ plots.
There is a correlation between Conc (or 1/Conc) and r$_c$,
while for other measures, there are no obvious correlations between our measures and $\beta$
model parameters.
The correlation between Conc and r$_c$ 
shows that
clusters with small core radii also show high concentrations.
Fig 8
illustrates that
our robust Conc measure is qualitatively similar 
to the classical morphological analysis based on $\beta$ model fitting,
and that 
Conc may be used as a robust measure of `photon expensive' r$_c$,
providing us with a possible alternative to extend the classical radial
profile analysis
to the faint high redshift universe.

\subsubsection{Power ratio}
\begin{figure}[h]
 \resizebox{\hsize}{!}{
 \includegraphics[bb=50 250 550 750,height=5cm,width=5cm,clip,angle=90]{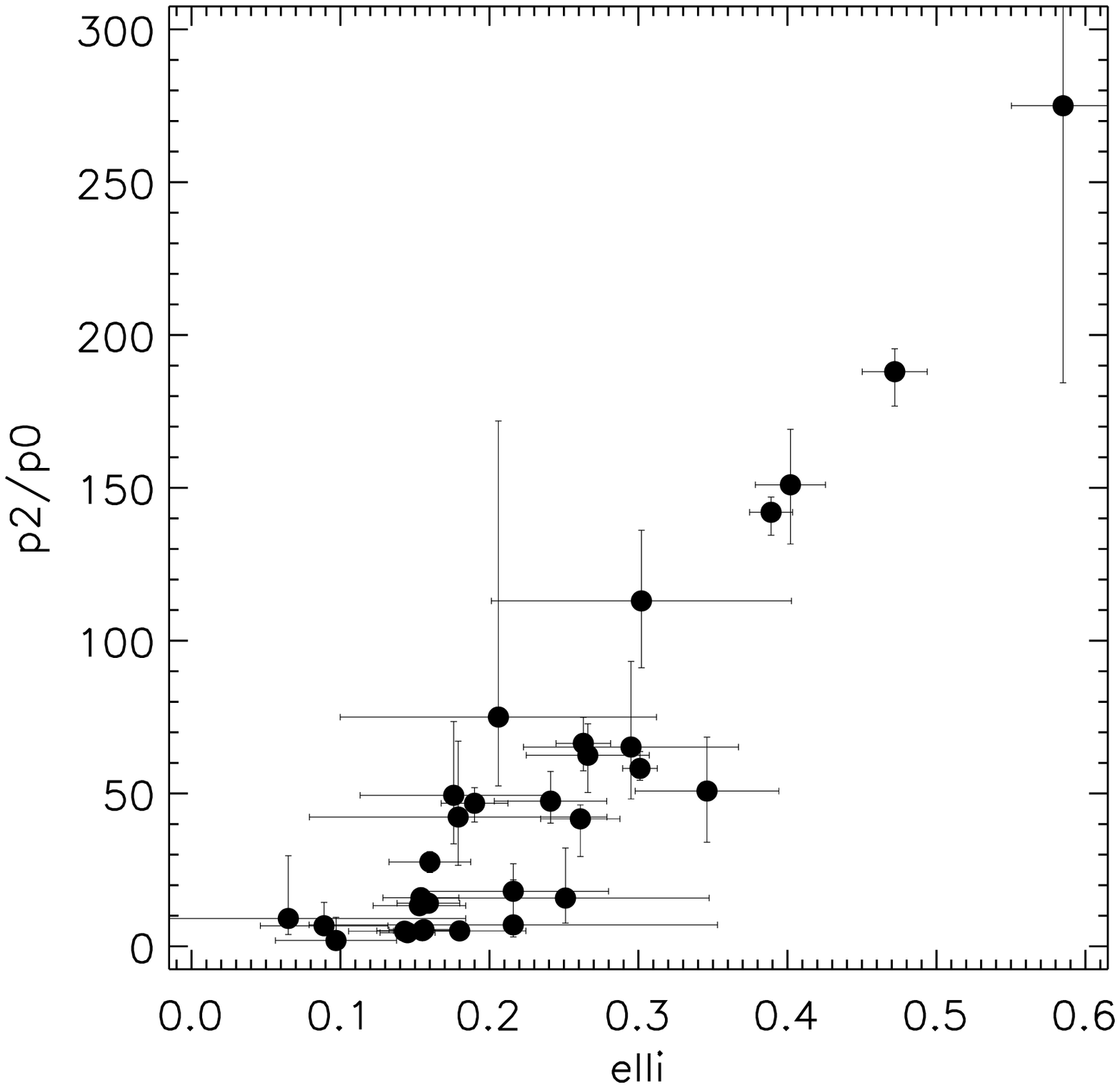}
 \includegraphics[bb=50 250 550 750,height=5cm,width=5cm,clip,angle=90]{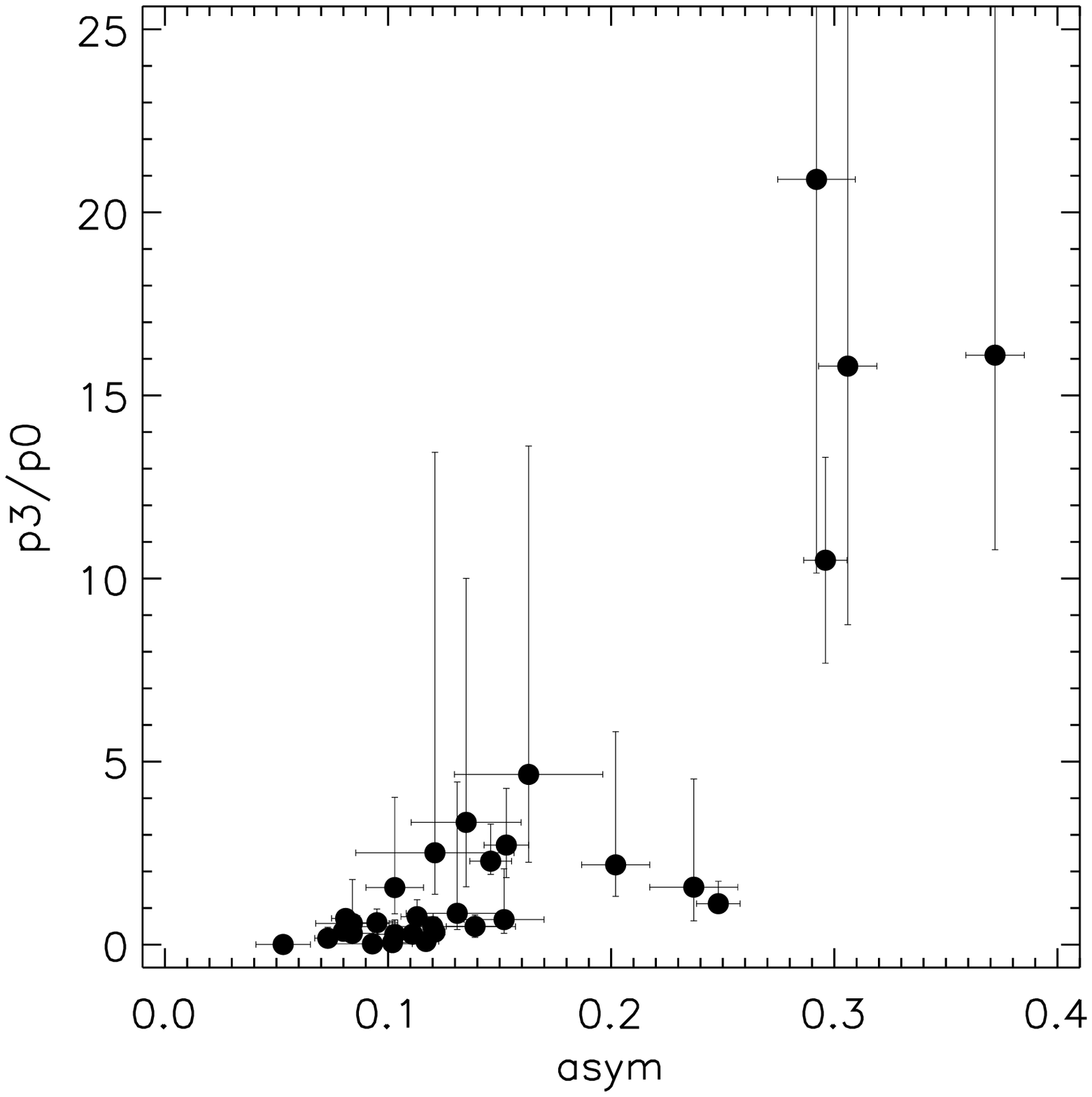}
 }
 \caption{
Our  X-ray morphological measures compared with
Power ratios.
Error bars for the power ratios are from Jeltema et al. (2005)
 estimated by Monte Carlo simulation for the 90\% confidence intervals.
The 90\% confidence intervals are then multiplied by a factor of 1/1.6 to
 be roughly comparable to our 1-sigma errors.
}
\label{FigTemp}
\end{figure}

 In Fig. 10, we compare our morphological measures with the power ratios
 (Buote \& Tsai 1995) taken from the literature.
 The power ratio method calculates the multipole moments of the
X-ray surface brightness in a circular aperture centered on the
cluster centroid. The powers are then normalized by P$_0$. 
We refer to Buote \& Tsai 1995 for more detail.
 In Jeltema et al. (2005), the power ratios are measured for 30 out of 101 clusters of our sample.
 In Fig. 10, the left panel shows a comparison between  P$_2$/P$_0$
 vs. our ellipticity, while the right panel shows P$_3$/P$_0$ vs. our
 asymmetry. Error bars for the power ratios are from Jeltema et al. (2005) 
 estimated by Monte Carlo simulation for the 90\% confidence intervals.
 (The errors from the normalization of the background are not included.)
 Error bars for our morphological measures, which are estimated
by similar Monte Carlo simulation except that we additionally
 include the effects of point sources, are plotted for the comparison.
 The 90\% confidence intervals are multiplied by a factor of 1/1.6 to 
 be roughly comparable to our one sigma errors.
 The  P$_2$/P$_0$ vs. Elli plot shows a tight correlation 
 (Spearman $\rho$ is 0.85), which 
 is expected by their similar definitions. Note also that the 
 sizes of errors are also comparable. 
 The P$_3$/P$_0$ vs. Asym plot shows a weaker correlation
 (Spearman $\rho$ is 0.79). The weaker correlation is probably 
  due to very large error bars for  P$_3$/P$_0$,
  (unfortunately much larger compared to our robust Asym,) 
  particularly for the clusters showing intrinsically high P$_3$/P$_0$ values.
  Unfortunately, almost all of  these high P$_3$/P$_0$ clusters 
   are at high redshift (z $>$ 0.4).


\subsection{Distributions of the measures}

\begin{figure}[h]
\resizebox{\hsize}{!}{\includegraphics[height=4cm,width=2cm,clip,angle=90]{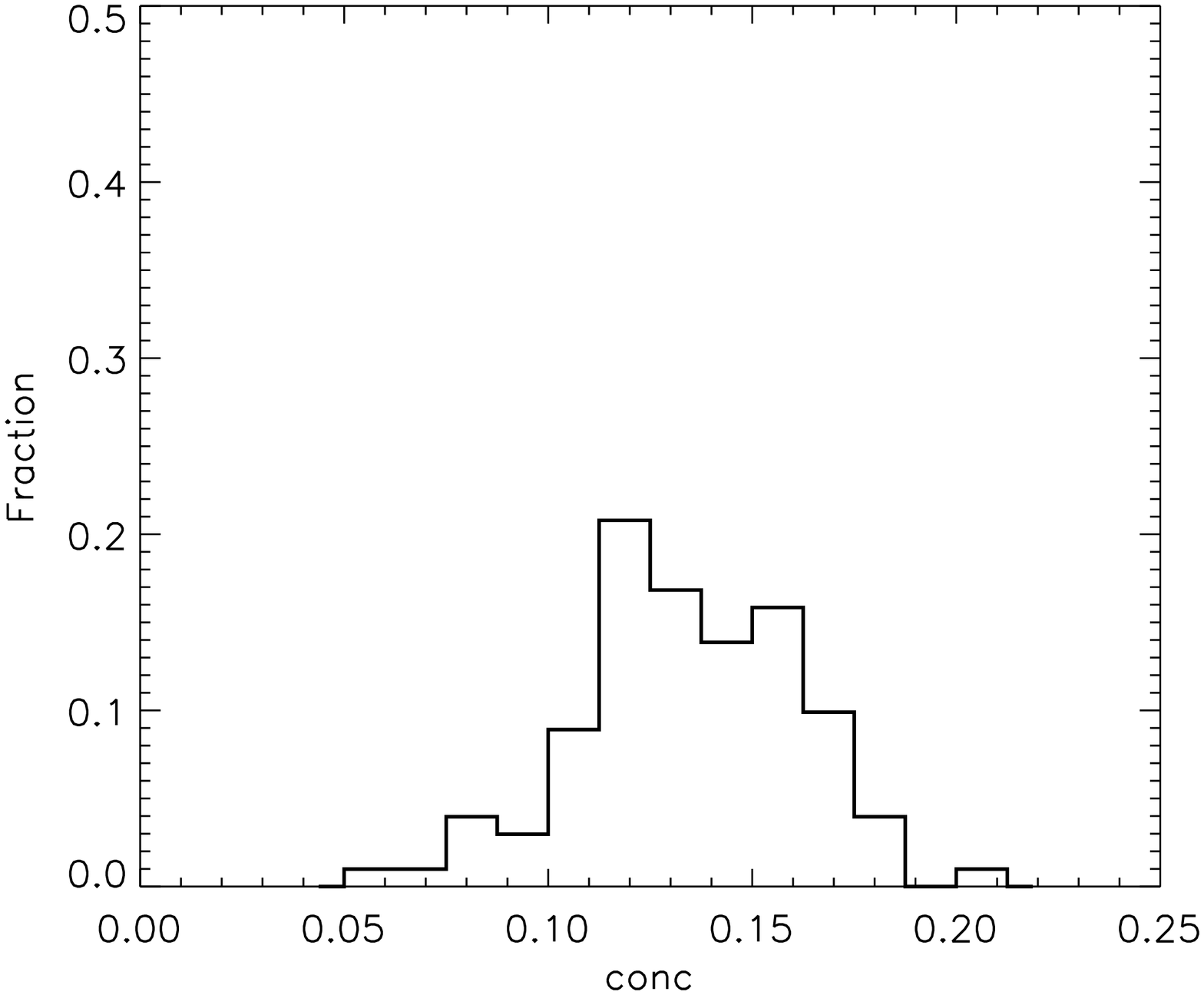}}
\resizebox{\hsize}{!}{\includegraphics[height=4cm,width=2cm,clip,angle=90]{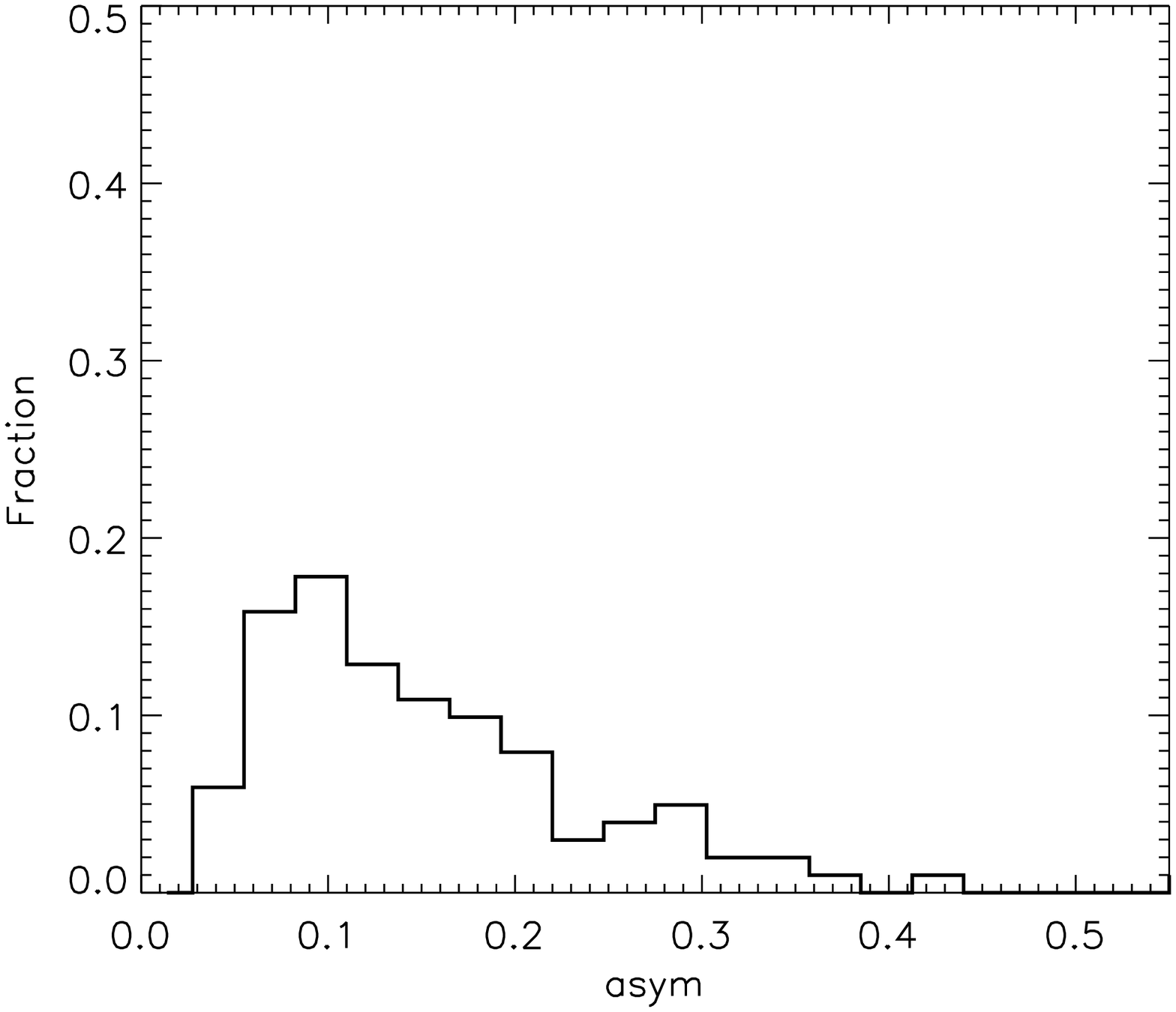}}
\resizebox{\hsize}{!}{\includegraphics[height=4cm,width=2cm,clip,angle=90]{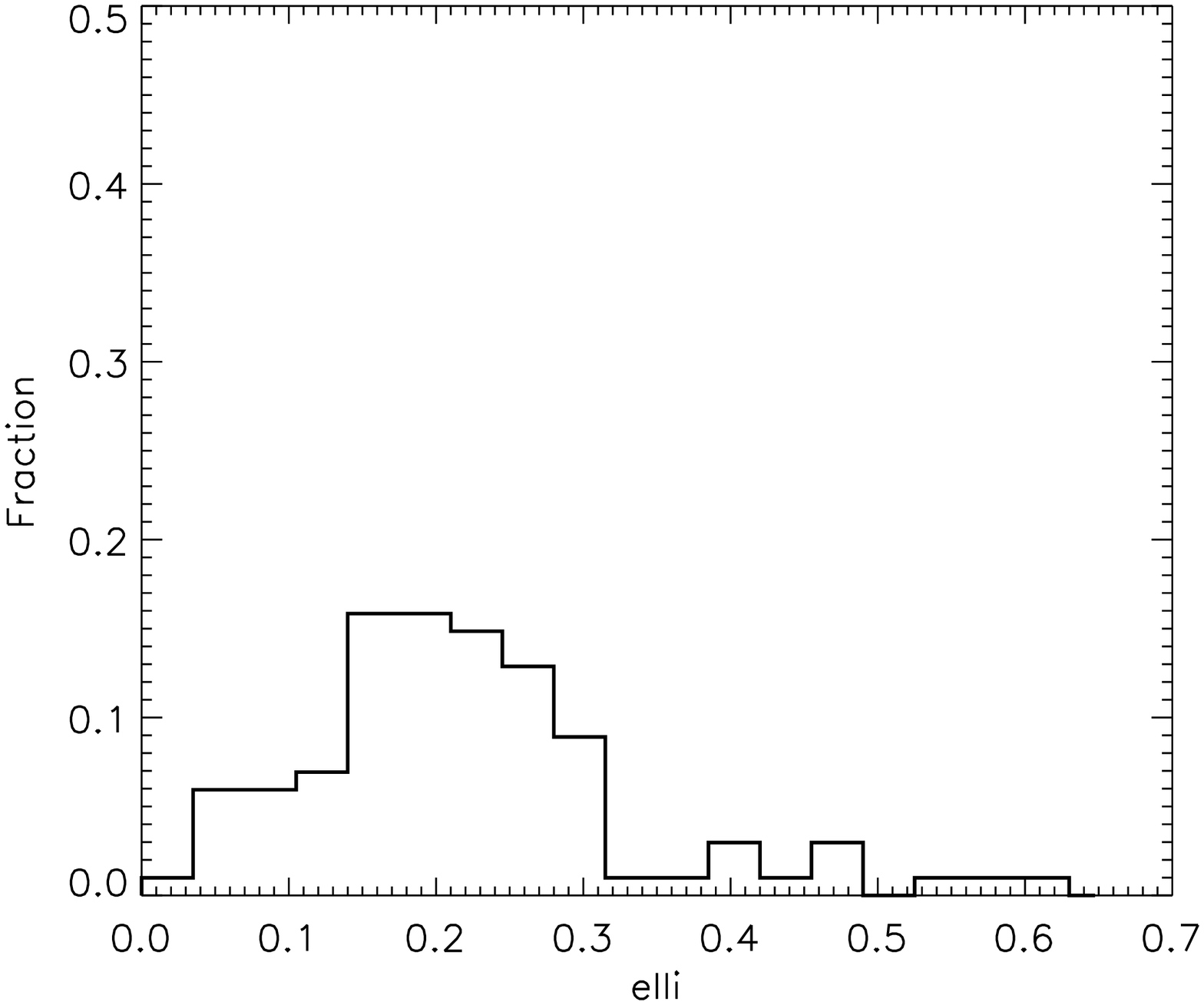}}
\resizebox{\hsize}{!}{\includegraphics[height=4cm,width=2cm,clip,angle=90]{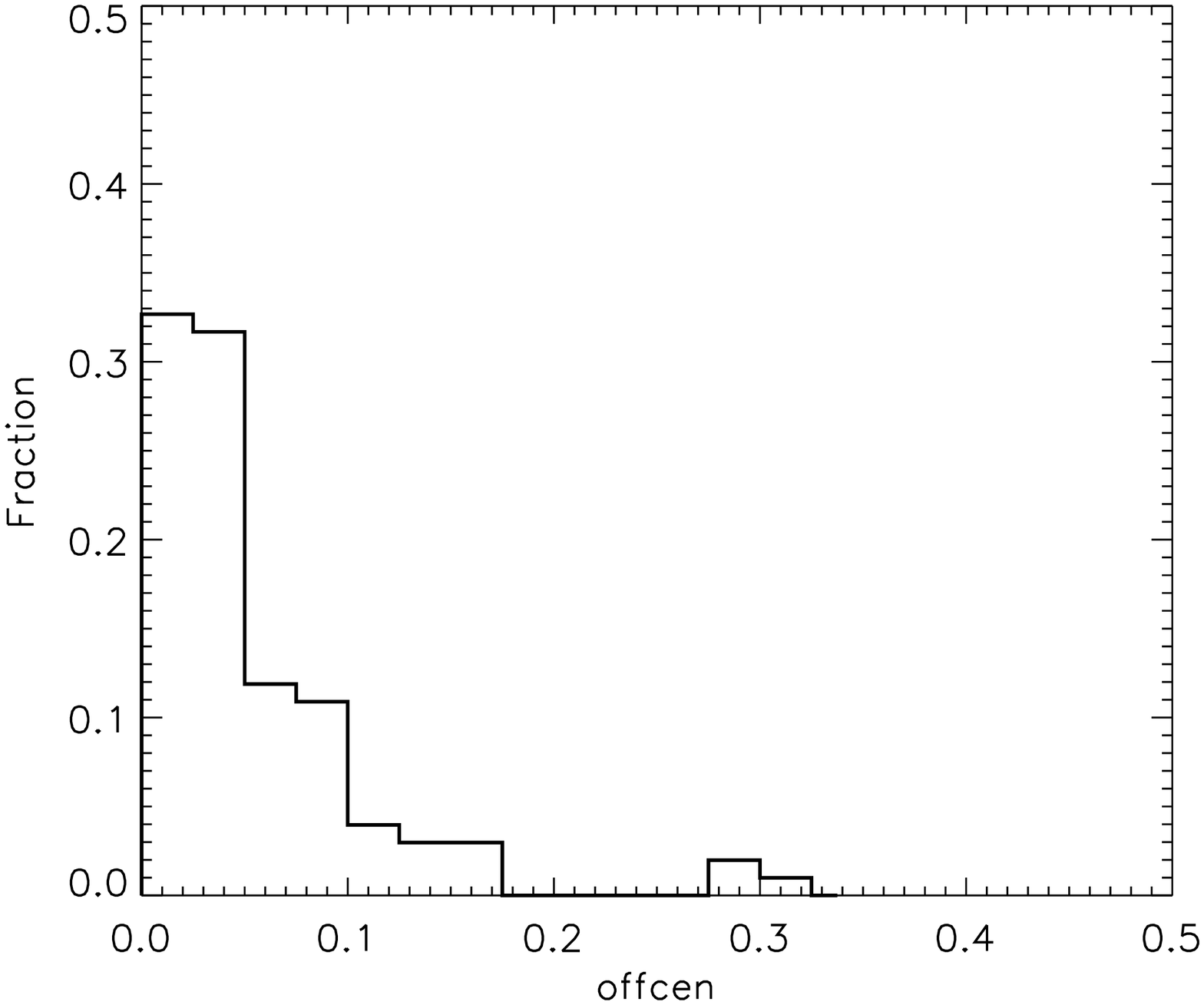}}
 \caption{
Distributions of X-ray morphological measures for the entire cluster sample.
}
\label{FigTemp}
\end{figure}

In Table 2, 
our morphological measures with 1-sigma error for the entire cluster sample
are listed, while
Fig. 11 shows  
the distributions of these measures for the entire sample.
An average value for each measure is
0.13, 0.16, 0.22, and 0.05,
while median is 0.13, 0.13, 0.21, and 0.04,
respectively for Conc, Asym, Elli, and Offcen.
Figure 12 
shows distributions of cluster morphology in Conc-Asym plane.
Each cluster point is represented by an X-ray image of the cluster.
The X-ray image is identical to the image used for the measurements of
morphology. 
The same Conc-Asym plot with 1-sigma error
is plotted in 
Fig. 13, 
together with other measure-measure planes.
In Fig. 13, 
we can see that clusters 
are scattered and occupying various places in these morphological planes,
showing various morphological characteristics.
However, there is a weak to strong trend between each set of two measures;
the value of
Spearman $\rho$
is -0.62, -0.78, 0.39 and 0.79 for Conc-Asym, Offcen-Conc, Elli-Offcen, and Elli-Asym, respectively.
 The correlation is relatively strong in the Asym-Conc and Offcen-Conc plots.
  This indicates
     that 
     low concentration clusters generally show high degree of asymmetry or 
    skewness, illustrating
the fact that there are not many highly-extended smooth symmetric clusters.  
Similarly, a correlation between Asym and Elli
may imply that
there are not many highly-elongated but otherwise smooth
symmetric clusters.
The correlation between Asym and Elli is consistent with the power ratio
analysis by Buote \& Tsai (1996).

\begin{table}[t]
\tiny
\caption[]{Morphological measures}
\begin{tabular}{lccccccc}
\hline
\hline
\hline
\noalign{\smallskip}
Name     &   Conc   & Elli & Asym      & Offcen  \\
\noalign{\smallskip}
\hline
\noalign{\smallskip}
A3562   &       0.143$\pm0.010$&0.152$\pm0.028$&0.159$\pm0.012$&0.012$\pm0.036$\\
A85     &       0.161$\pm0.004$&0.053$\pm0.013$&0.124$\pm0.005$&0.039$\pm0.014$\\
HydraA  &       0.171$\pm0.007$&0.134$\pm0.022$&0.043$\pm0.008$&0.003$\pm0.011$\\
A754    &       0.093$\pm0.004$&0.354$\pm0.008$&0.323$\pm0.008$&0.153$\pm0.032$\\
A2319   &       0.105$\pm0.006$&0.284$\pm0.021$&0.250$\pm0.009$&0.098$\pm0.023$\\
A3158   &       0.120$\pm0.005$&0.216$\pm0.016$&0.075$\pm0.007$&0.026$\pm0.033$\\
A3266   &       0.104$\pm0.005$&0.180$\pm0.037$&0.296$\pm0.023$&0.106$\pm0.021$\\
A2256   &       0.098$\pm0.007$&0.280$\pm0.025$&0.117$\pm0.021$&0.129$\pm0.049$\\
A1795   &       0.138$\pm0.003$&0.130$\pm0.008$&0.187$\pm0.004$&0.066$\pm0.011$\\
A399    &       0.100$\pm0.006$&0.161$\pm0.026$&0.194$\pm0.012$&0.102$\pm0.032$\\
A2065   &       0.123$\pm0.006$&0.309$\pm0.014$&0.207$\pm0.010$&0.076$\pm0.025$\\
A401    &       0.118$\pm0.005$&0.189$\pm0.023$&0.110$\pm0.014$&0.054$\pm0.033$\\
Zw1215 &       0.126$\pm0.012$&0.267$\pm0.046$&0.110$\pm0.012$&0.036$\pm0.022$\\
A2029   &       0.161$\pm0.007$&0.230$\pm0.026$&0.070$\pm0.029$&0.018$\pm0.033$\\
A2255   &       0.106$\pm0.010$&0.235$\pm0.036$&0.142$\pm0.025$&0.071$\pm0.041$\\
A1651   &       0.145$\pm0.011$&0.205$\pm0.039$&0.098$\pm0.040$&0.013$\pm0.029$\\
A478    &       0.153$\pm0.007$&0.211$\pm0.026$&0.056$\pm0.007$&0.009$\pm0.014$\\
RXJ1844 &       0.158$\pm0.044$&0.385$\pm0.071$&0.202$\pm0.022$&0.061$\pm0.019$\\
A2244   &       0.140$\pm0.005$&0.094$\pm0.014$&0.040$\pm0.004$&0.008$\pm0.014$\\
RXJ0820 &       0.122$\pm0.046$&0.044$\pm0.074$&0.085$\pm0.024$&0.017$\pm0.020$\\
A2034   &       0.124$\pm0.008$&0.159$\pm0.021$&0.120$\pm0.037$&0.028$\pm0.042$\\
A2069   &       0.127$\pm0.013$&0.466$\pm0.033$&0.214$\pm0.010$&0.036$\pm0.069$\\
RXJ0819 &       0.151$\pm0.023$&0.083$\pm0.073$&0.200$\pm0.019$&0.050$\pm0.026$\\
A1068   &       0.161$\pm0.010$&0.279$\pm0.043$&0.076$\pm0.007$&0.029$\pm0.011$\\
A2409   &       0.131$\pm0.016$&0.129$\pm0.054$&0.094$\pm0.013$&0.041$\pm0.036$\\
A2204   &       0.180$\pm0.012$&0.090$\pm0.043$&0.073$\pm0.009$&0.022$\pm0.014$\\
Hercu.A       &       0.147$\pm0.012$&0.212$\pm0.065$&0.073$\pm0.013$&0.007$\pm0.021$\\
A750    &       0.134$\pm0.013$&0.141$\pm0.073$&0.074$\pm0.043$&0.028$\pm0.026$\\
A2259   &       0.131$\pm0.019$&0.279$\pm0.061$&0.126$\pm0.017$&0.031$\pm0.040$\\
RXJ1720 &       0.158$\pm0.006$&0.156$\pm0.023$&0.073$\pm0.006$&0.018$\pm0.013$\\
A1201   &       0.122$\pm0.014$&0.457$\pm0.061$&0.292$\pm0.011$&0.077$\pm0.019$\\
A586    &       0.144$\pm0.016$&0.103$\pm0.056$&0.075$\pm0.014$&0.018$\pm0.026$\\
A2218   &       0.117$\pm0.007$&0.160$\pm0.027$&0.103$\pm0.008$&0.041$\pm0.028$\\
A1914   &       0.108$\pm0.006$&0.154$\pm0.025$&0.113$\pm0.007$&0.089$\pm0.029$\\
A2294   &       0.132$\pm0.024$&0.062$\pm0.074$&0.176$\pm0.018$&0.064$\pm0.038$\\
A1689   &       0.161$\pm0.005$&0.177$\pm0.019$&0.061$\pm0.007$&0.020$\pm0.016$\\
A1204   &       0.156$\pm0.011$&0.166$\pm0.040$&0.049$\pm0.010$&0.005$\pm0.013$\\
MS0839 &       0.164$\pm0.015$&0.064$\pm0.063$&0.078$\pm0.014$&0.007$\pm0.015$\\
A115    &       0.083$\pm0.006$&0.153$\pm0.015$&0.424$\pm0.010$&0.126$\pm0.011$\\
A520    &       0.086$\pm0.009$&0.261$\pm0.027$&0.153$\pm0.010$&0.157$\pm0.052$\\
A963    &       0.151$\pm0.008$&0.155$\pm0.031$&0.080$\pm0.008$&0.014$\pm0.015$\\
RXJ0439 &       0.185$\pm0.020$&0.113$\pm0.060$&0.092$\pm0.023$&0.018$\pm0.025$\\
A2111   &       0.119$\pm0.024$&0.302$\pm0.101$&0.152$\pm0.018$&0.101$\pm0.061$\\
A1423   &       0.168$\pm0.022$&0.265$\pm0.086$&0.114$\pm0.025$&0.017$\pm0.032$\\
Zw0949 &       0.167$\pm0.014$&0.243$\pm0.065$&0.062$\pm0.013$&0.001$\pm0.015$\\
MS0735  &       0.206$\pm0.009$&0.259$\pm0.025$&0.045$\pm0.009$&0.001$\pm0.009$\\
A773    &       0.121$\pm0.010$&0.241$\pm0.038$&0.102$\pm0.011$&0.030$\pm0.027$\\
A2261   &       0.159$\pm0.009$&0.143$\pm0.038$&0.084$\pm0.010$&0.016$\pm0.015$\\
A1682   &       0.103$\pm0.030$&0.267$\pm0.112$&0.349$\pm0.024$&0.093$\pm0.081$\\
A1763   &       0.123$\pm0.014$&0.302$\pm0.050$&0.136$\pm0.014$&0.036$\pm0.038$\\
A2219   &       0.119$\pm0.007$&0.389$\pm0.014$&0.081$\pm0.006$&0.048$\pm0.028$\\
A267    &       0.128$\pm0.011$&0.266$\pm0.041$&0.139$\pm0.013$&0.023$\pm0.034$\\
A2390   &       0.170$\pm0.005$&0.301$\pm0.012$&0.121$\pm0.005$&0.031$\pm0.013$\\
RXJ2129  &       0.173$\pm0.016$&0.216$\pm0.064$&0.093$\pm0.018$&0.025$\pm0.025$\\
RXJ0439  &       0.147$\pm0.015$&0.199$\pm0.047$&0.155$\pm0.011$&0.037$\pm0.024$\\
A2125   &       0.140$\pm0.021$&0.289$\pm0.087$&0.247$\pm0.014$&0.026$\pm0.044$\\
A68     &       0.131$\pm0.023$&0.265$\pm0.057$&0.153$\pm0.017$&0.058$\pm0.045$\\
Zw1454 &       0.163$\pm0.006$&0.143$\pm0.019$&0.074$\pm0.005$&0.017$\pm0.008$\\
A1835   &       0.170$\pm0.006$&0.106$\pm0.018$&0.067$\pm0.005$&0.022$\pm0.012$\\
A1758   &       0.084$\pm0.011$&0.472$\pm0.022$&0.248$\pm0.010$&0.298$\pm0.061$\\
A697    &       0.123$\pm0.011$&0.278$\pm0.039$&0.089$\pm0.013$&0.019$\pm0.025$\\
Zw1021 &       0.158$\pm0.007$&0.145$\pm0.018$&0.117$\pm0.006$&0.039$\pm0.013$\\
A781    &       0.060$\pm0.048$&0.215$\pm0.119$&0.568$\pm0.026$&0.143$\pm0.080$\\
A2552   &       0.128$\pm0.020$&0.187$\pm0.073$&0.140$\pm0.015$&0.039$\pm0.028$\\
A1722   &       0.127$\pm0.023$&0.272$\pm0.058$&0.211$\pm0.016$&0.079$\pm0.035$\\
MS1358   &       0.177$\pm0.011$&0.153$\pm0.031$&0.111$\pm0.009$&0.021$\pm0.013$\\
RXJ1158  &       0.177$\pm0.013$&0.217$\pm0.049$&0.083$\pm0.013$&0.006$\pm0.012$\\
A370    &       0.113$\pm0.009$&0.377$\pm0.026$&0.074$\pm0.012$&0.039$\pm0.048$\\
RXJ1532  &       0.161$\pm0.012$&0.180$\pm0.044$&0.053$\pm0.012$&0.004$\pm0.016$\\
MS1512   &       0.131$\pm0.014$&0.176$\pm0.043$&0.037$\pm0.015$&0.011$\pm0.018$\\
RXJ0850  &       0.115$\pm0.015$&0.243$\pm0.054$&0.144$\pm0.018$&0.060$\pm0.034$\\
RXJ0949  &       0.145$\pm0.023$&0.125$\pm0.052$&0.108$\pm0.015$&0.035$\pm0.026$\\
Zw0024 &       0.135$\pm0.037$&0.030$\pm0.092$&0.184$\pm0.031$&0.054$\pm0.025$\\
RXJ1416  &       0.136$\pm0.029$&0.240$\pm0.081$&0.168$\pm0.027$&0.056$\pm0.027$\\
RXJ2228 &       0.120$\pm0.017$&0.215$\pm0.060$&0.180$\pm0.019$&0.114$\pm0.038$\\
MS1621  &       0.103$\pm0.022$&0.153$\pm0.063$&0.225$\pm0.016$&0.078$\pm0.048$\\
RXJ1347   &       0.157$\pm0.005$&0.209$\pm0.015$&0.110$\pm0.005$&0.035$\pm0.013$\\
RXJ1701  &       0.171$\pm0.026$&0.204$\pm0.076$&0.181$\pm0.019$&0.041$\pm0.021$\\
3c295   &       0.165$\pm0.021$&0.039$\pm0.059$&0.087$\pm0.017$&0.005$\pm0.015$\\
\noalign{\smallskip}
\hline
\end{tabular}
\end{table}

\begin{table}[t]
\tiny
{\bf Table 2} (continued)
\\~
\\~
\\~
\begin{tabular}{lccccccc}
\hline
\hline
\noalign{\smallskip}
Name     &   Conc   & Elli & Asym      & Offcen  \\
\noalign{\smallskip}
\hline
\noalign{\smallskip}
RXJ1641  &       0.156$\pm0.032$&0.194$\pm0.101$&0.282$\pm0.028$&0.025$\pm0.031$\\
CRSSJ0030 &       0.125$\pm0.011$&0.230$\pm0.013$&0.256$\pm0.007$&0.046$\pm0.018$\\
RXJ1525 &       0.102$\pm0.022$&0.309$\pm0.071$&0.204$\pm0.015$&0.092$\pm0.051$\\
MS0451   &       0.117$\pm0.008$&0.263$\pm0.018$&0.146$\pm0.009$&0.051$\pm0.029$\\
MS0016  &       0.125$\pm0.009$&0.190$\pm0.022$&0.095$\pm0.009$&0.029$\pm0.026$\\
RXJ1121  &       0.100$\pm0.028$&0.176$\pm0.063$&0.372$\pm0.013$&0.086$\pm0.070$\\
RXJ0848   &       0.116$\pm0.081$&0.425$\pm0.127$&0.185$\pm0.014$&0.050$\pm0.062$\\
MS2053    &       0.134$\pm0.028$&0.251$\pm0.096$&0.135$\pm0.025$&0.039$\pm0.020$\\
RXJ0542  &       0.118$\pm0.019$&0.295$\pm0.072$&0.237$\pm0.020$&0.055$\pm0.042$\\
RXJ1221  &       0.117$\pm0.014$&0.346$\pm0.048$&0.202$\pm0.015$&0.091$\pm0.041$\\
RXJ1113 &       0.143$\pm0.097$&0.216$\pm0.137$&0.121$\pm0.035$&0.011$\pm0.035$\\
RXJ2302 &       0.145$\pm0.036$&0.065$\pm0.119$&0.163$\pm0.033$&0.030$\pm0.027$\\
MS1137   &       0.139$\pm0.012$&0.089$\pm0.043$&0.084$\pm0.016$&0.018$\pm0.020$\\
RXJ1350  &       0.115$\pm0.031$&0.206$\pm0.106$&0.292$\pm0.017$&0.086$\pm0.037$\\
RXJ1716  &       0.150$\pm0.021$&0.179$\pm0.100$&0.131$\pm0.023$&0.024$\pm0.027$\\
MS1054   &       0.092$\pm0.009$&0.402$\pm0.024$&0.296$\pm0.010$&0.318$\pm0.063$\\
RXJ0152   &       0.066$\pm0.013$&0.585$\pm0.035$&0.306$\pm0.013$&0.293$\pm0.060$\\
WGA1226  &       0.138$\pm0.012$&0.097$\pm0.041$&0.103$\pm0.013$&0.031$\pm0.021$\\
RXJ0910  &       0.135$\pm0.000$&0.163$\pm0.000$&0.268$\pm0.000$&0.027$\pm0.000$\\
RXJ1053 &       0.085$\pm0.014$&0.601$\pm0.068$&0.330$\pm0.008$&0.174$\pm0.019$\\
RXJ1252  &       0.144$\pm0.000$&0.208$\pm0.000$&0.184$\pm0.000$&0.033$\pm0.000$\\
RXJ0849   &       0.121$\pm0.021$&0.113$\pm0.098$&0.168$\pm0.007$&0.036$\pm0.067$\\
\noalign{\smallskip}
\hline
\end{tabular}
\end{table}

\begin{figure}
\resizebox{\hsize}{!}{\includegraphics[height=4cm,width=3cm,clip,angle=90]{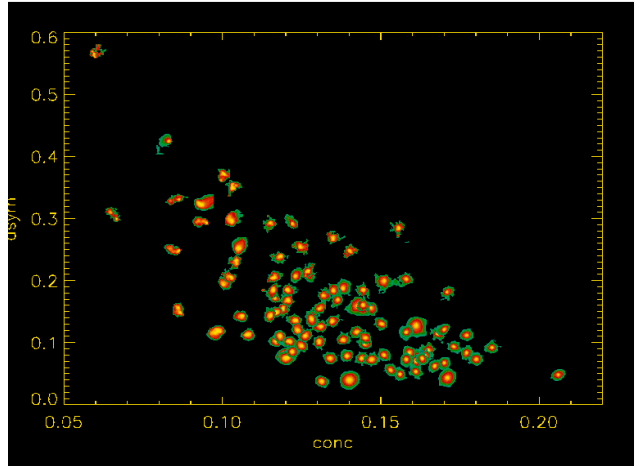}}
 \caption{
Distributions of cluster X-ray morphology in the Conc-Asym plane.
Each cluster point is represented by a X-ray image of the cluster.
The X-ray image is identical to the image used for the measurements of
morphology. 
}
\label{FigTemp}
\end{figure}

\begin{figure}
 \resizebox{\hsize}{!}{
 \includegraphics[height=4cm,width=2cm,clip,angle=90]{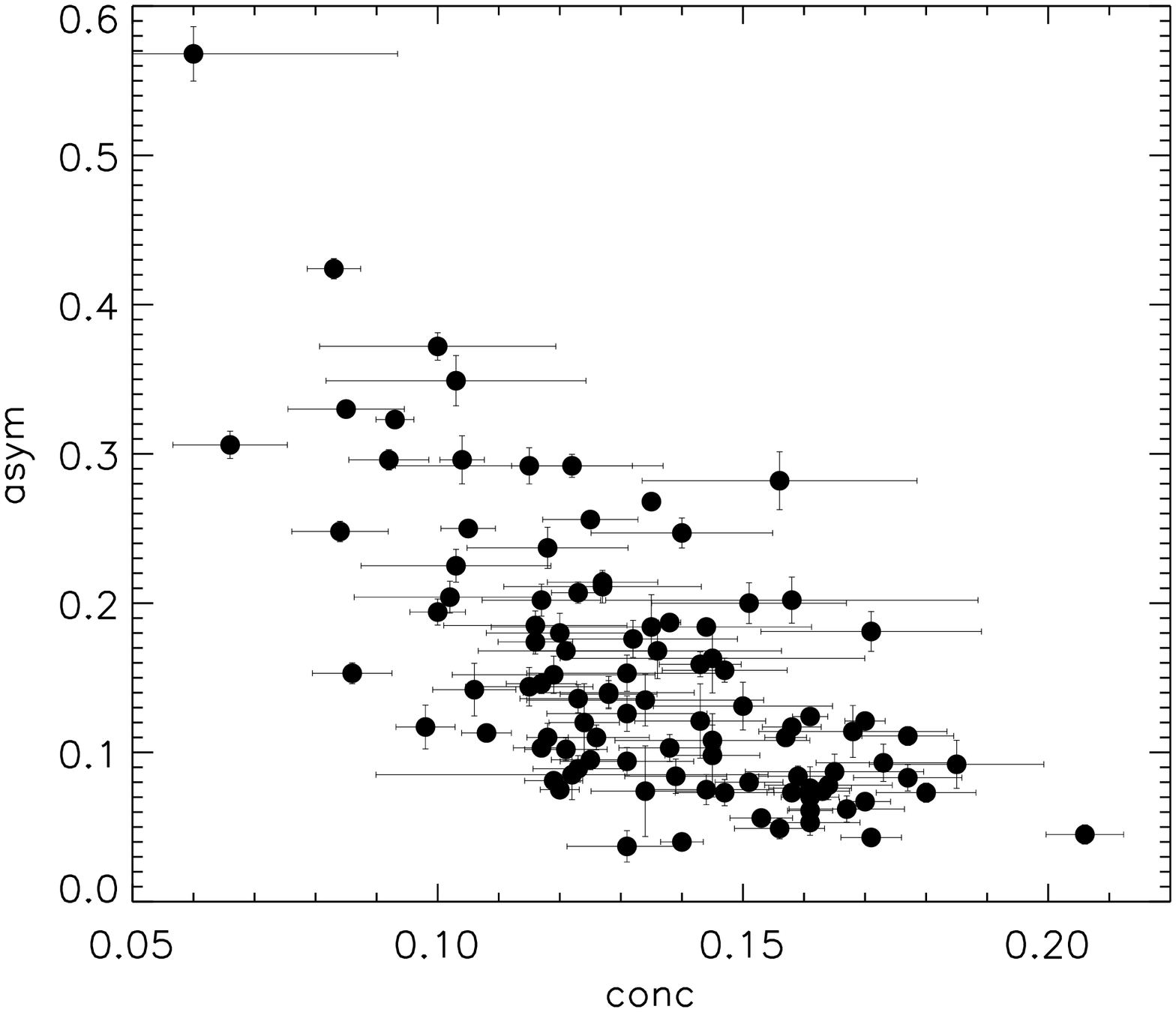}
  }
 \resizebox{\hsize}{!}{
 \includegraphics[height=4cm,width=2cm,clip,angle=90]{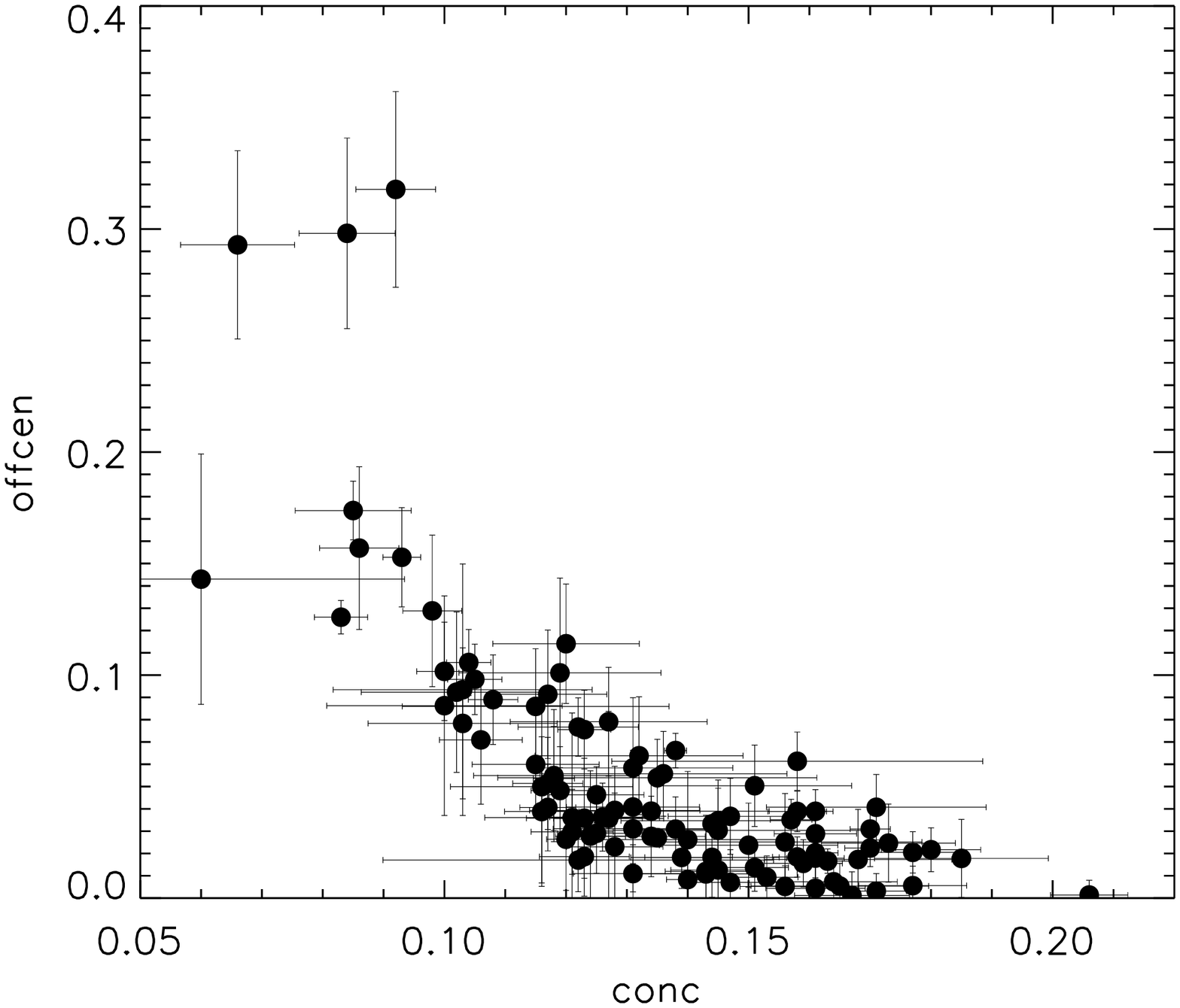}
  }
 \resizebox{\hsize}{!}{
 \includegraphics[height=4cm,width=2cm,clip,angle=90]{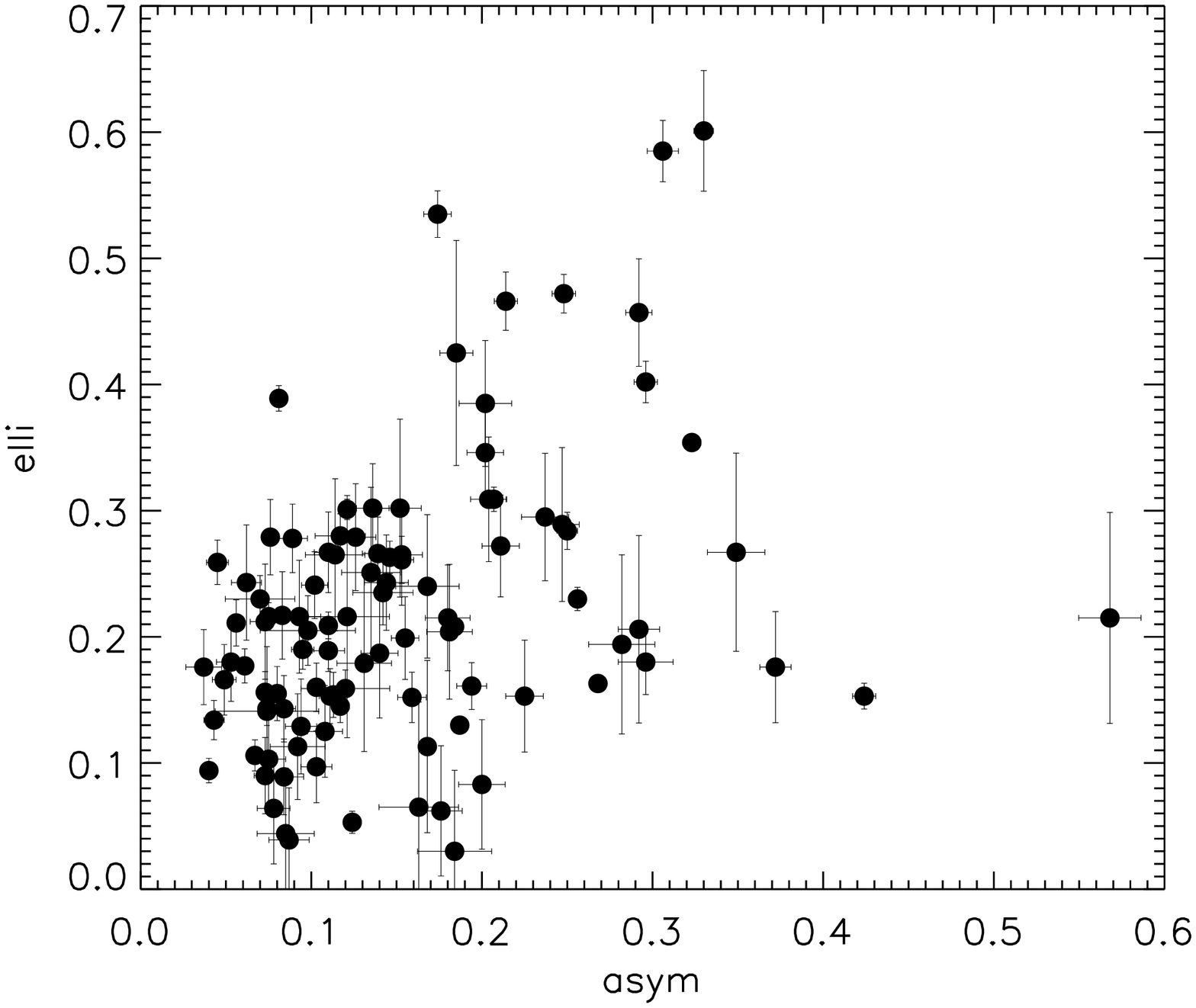}
  }
 \resizebox{\hsize}{!}{
 \includegraphics[height=4cm,width=2cm,clip,angle=90]{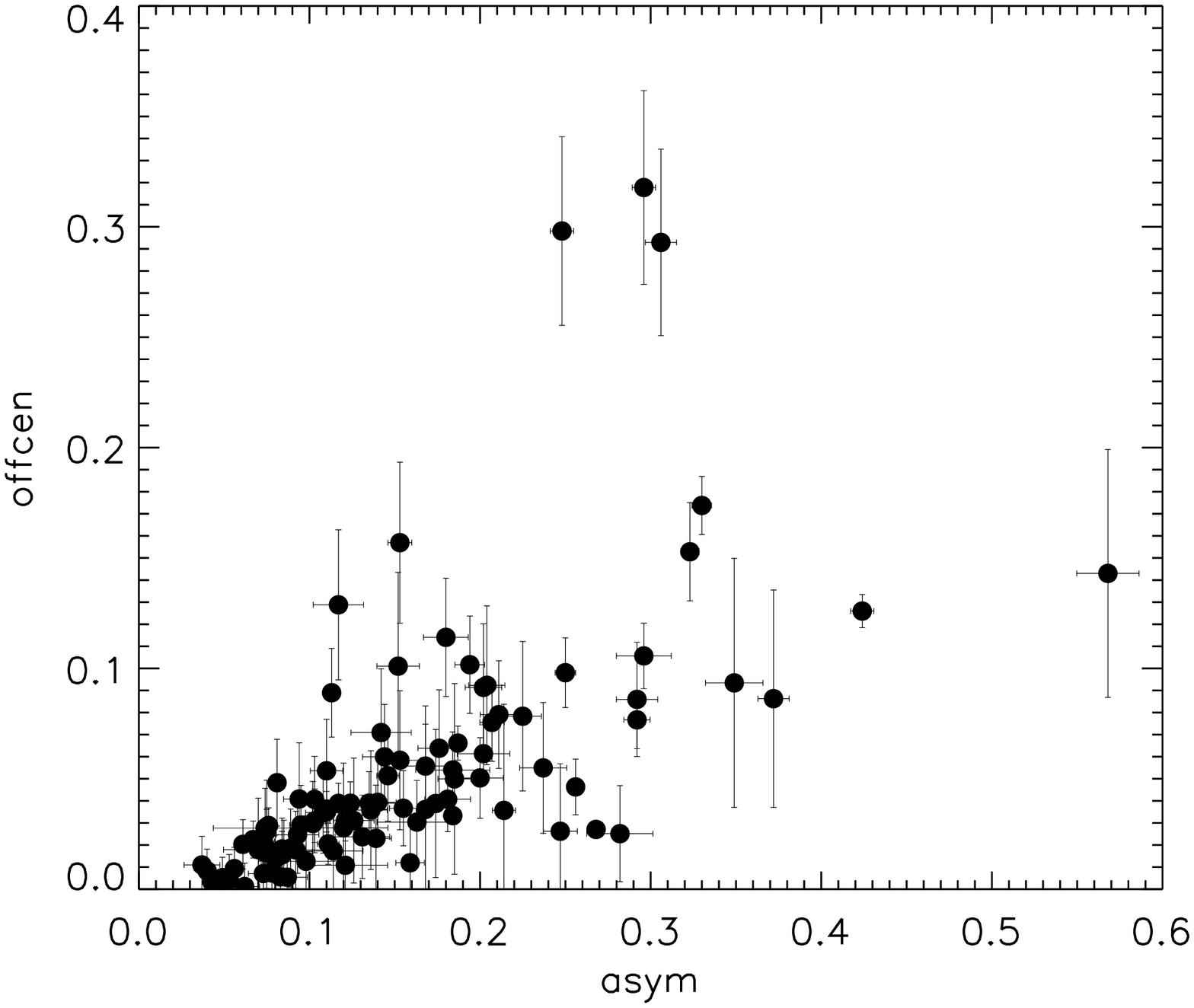}
  }
 \caption{
Distributions of clusters in various X-ray  measure-measure planes 
with 1-sigma error.
}
\label{FigTemp}
\end{figure}

\subsection{L-T relation}
\begin{figure}
\resizebox{\hsize}{!}{\includegraphics[height=4cm,width=3cm,clip,angle=90]{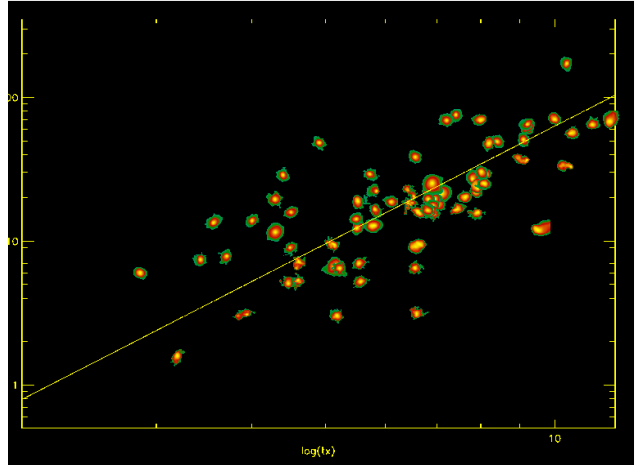}}
 \caption{
Distribution of cluster X-ray morphology in the L-T plane.
The solid line is the fitted line by Wu et al. (1999) based on  the 
local cluster sample.
}
\label{FigTemp}
\end{figure}

\begin{figure}[h]
 \resizebox{\hsize}{!}{\includegraphics[height=4cm,width=2cm,clip,angle=90]{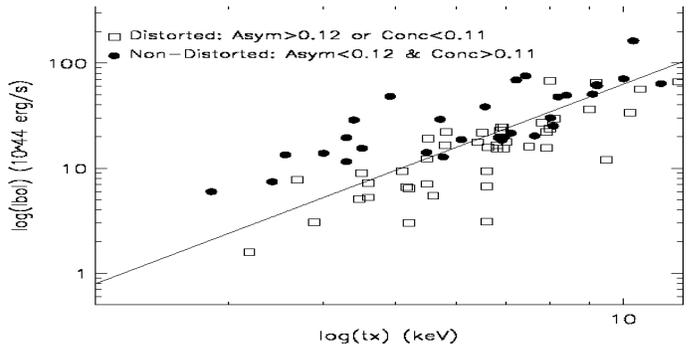}}
 \caption{
The L-T relation with morphological subsets. 
The solid ovals are  the ``non-distorted'' clusters and open rectangles are
the ``distorted'' clusters. 
}
\label{FigTemp}
\end{figure}

\begin{figure}[h]
 \resizebox{\hsize}{!}{\includegraphics[height=4cm,width=2cm,clip,angle=90]{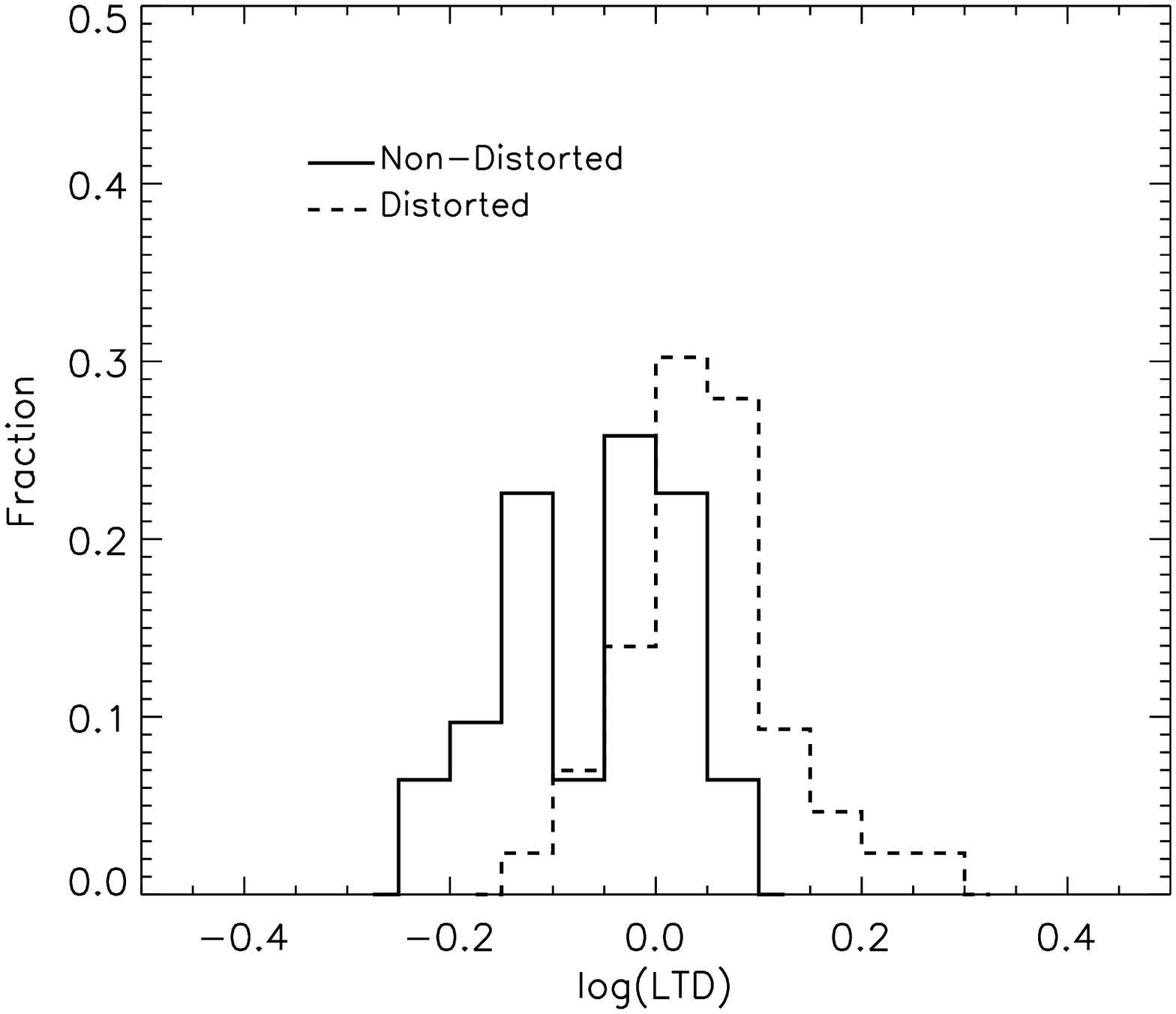}}
 \caption{
Distributions of 
the shortest distance to the L-T line (LTD) for 
``non-distorted'' and ``distorted'' clusters.
}
\label{FigTemp}
\end{figure}

In Fig 14, 
we plot a distribution of cluster morphology in 
the bolometric luminosity (Lbol) and X-ray temperature (Tx) plane 
using a subset of our sample
with available literature data.
The straight line is a fit from Wu et al. (1999) 
based on 256 low redshift clusters
in the form of log(Lbol)=2.72log(Tx)-0.92 
showing
that our sample  follows the standard L-T relationship 
(Spearman $\rho$ = 0.73)
with some scatter. 
To be consistent with the fitted line, here we used H$_o$=50, $\Omega_m$=1, $\Omega_{\Lambda}$=0, cosmology.
When we subdivide the sample according to their apparent distortions,
using quantitative definition: ``distorted'' to be Asym $>$ 0.12 or Conc $<$ 0.11, and
``non-distorted'' to be Asym $<$ 0.12 and Conc $>$ 0.11,
and separately plot using
open rectangles 
for the ``distorted'' clusters 
and solid ovals for the ``non-distorted" clusters 
(Fig. 15), 
we see that the distorted and non-distorted clusters occupy 
well-defined loci in the L-T plane.
If we plot  
distributions of the 
shortest
distance from the point to the L-T line (LTD) 
in the log(Lbol) and log(Tx) plane 
for the two subsets 
(Fig. 16), with
positive sign meaning a lower log(Lbol) than the L-T line, 
we see the same trend.
For Fig. 16, 
a K-S test shows the probability that the two
distributions are drawn from the same parent distribution is 
only 9.56 $\times$ 10$^{-5}$.
Meanwhile, we do not detect any significant difference in the width of 
the distributions between two subsets 
($\sigma$= 0.078 and 0.081, for the ``distorted'' and 
``non-distorted'' clusters, respectively).

\subsection{Evolutionary effects}

\begin{figure}
 \resizebox{\hsize}{!}{\includegraphics[height=2cm,width=3.5cm,clip,angle=0]{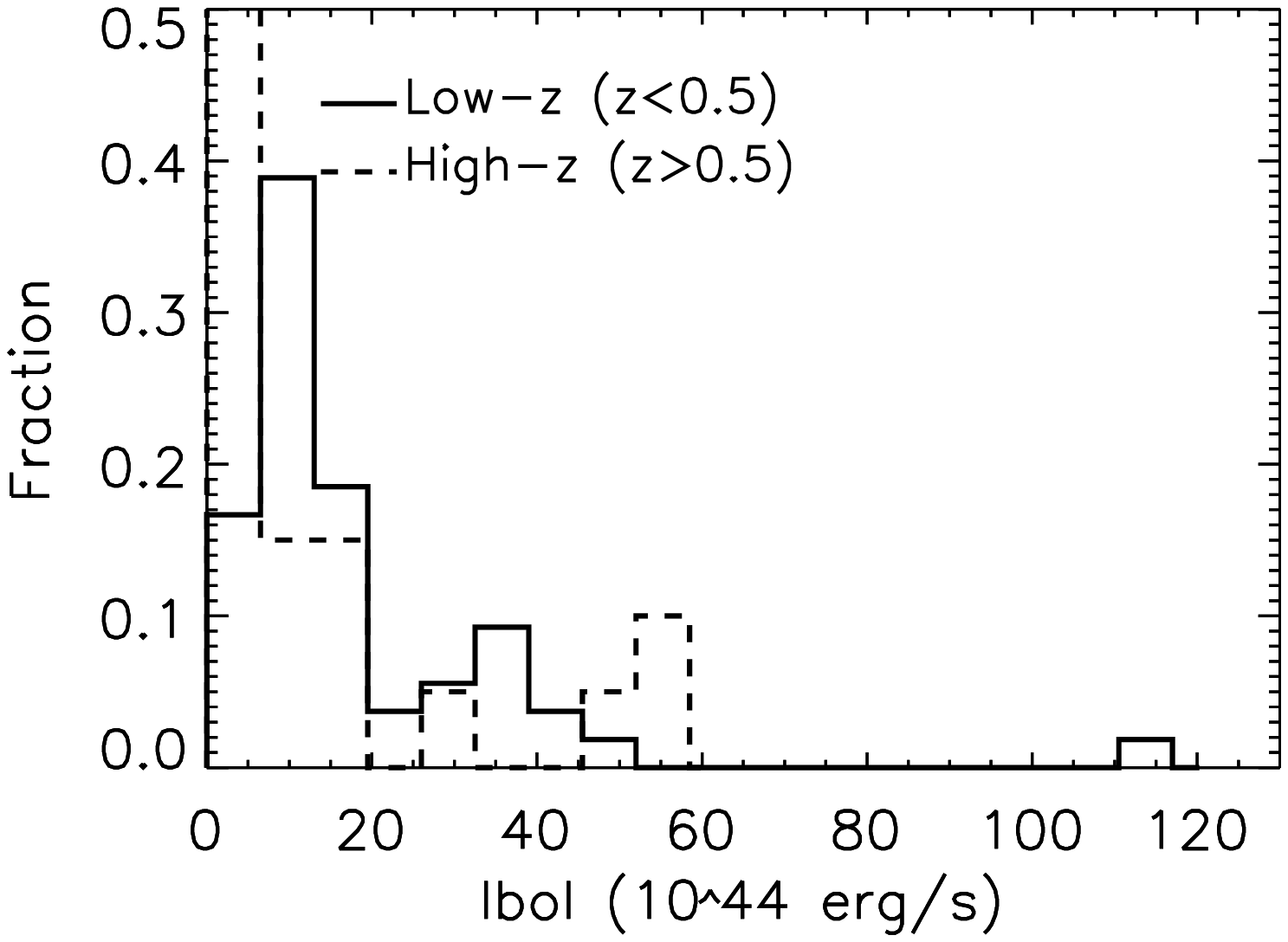}}
 \caption{
Distributions of Lbol for high and low redshift cluster subsets. 
}
\label{FigTemp}
\end{figure}

To investigate evolutionary effects, we subdivide our sample into low-z and
high-z subsets using a border z=0.5.
Fig. 17 
shows distributions of Lbol for the high-z and low-z samples,
while 
Fig. 18 
shows distributions of our measures for high-z and low-z samples.
A K-S test shows the probability that the two
  distributions are drawn from the same parent distribution is 
 0.01, 7$\times$10$^{-3}$, 0.50, and 0.15
  for Conc, Asym, Elli, and  Offcen, respectively.
 Table 3 summarizes K-S statistics for various redshift subsets,
including the redshift border other than z = 0.5. 
In Table 3, overall, we see no strong evolution in any of our measures
for any redshift subset.
 There is a hint of weak evolution for  Conc and Asym,
 in such a way that high-z clusters show more distorted morphology,
 consistent with the sophisticated power ratio method of 
 Jeltema et al. (2005), but it is unfortunately not statistically 
 very significant.
 Note also that the possible weak trend in Conc becomes 
 increasingly
 insignificant with the high-z (z$>$0.3) vs. low-z (z$<$0.3)
 comparison in Table 3.

\begin{table}[h]
\scriptsize
\caption[]{K-S probability with various redshift subsets.}
\begin{tabular}{lccc}
\hline
\hline
\noalign{\smallskip}
Measure & z$>$0.5 vs z$<$0.5 & z$>$0.5 vs z$<$0.3 & z$>$0.3 vs z$<$0.3 \\
\noalign{\smallskip}
\hline
\noalign{\smallskip}
Conc & 0.01 & 0.03 & 0.33 \\
Asym & 0.007 & 0.005 & 0.005 \\
Elli & 0.50 & 0.65 & 0.50 \\
Offcen & 0.15 & 0.14 & 0.28 \\
\noalign{\smallskip}
\hline
\end{tabular}
\end{table}

\begin{figure}
 \resizebox{\hsize}{!}{\includegraphics[height=2cm,width=4cm,clip,angle=0]{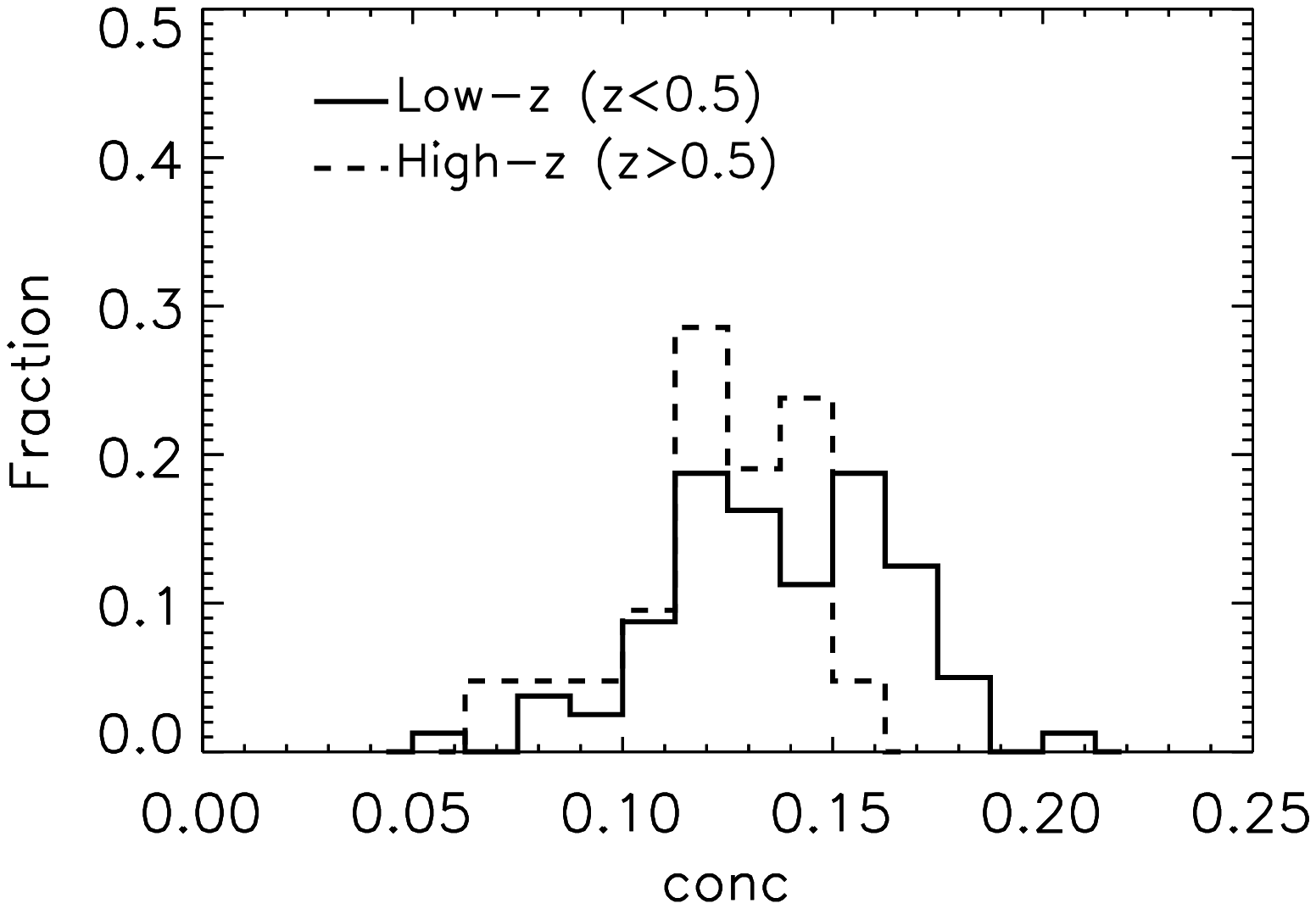}}
 \resizebox{\hsize}{!}{\includegraphics[height=2cm,width=4cm,clip,angle=0]{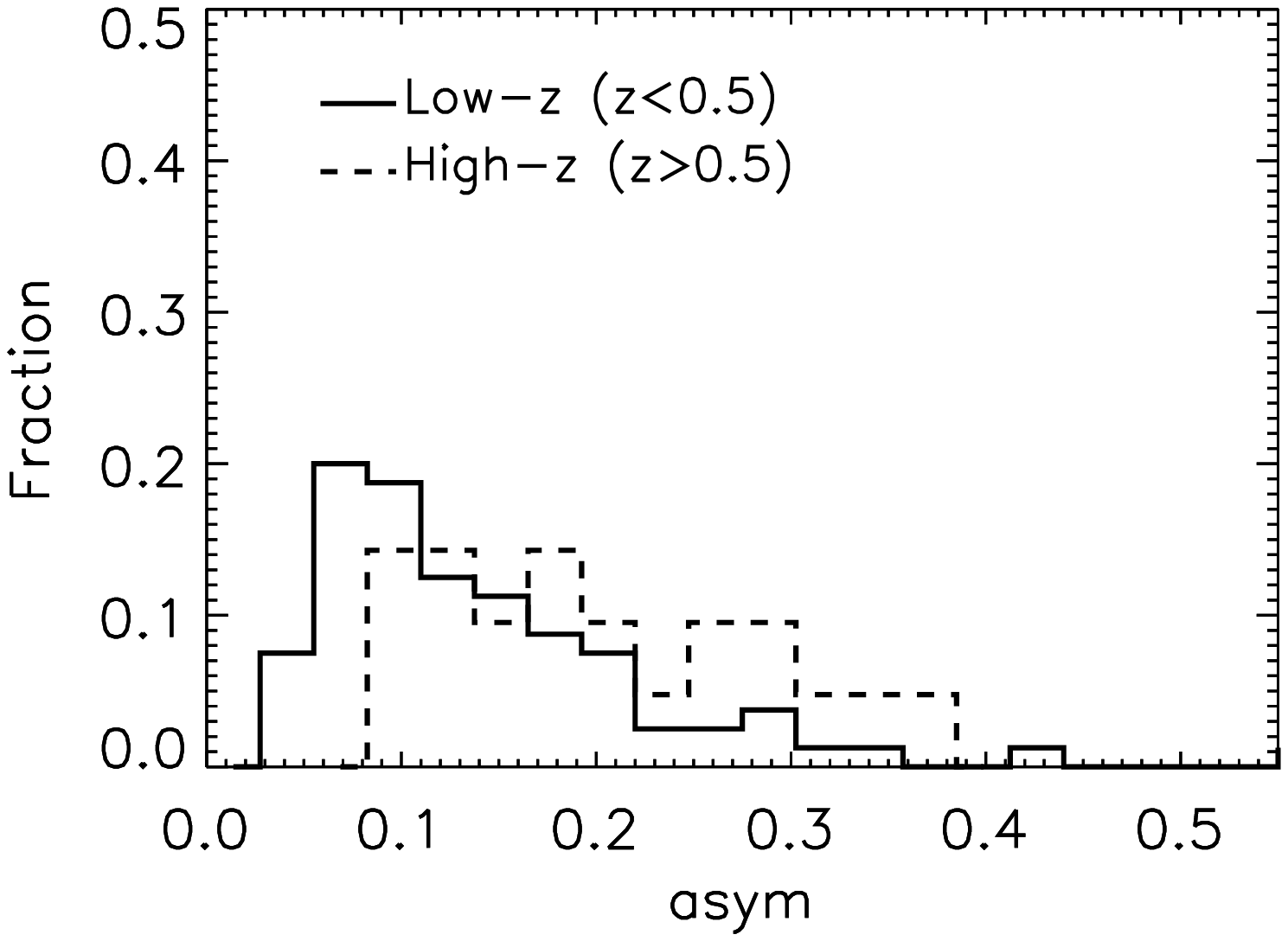}}
 \resizebox{\hsize}{!}{\includegraphics[height=2cm,width=4cm,clip,angle=0]{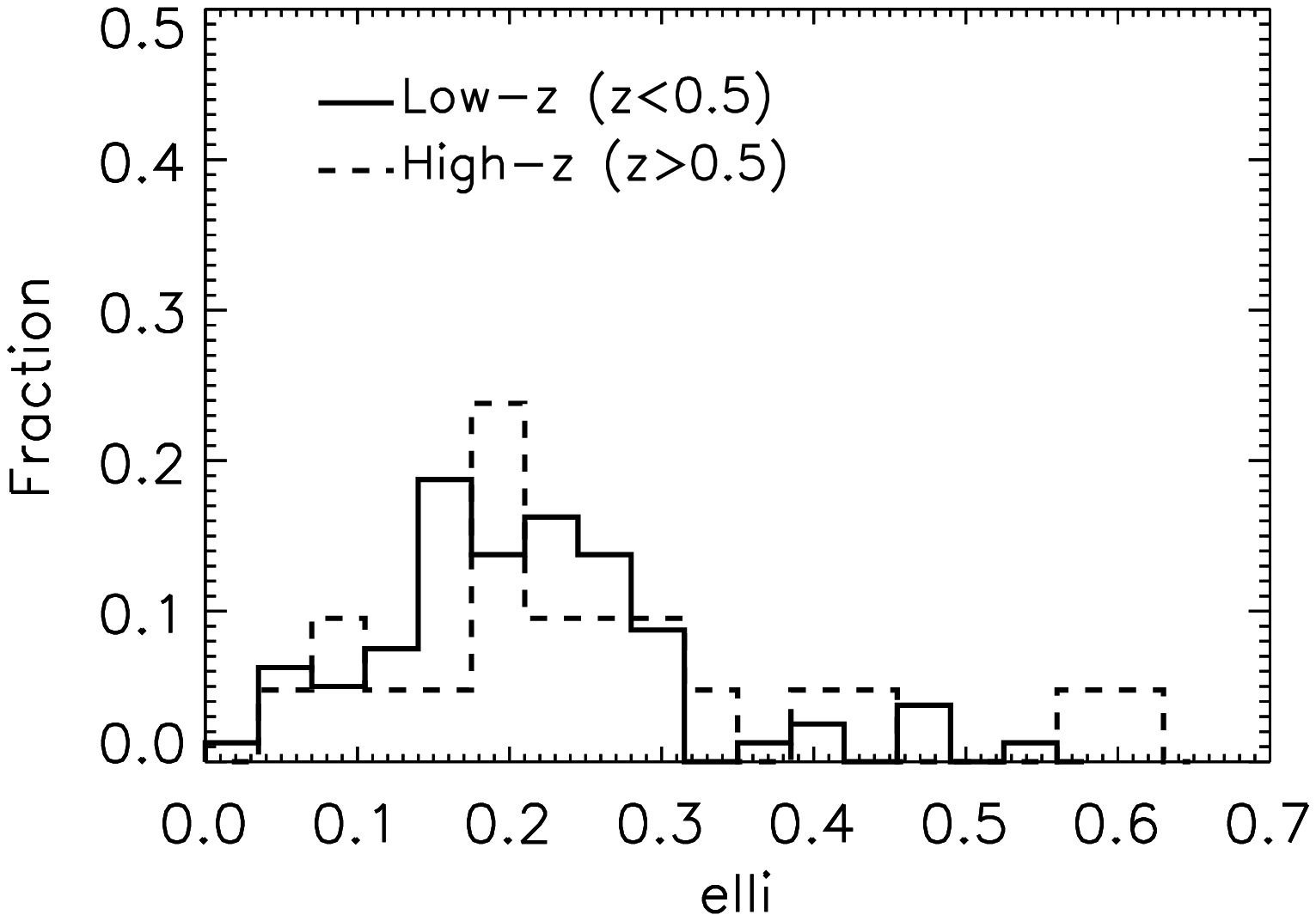}}
 \resizebox{\hsize}{!}{\includegraphics[height=2cm,width=4cm,clip,angle=0]{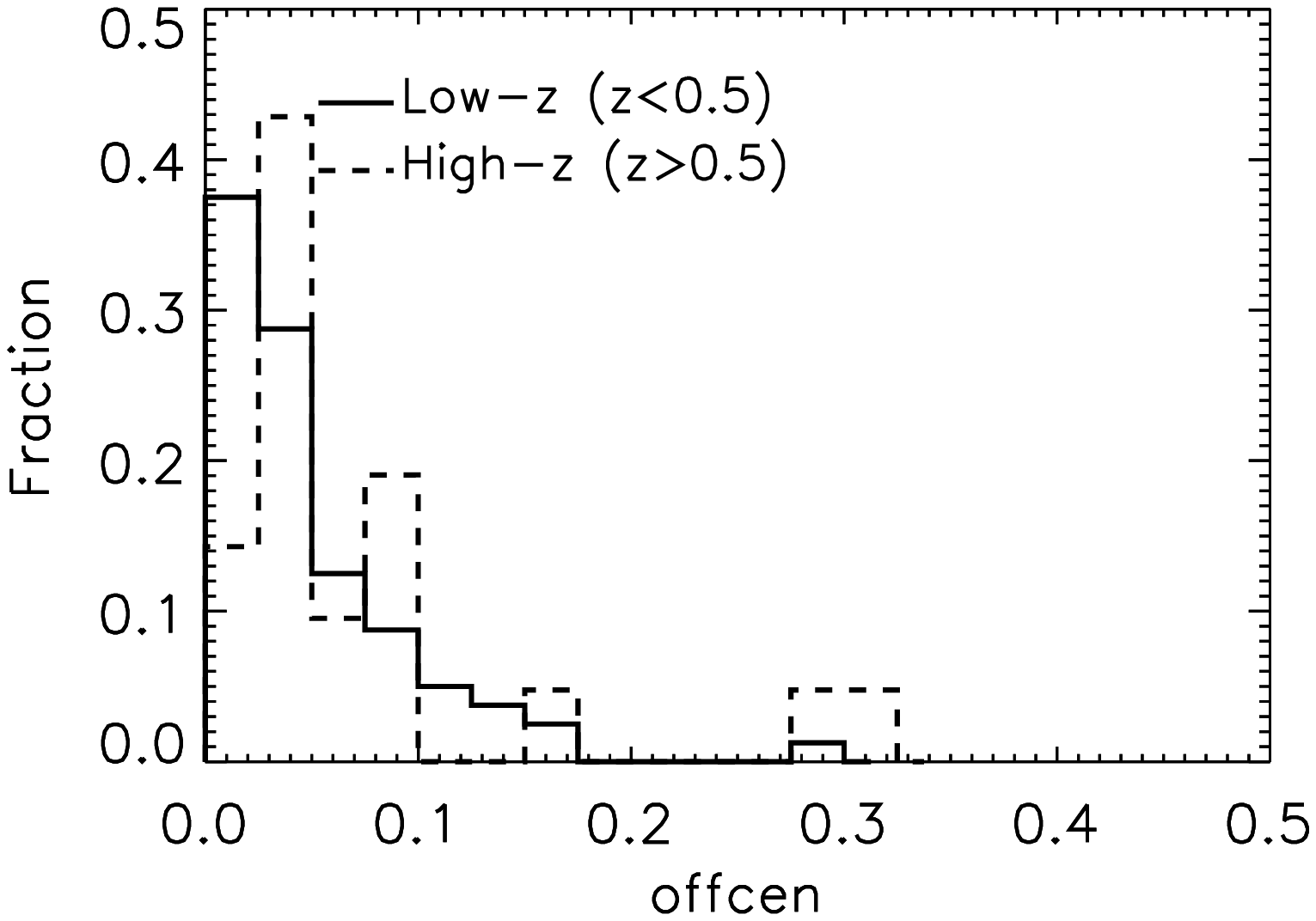}}
 \caption{
Distributions of various morphological  measures for high-z and low-z cluster subsets. 
}
\label{FigTemp}
\end{figure}

\section{Summary and discussions}

Using a sample of 101 clusters at redshift z$\sim$0.05-1 taken from the
Chandra archive,
we quantitatively investigated the relationships between
   the cluster X-ray morphology and various other cluster characteristics.
The X-ray morphology is characterized by
a series of
objectively-measured simple statistics of
X-ray surface brightness,
which are designed to be  robust against 
variations of image quality
  caused by various exposure times
   and various cluster redshifts.

   We found:
   (1) Our measures are robust against variations of image quality
         effects
        introduced by exposure time difference,
        and various cluster redshifts.
   (2) The distorted and non-distorted clusters occupy
      well-defined loci in the L-T plane.
  (3) Ellipticity and Offcenter show no evolutionary effects between high and low redshift cluster subsets,
 while
 there may be a hint of weak evolutions for Conc and Asym,
 in such a way that high-z clusters show more distorted morphology.
   (4) No correlation is found between X-ray morphology and X-ray luminosity, or
       X-ray morphology and X-ray temperature of clusters, 
       implying that interaction of clusters
       may {\it not} enhance or decrease the luminosity or temperature of clusters
       for extended period of time.
   (5) Clusters are scattered and occupy various places in the plane composed of
       two X-ray morphological measures, showing a wide variety of characteristics.
   (6) Relatively strong correlations in the Asym-Conc and Offcen-Conc plots 
     indicate 
     that 
     low concentration clusters generally show high degree of asymmetry or 
    skewness, illustrating
the fact that there are not many highly-extended smooth symmetric clusters.  
Similarly, a correlation between Asym and Elli
may imply that
there are not many highly-elongated but otherwise smooth
symmetric clusters.

   During mergers, clusters are expected to follow a
   complex track in the L-T plane as shown in several 
   numerical simulations (e.g. Ricker \& Sarazin 2001).
   Meanwhile, simulations by  Randall et al. (2002) and Rowley et al. (2004) find that
   even though the temperature and luminosity of a cluster varies significantly
   during a merger, it still follows an L $\sim$ T$^2$ relation.
   Apart from simulations, the actual observational evidence also shows  mixed results.
   For example, a study of a `major merger' cluster A2319 
   (e.g. O'Hara et al. 2004) measuring L \& Tx
   shows  little deviation from the self-similar L-T relation, which 
   is consistent with the fact that bulk properties of clusters either do not change much 
   as a result of mergers or change in a correlated way that maintains the
    small scatter of scaling relations.
   Meanwhile, a study of `merging double' cluster RX J1053.7+5735 (e.g. Hashimoto et al. 2002)
   shows  significant deviation from the L-T relation. 

   One of the major problem of these studies is
   their ambiguous definition of `merger', and resulting heterogeneous sample of
   dynamically unrelaxed clusters. 
   These results are further complicated by 
    possible bias in measuring  L or Tx   
    of clusters with various degree of morphological distortion.
   For example, Lbol 
  may be systematically underestimated for distorted clusters, 
  because the flux from distorted feature
  are typically diffuse and extended, and may be excluded from a 
  luminosity measurement 
 using a finite-sized aperture.
 Tx, on the other hand, can be estimated high without the extended outer part
of distorted clusters. Because it is related to the faint signal from the
outskirt, these biases can  be more prominent in
the measurements with small effective area satellites, such as $ROSAT$ and $ASCA$.
Similarly,
the presence of a `cool core' (e.g. Allen et al. 2001) may bias low 
the cluster global temperature measurements and/or bias high the cluster global luminosity. If the cool core tends to occur in the
dynamically relaxed cluster,
the observed trend
in Fig. 14, 15, and 16 
may partially be the result   
of this underestimated global temperature and/or overestimated luminosity 
of the cool core clusters.
Indeed, several studies (e.g. Fabian et al. 1994; 
Allen \& Fabian 1998;
McCarthy et al. 2004) showed that the clusters with
very large values of `mass deposition rate'
may preferentially lie on the high-luminosity side of the L-T relation.
The cool core clusters may also 
partially be attributed to the lack of difference in the 
width of the L-T distributions
between the distorted and non-distorted clusters 
(Fig. 16),
because of possible enlarged scatter of the L-T distribution for
the non-distorted clusters due to the existence of the 
cool core clusters (e.g.  
Fabian et al. 1994; Markevitch 1998; 
Allen \& Fabian 1998; Reichart et al. 1999),
while the distorted clusters may intrinsically have the large scatter. 

Unfortunately,
most of these studies used the mass deposition rate
to identify their `cooling flow' clusters.
The mass deposition rate is, however, mainly 
based on the information from the surface brightness profile,
roughly equivalent of measuring the morphology
of clusters, similar to  our concentration index.
Thus, previously reported behaviors of the `cooling flow' 
clusters in the
L-T plane may be simply reflecting the `morphology effect'
which is qualitatively consistent with our result,
and may be 
showing nothing more than 
the fact that the clusters with high central surface brightness 
behave differently from those with low central surface brightness.
Indeed McCarthy et al. (2004) reported if they restricted themselves
only to the presence of  central positive temperature gradient
to select their `cooling flow' clusters,
the correlation between the presence of the `cooling flow'  and 
the L-T distribution became  unclear.
Meanwhile studies using simple X-ray core radius of clusters, 
rather than the mass deposition rate,
(e.g. Ota and Mitsuda 2004) reported possible dichotomy in the 
core radius distribution, which may be consistent with 
the fact that high central surface brightness clusters behave
differently from those with low central surface brightness.

Regardless of the influence of the cooling core,
a simple comparison between our morphological measures and the luminosity (or temperature) 
(Fig. 5 \& 6) shows the luminosity or temperature in the literature
is not correlated to the morphological measures,
indicating that 
we have no obvious bias in the measurements of L, or Tx  
for clusters with different degree of morphological distortions.
We have also tested systematics for  various satellites
and various literatures, and found  no apparent systematic bias 
among them.

No matter what causes the shift in the  L-T plane,
the fact that distorted and non-distorted clusters occupy
well-defined loci in the L-T plane
    demonstrates that
    the measurements of the global luminosity and temperature  for distorted
   clusters 
   should be interpreted with caution, or alternatively,
  a rigorous morphological characterization is  necessary
  when we use a sample of clusters with
  heterogeneous morphological characteristics
  to investigate the L-T and other related  scaling relations.
   There is much left to learn about the effect of the cluster dynamical state
   on the bulk properties of clusters of galaxies

\begin{acknowledgements}
We thank Peter Schuecker for useful discussions. 
JPH thanks the Alexander v. Humboldt Foundation for its
generous support.
We acknowledge the referee's comments, which improved the manuscript.
\end{acknowledgements}

\appendix
\section{Derivation of Adaptive Scalings}
 
\subsection{Exposure time effect}
 Intrinsic noise N$_{0}$, assuming Poisson noise, contained  
 in the original unsmoothed image I$_{0}$ (with
background) is:
\begin{eqnarray}
   N_{0} & = & \sqrt{I_{0}}
\end{eqnarray}
 In the intermediate scaled image I$_{1}$ after the scaling, 
 the intrinsic noise contained originally in I$_{0}$
 will be scaled linearly, and then  new Poisson noise
 will be added in quadrature.  Thus the total
 noise N$_{1}$ in the intermediate scaled image I$_{1}$ after
adding the new noise is, 
\begin{eqnarray}
   N_{1} & = & \sqrt{I_{0}I_{1}^{2}/I_{0}^{2} + I_{1}} \nonumber \\
         & = & \sqrt{I_{1}^{2}/I_{0} + I_{1}}
\end{eqnarray}
 Meanwhile, in the final scaled image I$_{2}$,
 the total noise N$_{2}$ will be:
\begin{eqnarray}
   N_{2} & = & \sqrt{I_{2}} 
\end{eqnarray}
Now, the signal-to-noise ratio in the intermediate scaled image
I$_{1}$ (after adding the noise) and the final scaled image I$_{2}$ should be the same, thus,
\begin{eqnarray}
  I_{1}/N_{1} & = & I_{2}/N_{2} \nonumber \\ 
  I_{1}/\sqrt{I_{1}^{2}/I_{0} + I_{1}} & = & I_{2}/\sqrt{I_{2}} 
\end{eqnarray}
Solving for I$_{1}$ gives:
\begin{eqnarray}
  I_{1} &=& \frac{I_{0}^2I_{2}^2}{(I_{0}^2I_{2}-I_{2}^2I_{0})} \nonumber \\     
        &=& \frac{I_{0}I_{2}}{(I_{0}-I_{2})}     
\end{eqnarray}
 However, 
 the scaling from I$_{0}$ to I$_{2}$ is the 
 scaling factor R$_{0}$ described in sec. 4.2.1: 
\begin{eqnarray}
  R_0&=& I_{2}/I_{0} \nonumber\\
     &=&  t1/t0 
\end{eqnarray}
 thus, I$_{1}$ can be obtained from I$_{0}$ and  R$_{0}$:  
\begin{eqnarray}
  I_{1} &=& \frac{I_{0}I_{0}R_0}{(I_{0}-I_{0}R_0)} \nonumber      \\
        &=& I_0\frac{R_0}{(1-R_0)}     
\end{eqnarray}

\subsection{Redshift effect}
\subsubsection{Dimming effect}
 Intrinsic noise N$_{0}$ contained  
 in the original (this time, background subtracted) image I$_{0}$
 is not proportional to I$_{0}$, but proportional 
 to I$_{0}$ + B, even if the background B 
 is already subtracted. 
 Therefore, assuming the noise to be Poissonian,  
 
\begin{eqnarray}
   N_{0} & = & \sqrt{I_{0}+B}
\end{eqnarray}
 In the intermediate scaled image I$_{1}$, 
 the intrinsic noise contained originally in I$_{0}$
 will be scaled linearly, and then  new Poissonian noise
 will be added in quadrature.  Thus the total
 noise N$_{1}$ in the intermediate scaled image I$_{1}$ after
adding the new noise is,
\begin{eqnarray}
   N_{1} & = & \sqrt{(I_{0}+B)I_{1}^{2}/I_{0}^{2} + I_{1}} 
\end{eqnarray}

 Meanwhile, in the final scaled image I$_{2}$,
 the total noise N$_{2}$ should include the contribution from the background B,
 even if the background B is remained subtracted, thus
\begin{eqnarray}
   N_{2} & = & \sqrt{I_{2}+B} 
\end{eqnarray}

Now, the signal-to-noise ratio in the intermediate scaled image
I$_{1}$ (after adding the noise) and the final scaled image I$_{2}$ should be the same, thus,
\begin{eqnarray}
  I_{1}/N_{1} & = & I_{2}/N_{2} \\ 
  I_{1}/\sqrt{(I_{0}+B)I_{1}^{2}/I_{0}^{2} + I_{1}} & = & I_{2}/\sqrt{I_{2}+B} 
\end{eqnarray}
Solving for I$_{1}$ gives,  
\begin{eqnarray}
  I_{1} &=& \frac{I_{0}^2I_{2}^2}{[I_{0}^2(I_{2}+B)-I_{2}^2(I_{0}+B)]}     
\end{eqnarray}
 However, 
 the scaling from I$_{0}$ to I$_{2}$ is the 
 scaling factor R$_{1}$ described in sec. 4.2.2: 
\begin{eqnarray}
  R_1&=& I_{2}/I_{0} \nonumber\\
     &=&[(1+z0)/(1+z1)]^4 
\end{eqnarray}
 thus, I$_{1}$ can be obtained from I$_{0}$, R$_{1}$, and B:  

\begin{eqnarray}
  I_{1} &=& \frac{I_{0}^2R_1^2}{[I_{0}R_1+B-R_1^2(I_{0}+B)]}    
\end{eqnarray}


\subsubsection{Angular-size effect}

 The noise N$_{2}$ contained  
 in the dimmed, background re-added, and rebinned image I$^{''}_{2}$
 is proportional to I$^{''}_{2}$. 
 Therefore, similarly to the appendix A.1., 
the total noise N$_3$ 
 in the intermediate scaled image I$_3$ 
after adding the new Poisson noise will be:
\begin{eqnarray}
   N_{3} & = & \sqrt{I^{''}_{2}I_{3}^{2}/I_{2}^{''2} + I_{3}} \nonumber \\
         & = & \sqrt{I_{3}^{2}/I^{''}_{2} + I_{3}}
\end{eqnarray}
 Meanwhile, in the final scaled image I$_4$, the total noise N$_4$ 
 will be:
\begin{eqnarray}
   N_{4} & = & \sqrt{I_{4}}
\end{eqnarray}
The signal-to-noise ratio in the intermediate scaled image
I$_{3}$ (after adding the noise) and the final scaled image I$_{4}$ should be the same, thus,
\begin{eqnarray}
  I_{3}/N_{3} & = & I_{4}/N_{4} \nonumber \\ 
  I_{3}/\sqrt{I_{3}^{2}/I^{''}_{2} + I_{3}} & = & I_4/\sqrt{I_4}
\end{eqnarray}
Solving for I$_{3}$ gives:
\begin{eqnarray}
  I_{3} &=& \frac{I_{2}^{''}I_{4}}{(I_{2}^{''}-I_{4})}     
\end{eqnarray}
 Now,  in order 
to scale the final image I$_4$ while conserving the surface brightness with
respect to the `pre-rebinned' image (i.e. I$_2^{'}$), 
\begin{eqnarray}
  I_{4} &=&  I_2^{''}/R_2^2       
\end{eqnarray}
 thus, I$_{3}$ can be obtained from I$^{''}_{2}$ and  R$_{2}$:
\begin{eqnarray}
  I_{3} &=& \frac{I_{2}^{''}I_{2}^{''}/R_2^2}{(I_{2}^{''}-I_{2}^{''}/R_2^2)} \nonumber      \\
        &=& I_2^{''}\frac{1}{(R_2^2-1)}     
\end{eqnarray}

\subsection{Combining the exposure and redshift effects}
Starting from the equation A.19, namely,
\begin{eqnarray}
  I_{3} &=& \frac{I_{2}^{''}I_{4}}{(I_{2}^{''}-I_{4})} \nonumber 
\end{eqnarray}
Then, unlike the equation A.20, 
the final image I$_4$ is now related to the original pre-rebinned 
image I$_2^{'}$ by the new factor R$_3$: 
\begin{eqnarray}
  R_3 & = &  I_4/I_2^{'} \nonumber \\
      & = &  \frac {I_4 }{I_2^{''}/R_2^2} = t2/t0 \nonumber     
\end{eqnarray}
So,  
\begin{eqnarray}
  I_{4}/R_3 &=& I_2^{''}/R_2^2       
\end{eqnarray}
instead of I$_{4}$ = I$_2^{''}$/R$_2^2$  in the Eq. A.20.  
Thus, I$_{3}$ can be obtained from I$^{''}_{2}$, R$_3$, and  R$_{2}$:
\begin{eqnarray}
  I_{3} &=& \frac{I_{2}^{''}I_{2}^{''}R_3/R_2^2}{(I_{2}^{''}-I_{2}^{''}R_3/R_2^2)} \nonumber      \\
        &=& I_2^{''}\frac{R_3}{(R_2^2-R_3)}     
\end{eqnarray}

\end{document}